\documentclass[aoas,preprint]{imsart}

\RequirePackage[OT1]{fontenc}
\RequirePackage{amsthm,amsmath}
\RequirePackage{natbib}
\RequirePackage[colorlinks,citecolor=blue,urlcolor=blue]{hyperref}
\arxiv{arXiv:0000.0000}

\usepackage[margin=1.65in]{geometry}

\usepackage{amssymb}
\usepackage{amsmath}
\usepackage{graphicx}%
\usepackage{amsfonts}%
\usepackage{float}
\usepackage{natbib}
\usepackage{xr}
\usepackage{hyperref}
\newcommand{\bet}{ \mbox{\boldmath $ \eta $} }

\newcommand{\bbeta}{ \mbox{\boldmath $ \beta $} }

\newcommand{\balpha}{ \mbox{\boldmath $ \alpha $} }

\newcommand{\bepsilon}{ \mbox{\boldmath $\epsilon$}}

\newcommand{\btheta}{ \mbox{\boldmath $ \theta $} }

\newcommand{\bmu}{ \mbox{\boldmath $\mu$} }

\newcommand{\bLambda}{ \mbox{\boldmath $\Lambda$} }

\newcommand{\bgamma}{ \mbox{\boldmath $\gamma$} }

\newcommand{\boeta}{ \mbox{\boldmath $\eta$} }

\newcommand{\bzero}{\textbf{0}}
\newcommand{\bone}{\textbf{1}}

\newcommand{\bz}{\textbf{z}}
\newcommand{\ba}{\textbf{a}}
\newcommand{\bA}{\textbf{A}}

\newcommand{\bk}{\textbf{k}}
\newcommand{\bK}{\textbf{K}}

\newcommand{\bI}{\textbf{I}}

\newcommand{\bM}{\textbf{M}}

\newcommand{\bP}{\textbf{P}}
\newcommand{\bs}{\textbf{s}}

\newcommand{\bV}{\textbf{V}}
\newcommand{\bw}{\textbf{w}}

\newcommand{\bx}{\textbf{x}}
\newcommand{\bX}{\textbf{X}}

\newcommand{\bY}{\textbf{Y}}

\begin{document}

\begin{frontmatter}
\title{Spatial Functional Data Modeling of Plant Reflectances}
\runtitle{Spatial Modeling of Reflectance}

\begin{aug}
\author{\fnms{Philip A.} \snm{White}\thanksref{t1,m1}\ead[label=e1]{pwhite@stat.byu.edu}},
\author{\fnms{Henry} \snm{Frye}\thanksref{m2}},
\author{\fnms{Michael F.} \snm{Christensen}\thanksref{m3}},
\author{\fnms{Alan E.} \snm{Gelfand}\thanksref{m3}},
\and
\author{\fnms{John A.} \snm{Silander, Jr.}\thanksref{m2}}

\thankstext{t1}{Corresponding Author}
\runauthor{P. White et al.}

\affiliation{Brigham Young University \thanksmark{m1}, University of Connecticut\thanksmark{m2}, and Duke University\thanksmark{m3}}

%
\end{aug}

\vspace*{0.5cm}
\begin{abstract}

Plant reflectance spectra – the profile of light reflected by leaves across different wavelengths - supply the spectral signature for a species at a spatial location to enable estimation of functional and taxonomic diversity for plants. We consider leaf spectra as ``responses'' to be explained spatially. These spectra/reflectances are functions over a wavelength band that respond to the environment.

Our motivating data are gathered for several families from the Cape Floristic Region (CFR) in South Africa and lead us to develop rich novel spatial models that can explain spectra for genera within families.  Wavelength responses for an individual leaf are viewed as a function of wavelength, leading to functional data modeling.  Local environmental features become covariates. We introduce wavelength - covariate interaction since the response to environmental regressors may vary with wavelength, so may variance. Formal spatial modeling enables prediction of reflectances for genera at unobserved locations with known environmental features. We incorporate spatial dependence, wavelength dependence, and space-wavelength interaction (in the spirit of space-time interaction). We implement out-of-sample validation to select a best model, discovering that the model features listed above are all informative for the functional data analysis.  We then supply interpretation of the results under the selected model.

\end{abstract}


\end{frontmatter}

\section{Introduction}
\label{sec:Intro}

The reflectance of the surface of a material is the fraction of incident electromagnetic radiation reflected at the surface. It is a function of the wavelength (or frequency) of the light, its polarization, and the angle of incidence. The reflectance as a function of wavelength is called a reflectance spectrum. The literature on reflectances is substantial, with a large portion focused on the interaction of electromagnetic energy with the atmosphere and terrestrial objects, e.g., reflectances associated with different land cover/vegetation types. Typically, they are gathered by satellites, aircraft, and ground-level sensors.  The focus of this manuscript is on plant reflectances, i.e., data gathered for plants at leaf level.



The importance of leaf level reflectance modeling arises because the scales at which remote sensing devices detect reflectance spectra often do not match those relevant to ecological scales \citep{gamon_consideration_2020}. For example, a satellite imager can measure the reflectance signal for an entire $30 \textrm{m}^2$ pixel; the reflectance signal of this pixel would be a composite of all the different spectral signatures of the plant species within that area. To disentangle what plants are on the ground, remote sensing scientists use spectral unmixing techniques which rely on spectral libraries \citep{quintano_spectral_2012, shi_incorporating_2014}. These libraries are collections of pure endmembers, i.e., the pure reflectance spectra of leaf surfaces, which serve as representative spectra for different plant functional types, plant taxonomic groups, and/or individual plant species. Use of such leaf spectral libraries in ecology or biodiversity science has been termed as a ``spectranomic'' approach \citep{asner_spectranomics_2016}. 
Being able to statistically predict leaf reflectance spectra across environmental gradients constitutes a major advancement because it enables prediction-based spectral libraries that could be used in the validation and inference of remote sensing data at large spatial extents. Large hyperspectral remote sensing efforts are already under way, e.g., NASA’s Surface Biology Geology satellite mission \citep{cawse-nicholson_nasas_2021}, making pressing the need to predict spectral signals of plants.

Furthermore, leaf-level spectra have become an invaluable tool to capture the diversity in leaf traits that have accumulated over the course of seed plant evolution \citep{reich2003evolution,cornwell2014functional} enabling estimation of functional diversity \citep{kokaly2009,schneider2017} and taxonomic diversity  \citep{clark2005,cavender2016}.  They provide drivers for ecosystem processes \citep{schweiger2018} and guide conservation \citep{asner2017airborne}.

Statistical analyses of plant reflectance spectra have been limited to treating spectra as functional predictors of scalar variables such as plant traits. That is, we are in the realm of functional linear regression modeling. Modeling approaches rely on dimension reduction, e.g., spline basis representation \citep{ordonez2010functional}, partial least squares regression \citep{doughty_can_2017}, partial least squares-discriminant analysis \citep{cavender-bares_associations_2016}, or some form of machine learning \citep{feret_estimating_2019}. 

Some approaches generate hypothetical leaf reflectance spectra from physical first principles, i.e., radiative transfer models \citep{jacquemoud_prospect:_1990,jacquemoud_modeling_2019}.  However, these are not functional response models driven by environment.  That is, our contribution is to model plant reflectance curves (as a function of wavelength) as a functional response variable, in particular, at genus level within family. 
We incorporate local spatial environmental covariates as regressors. 

We introduce the following innovations, motivated by careful exploratory data analysis. We specify a functional data model incorporating spatial random effects to predict reflectance curves at genera level, at locations without collected plant samples. We also capture wavelength dependence through random effects. For further enrichment, we add space-wavelength interaction (in the spirit of space-time interaction) by constructing a space-wavelength random effect through wavelength kernel convolutions of spatial Gaussian processes. In general, this random effect has nonseparable covariance and is wavelength nonstationary. Additionally, we model the variance to be heterogeneous across wavelengths. Also, expecting that the reflectance response to environmental regressors may vary with wavelength, we include wavelength - covariate interactions. 
Lastly, because the rich space-wavelength modeling essentially annihilates the significance of the covariate/wavelength effects, we present a novel orthogonalization to remove spatial and functional confounding between random effects and environmental regressors.

Functional data analysis (FDA) is well established for analyzing data representing curves/surfaces varying over a continuum. 
The physical continuum over which these functions are defined is often time but here, it is wavelength.  Pioneering work for FDA is attributed to Ramsey and Silverman \citep[e.g.,][]{ramsay2005,ramsay2007applied}. The field has undergone rapid growth, and numerous applications have been found in areas such as imaging \citep{locantore1999robust} (including MRI brain imaging \citep{tian2010cortical}), finance \citep{laukaitis2008functional}, climatic variation \citep{besse2000autoregressive}, spectrometry data \citep{reiss2007functional}, and time-course
gene expression data \citep{leng2006classification}. For a more comprehensive overview of applications, see \cite{ullah2013}. 

Explicit modeling of functional data is usually carried out by specifying functions in one of two ways: (i) as finite linear combinations of some set of basis functions or (ii) as realizations of some stochastic process.  A key feature of functional data analysis implementation is some version of dimension reduction to specify functions.  Here, we have random functions over a wavelength span as well as over a spatial region.  We combine both approaches, using basis functions over wavelength with process realizations over space to build space by wavelength regressions over environment. 

We work with plant reflectances gathered from the Cape Floristic Region (CFR) in South Africa.  We present an extensive cross-validation study for model selection across a rich collection of models to demonstrate the ability of our space-wavelength modeling to predict reflectances well for genera within a family at unobserved locations.  We present and discuss our findings for three plant families found within the CFR.

The format of the paper is as follows.  Section \ref{sec:data} describes the collected data. Section \ref{sec:EDA} undertakes a broad exploratory data analysis to motivate the features we incorporate in our modeling.  Section \ref{sec:mod} explains our modeling, model comparison, and presents a novel orthogonalization for functional regression coefficients.  Section \ref{sec:analysis} presents the analysis of the CFR data while a brief Section \ref{sec:conc} offers a summary and suggestion for future work.  Substantial detail of our exploratory analysis, as well as model sensitivity analysis, has been placed in the Supplemental Material.

\section{The Dataset}\label{sec:data}


We work with plant reflectances gathered from the Cape Floristic Region (CFR) in South Africa, see Figure \ref{fig:locs}. Reflectances were measured with a USB-4000 Spectrometer (manufactured by Ocean Optics) using a leaf clip attachment. Sun leaves from the top of each selected canopy were measured. The spectrometer has a range of 450-950 nanometers (nm) with a total of 500 reflectance measurements. We study plant reflectance viewed as a function of wavelength $t$, across the window $t \in [450,950)$, typically referred to as a spectral signature.


With interest in a spatial model for plant reflectance that enables prediction of reflectance for genera within a family at unobserved locations, we work with adjacent subregions of the CFR characterized by a fynbos landscape, known as the Hantam-Tanqua-Rogeveld (HTR) and Cederberg. Three prevalent families that often characterize landscapes in this area are Aizoaceae, Asteraceae, and Restionaceae \citep{slingsby2014functions}.  These families have broad overlap in their reflectances (Figure \ref{fig:curves_aster}). However, a linear discriminant analysis (LDA) to predict these families based on their reflectances yields clear separation of the groups, demonstrating that reflectances can be used to effectively predict taxonomic differences.
More precisely, a structured classification by family using plant samples was conducted employing LDA and reveals the separation between families (See Supplemental Material for full details).  

Much of the observed variation across the three families is likely due to differences in composite leaf traits though isolating the relative impact of each trait on the reflectances is beyond our intentions here. However, it is established that reflectance variation is a signal of leaf trait variation (e.g., anatomical, physiological, and structural traits \citep{jacquemoud_ustin_2019b}) and can be influenced by the environmental factors (e.g., climate and soil) that the plants inhabit.


\section{Exploratory Data Analysis and Modeling}\label{sec:EDA}


We explore the characteristics of plant reflectances for the three families given above (Aizoaceae, Asteraceae, Restionaceae) in the HTR and Cederberg areas. We retain the entire dataset because it is somewhat small from a spatial perspective.  Note that the domains for the three families do not overlap well (Figure \ref{fig:locs}) so we will fit each family separately when implementing our spatial modeling.

The number of genera with observed reflectances within each family is:  Aizoaceae - 16, Asteraceae - 38, and Restionaceae - 10.  The Supplemental Material provides: (i) the proportion of sites where each family is present, (ii) the number of sites with one, two, or three families, and (iii) a more detailed breakdown of family co-occurrence. To summarize, Aizoaceae and Restionaceae rarely co-occur; in fact, Restionaceae is mostly limited to the Cederberg region apart from a few HTR observations. Replication at the genus level is uncommon and even more uncommon at the species level.  

\subsection{Data Locations}

\begin{figure}[H]
 \begin{center}
 \includegraphics[width = 0.48\textwidth]{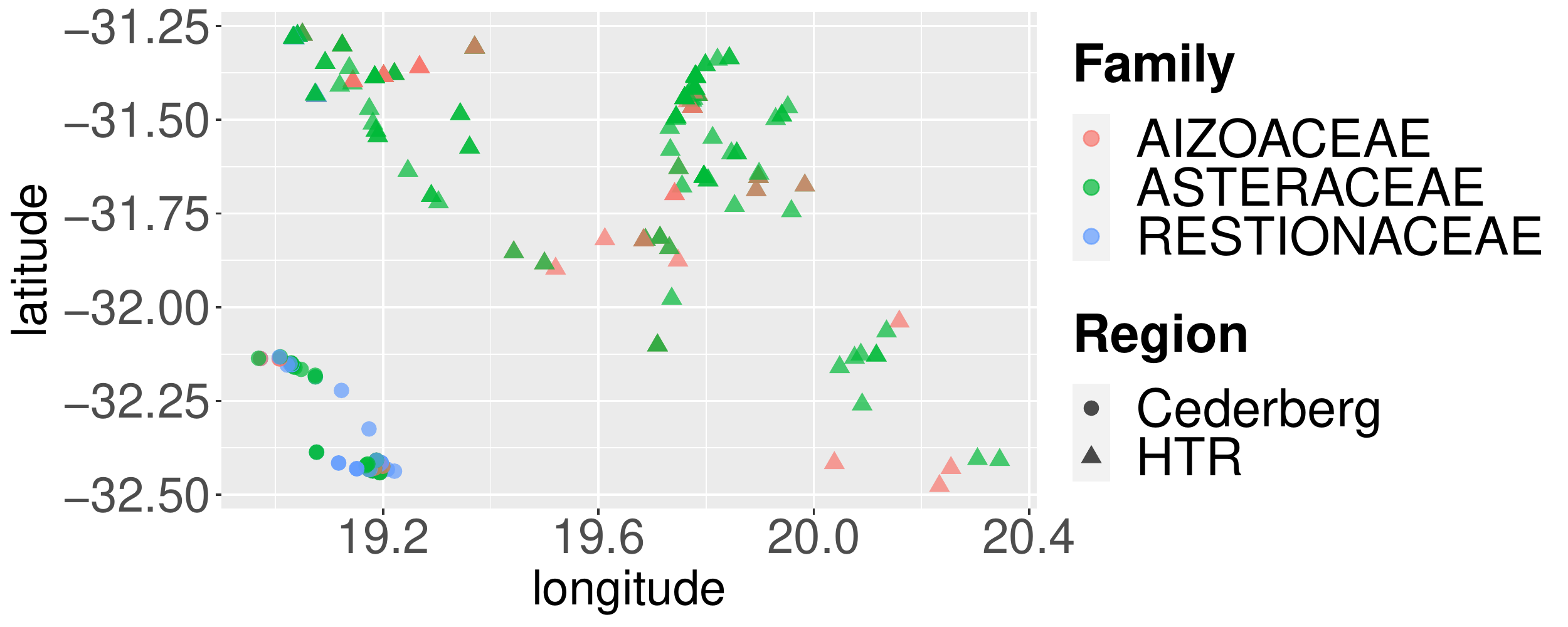}
  \includegraphics[width = 0.48\textwidth]{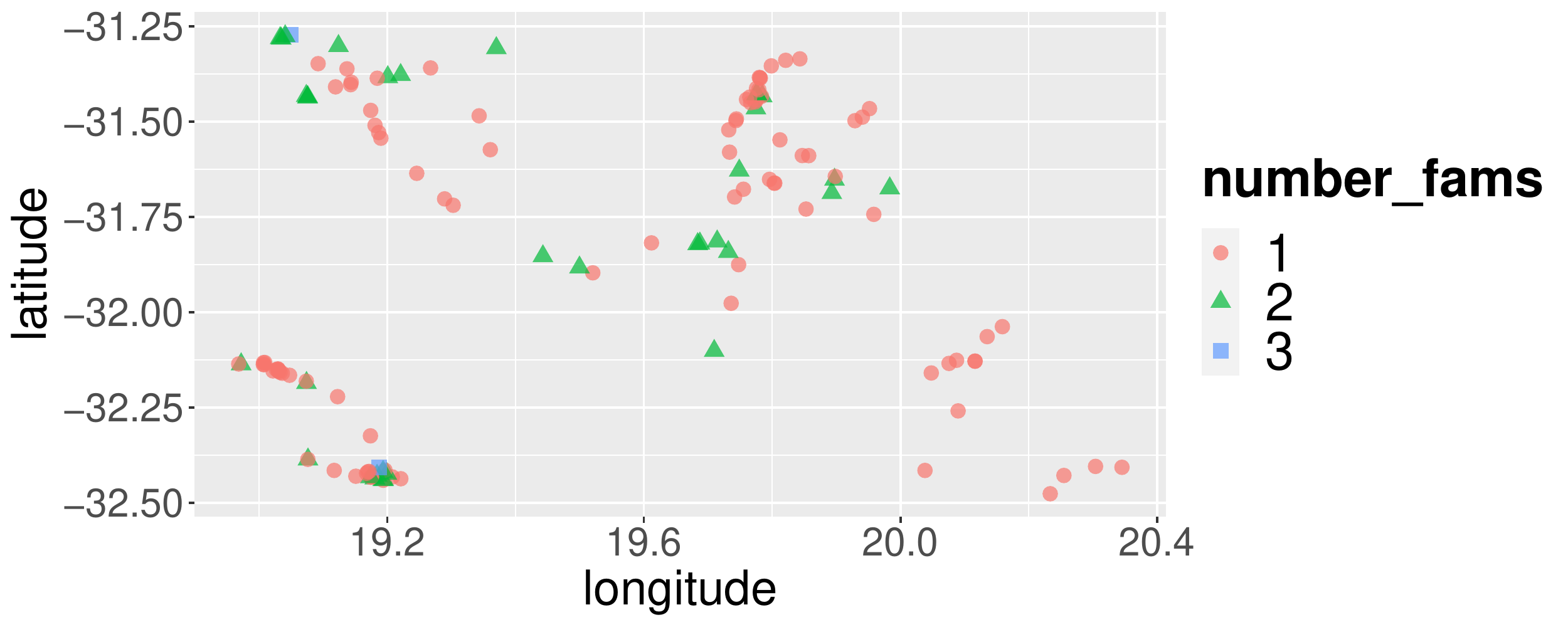}
 \end{center}
 
  \vspace{-4mm}
 
 \caption{Locations (Left) colored by family with region-specific shapes and (Right) colored by the number of families observed at the site.}\label{fig:locs}
 \end{figure}

In Figure \ref{fig:locs}, we show all locations, where reflectances are observed with sites coded by region (shape) and family (color). We also plot locations coded by the number of families observed at that site. 
In the HTR and Cederberg regions, only 22 of the 133 sites have more than one reflectance spectrum for a given species; only 27 of the 183 total species (across all families) are observed at more than one site. This suggests that species-level modeling is infeasible.
The Supplemental Material offers more commentary on data locations and duplication.  


\subsection{Reflectance spectra and Environmental Variables}

\begin{figure}[ht]
 \begin{center}
 \includegraphics[width = 0.4\textwidth]{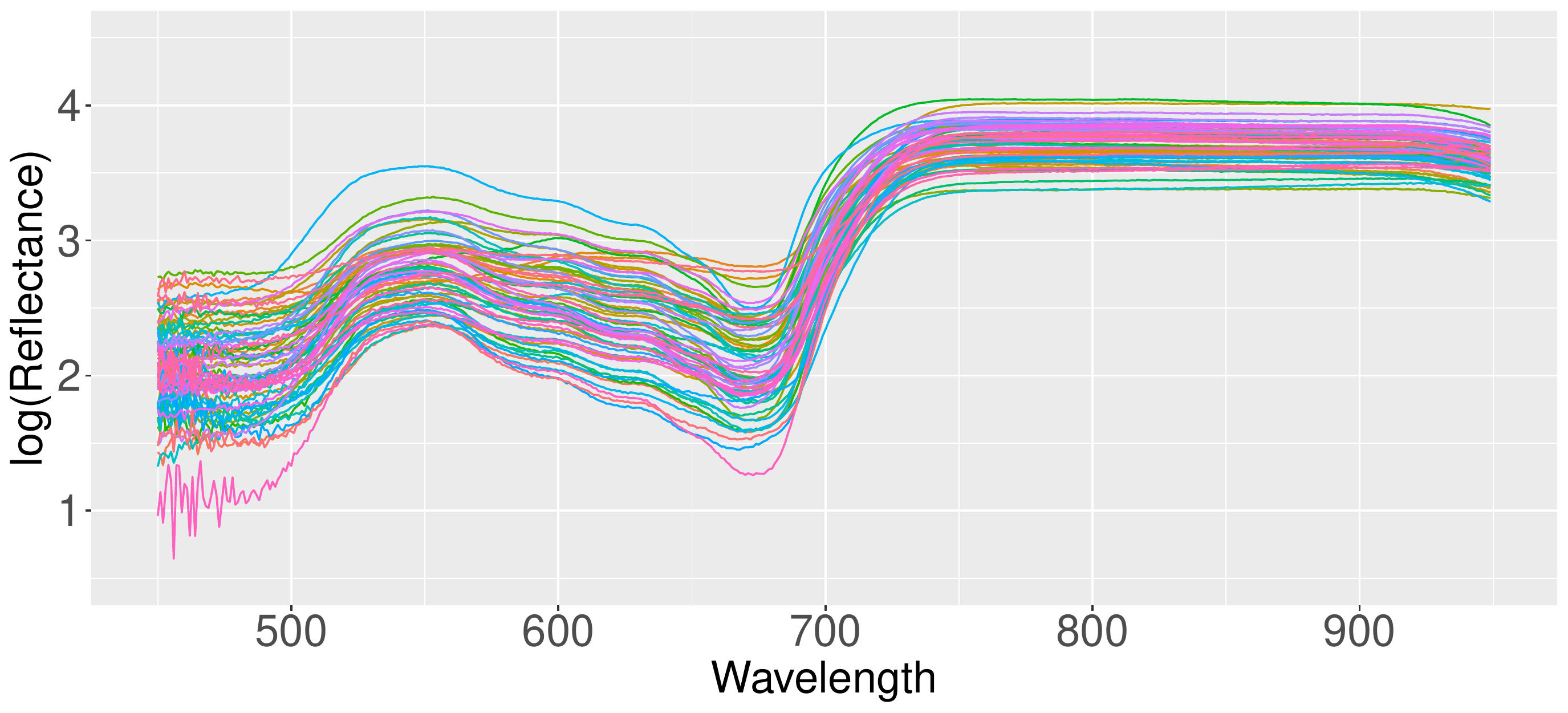}
  \includegraphics[width = 0.4\textwidth]{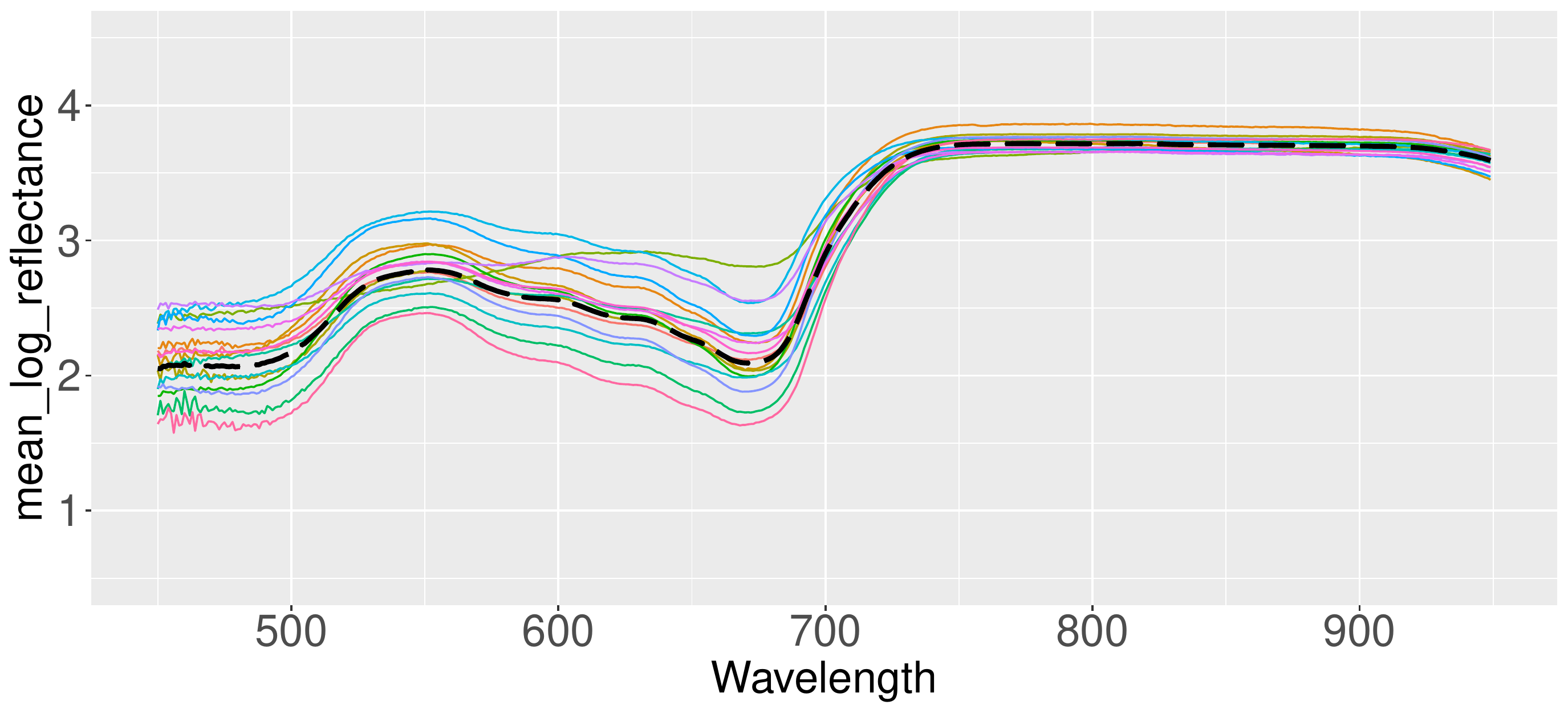}
 \includegraphics[width = 0.4\textwidth]{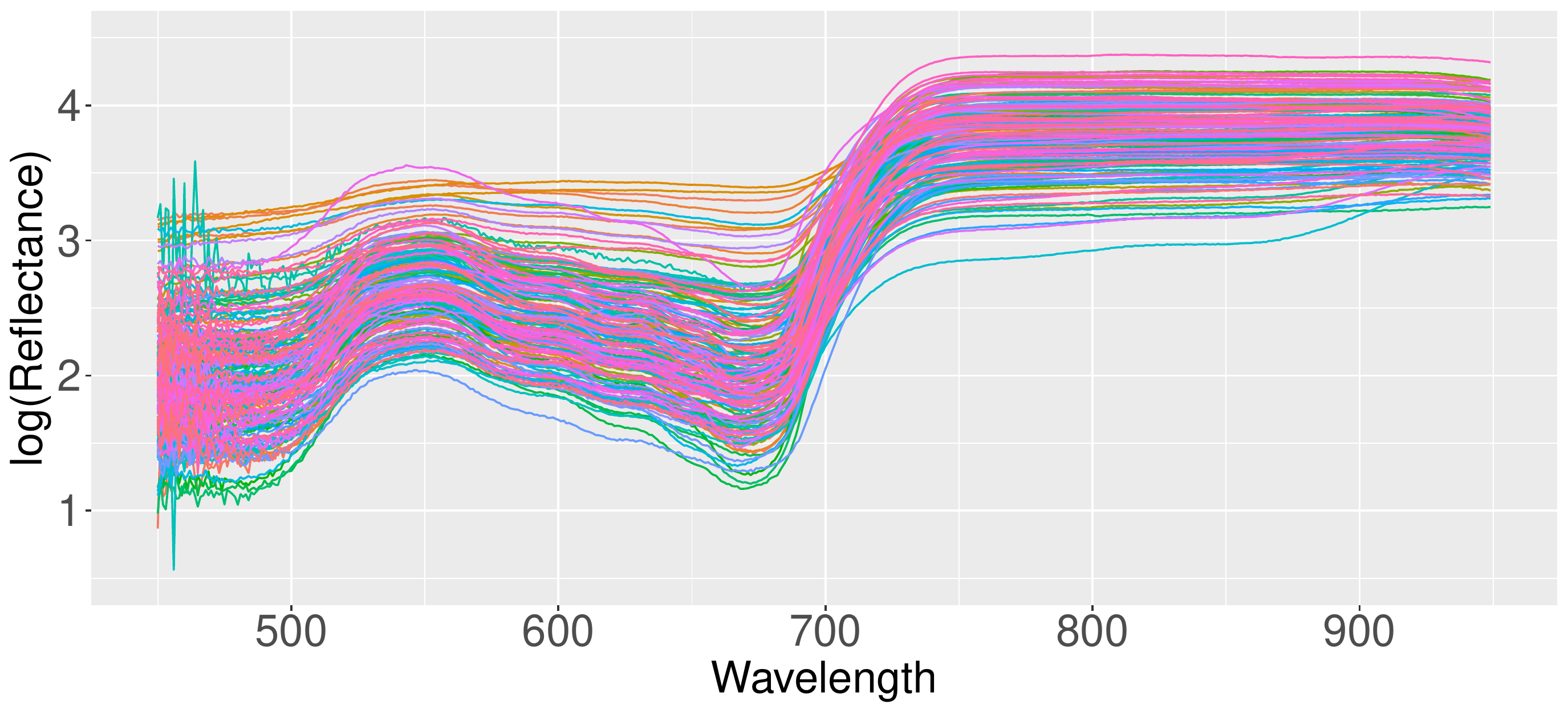}
  \includegraphics[width = 0.4\textwidth]{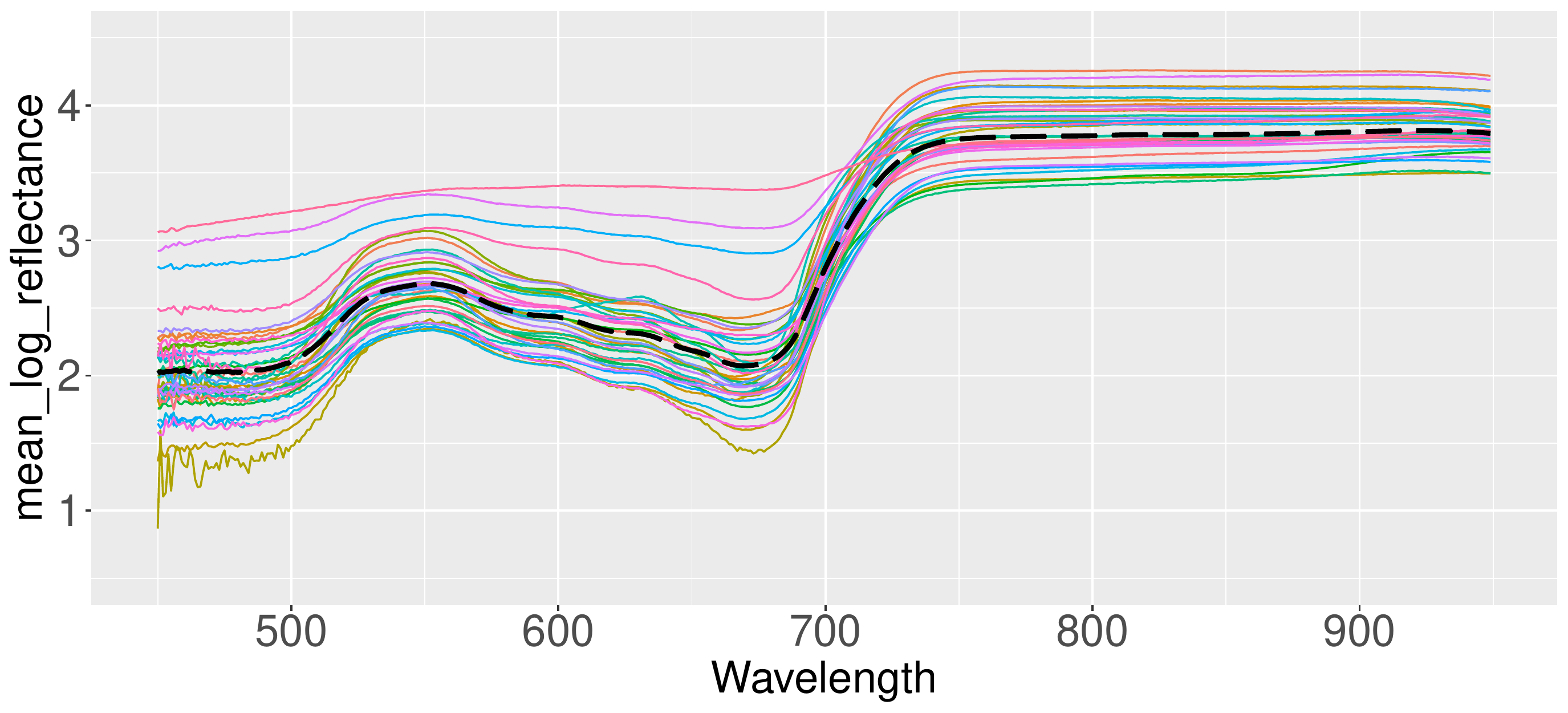}
 \includegraphics[width = 0.4\textwidth]{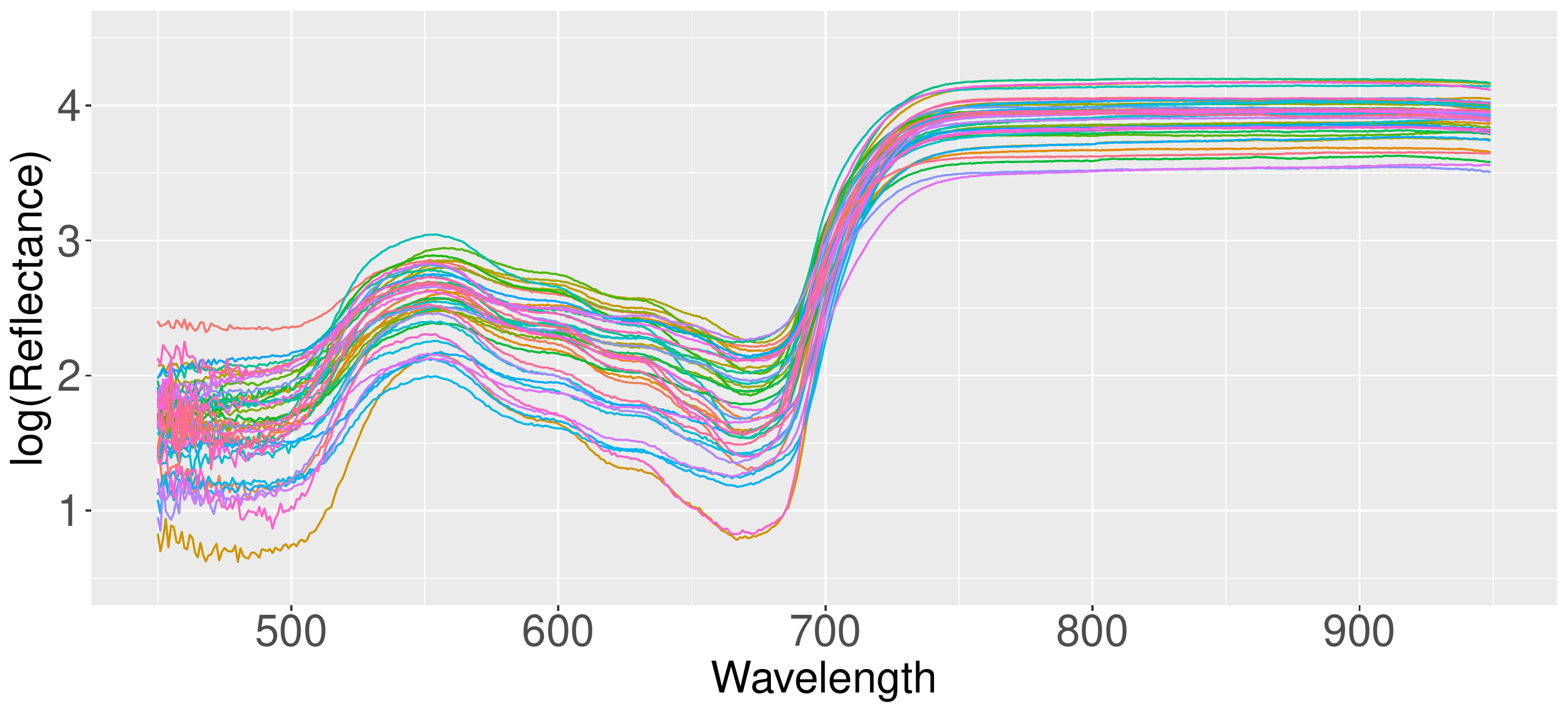}
  \includegraphics[width = 0.4\textwidth]{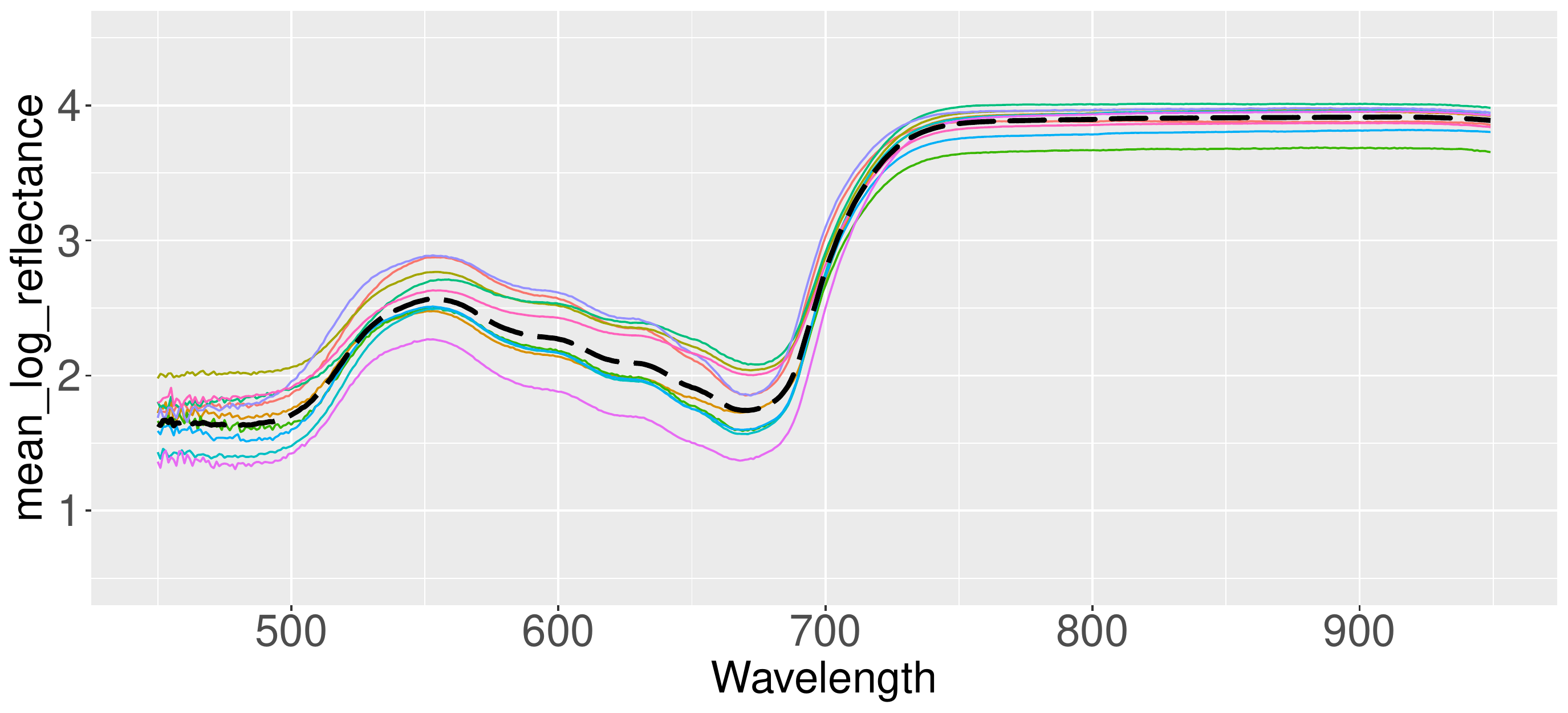}

 \end{center}
 
 \vspace{-4mm}
 
 \caption{Log reflectances for Aizoaceae (Top-Left) All Curves, (Top-Right) Genus-specific means; Asteraceae (Middle-Left) All Curves, (Middle-Right) Genus-specific means; Restionaceae (Bottom-Left) All Curves, (Bottom-Right) Genus-specific means.}\label{fig:curves_aster}
 \end{figure}

To visualize the form and variability in reflectance spectra, we plot all of the curves by family in Figure \ref{fig:curves_aster} along with plots of the genus-specific means. We can see that the family-specific means do not capture the spread of the variability seen in all the curves while the genus-specific means show nearly the same variability for all of the curves. 

\begin{figure}[ht]
 \begin{center}
 \includegraphics[width = 0.365\textwidth]{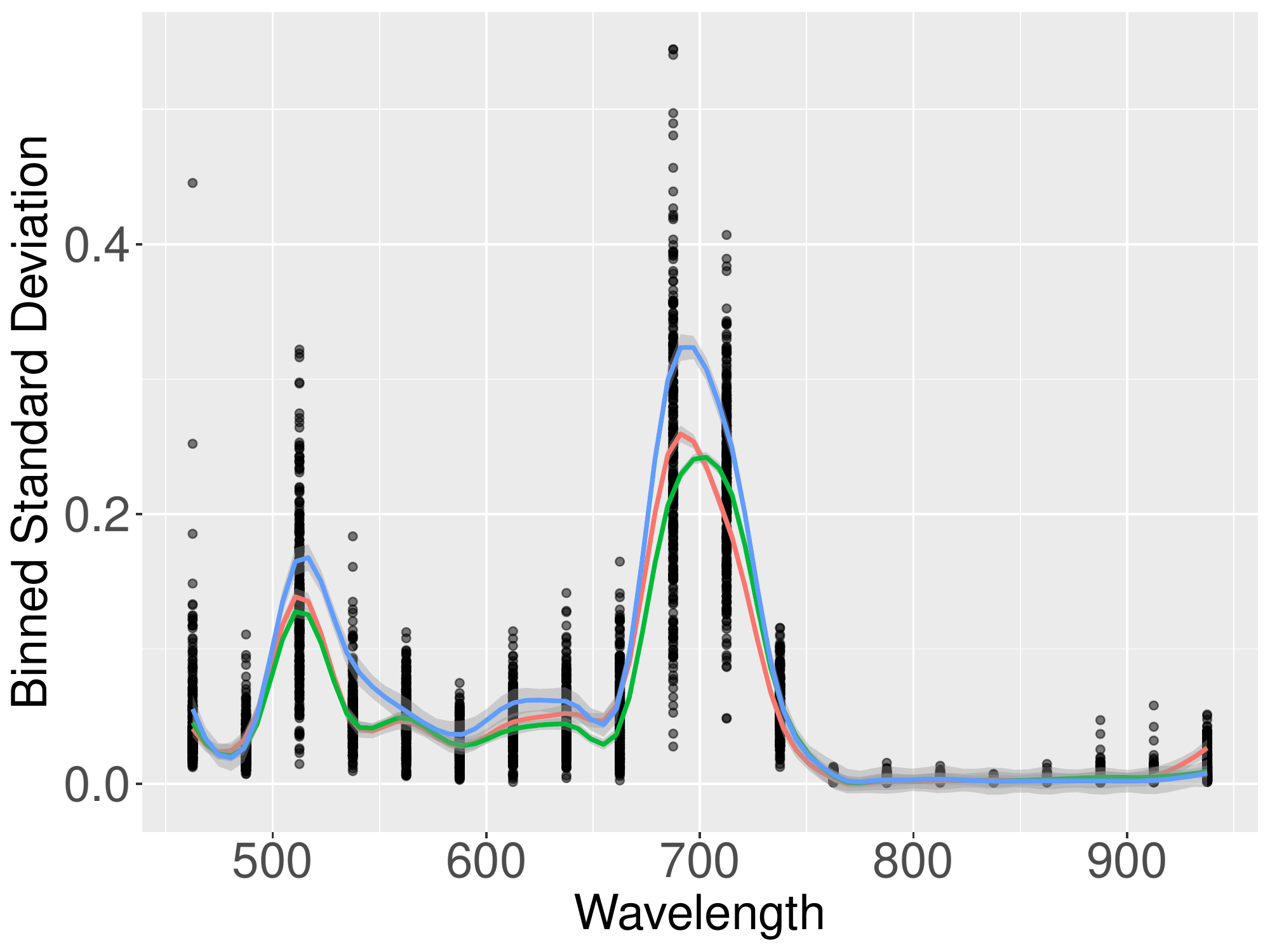}
  \includegraphics[width = 0.615\textwidth]{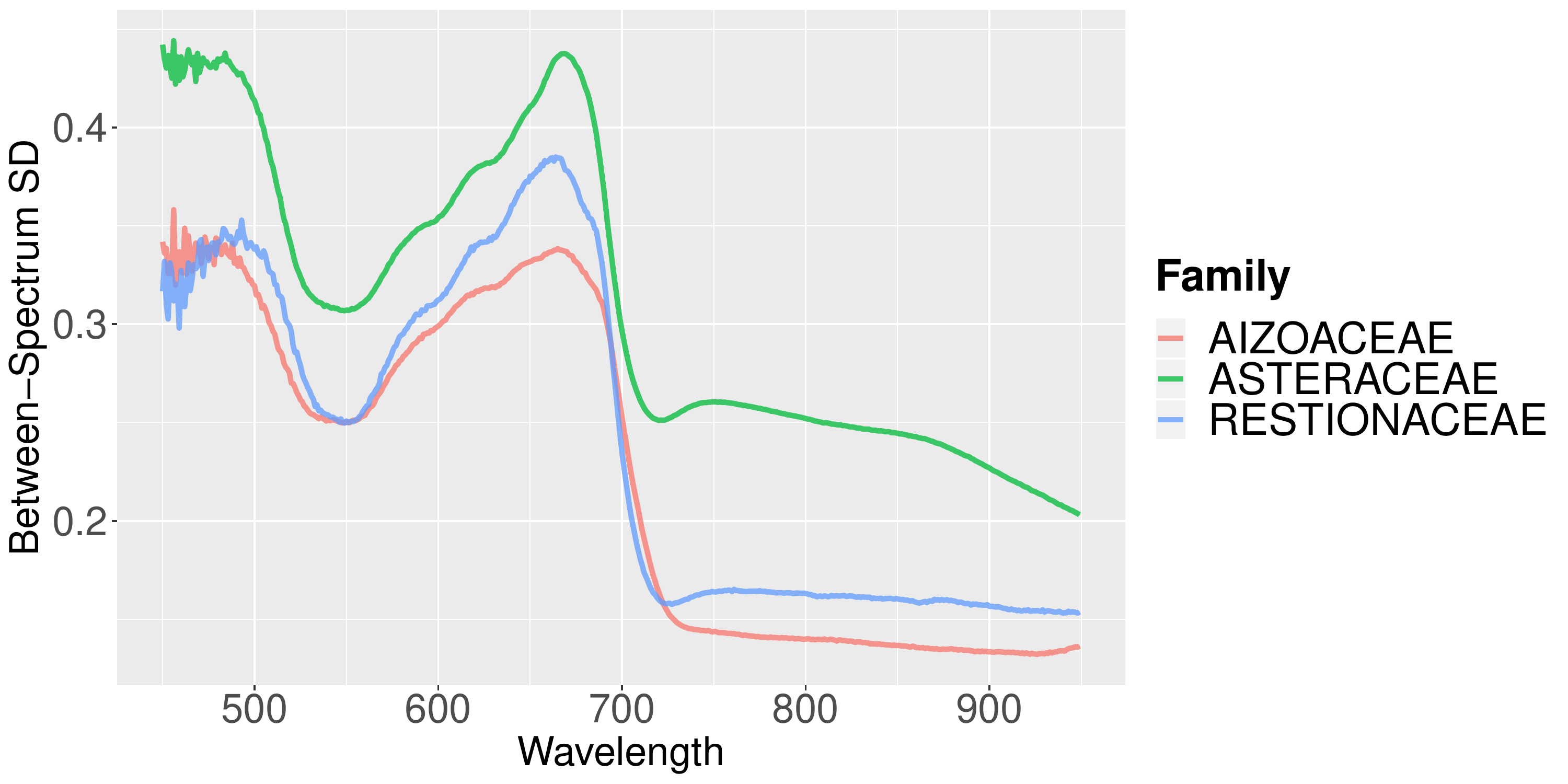}

 \end{center}
 
  \vspace{-4mm}
 
 \caption{(Left) 25-nm binned standard deviations for each reflectance spectrum with smoothed family-specific curves. (Right) Family-specific between-spectrum standard deviation as a function of wavelength.}\label{fig:within_between}
 \end{figure}

To assess within reflectance function variability as well as between-function variability, we calculate binned standard deviations for every curve. For these binned standard deviations, we estimate a smooth family-specific average standard deviation. Additionally, we calculate the family-specific between-curve standard deviation. These are plotted in Figure \ref{fig:within_between} and show that variability within reflectance spectrum changes with wavelength and, perhaps, with family. In addition, the variability between curves changes as a function of wavelength and differs by family. These findings lead us to impose heterogeneity in variance across wavelength, adopting wavelength varying variance curve models on the log scale.

Given these plots, we are led to four modeling needs: (i) to allow for family and genus differences, (ii) to model heterogeneity for the reflectance spectrum because within-curve variability changes across wavelength (iii) to capture between-curve variability through spatial modeling and/or environmental variables, and (iv) to adopt heteroscedastic errors since reflectances at lower wavelengths ($<500$ nm) appear to be more volatile.

\begin{figure}[ht]
 \begin{center}
 \includegraphics[width = 0.48\textwidth]{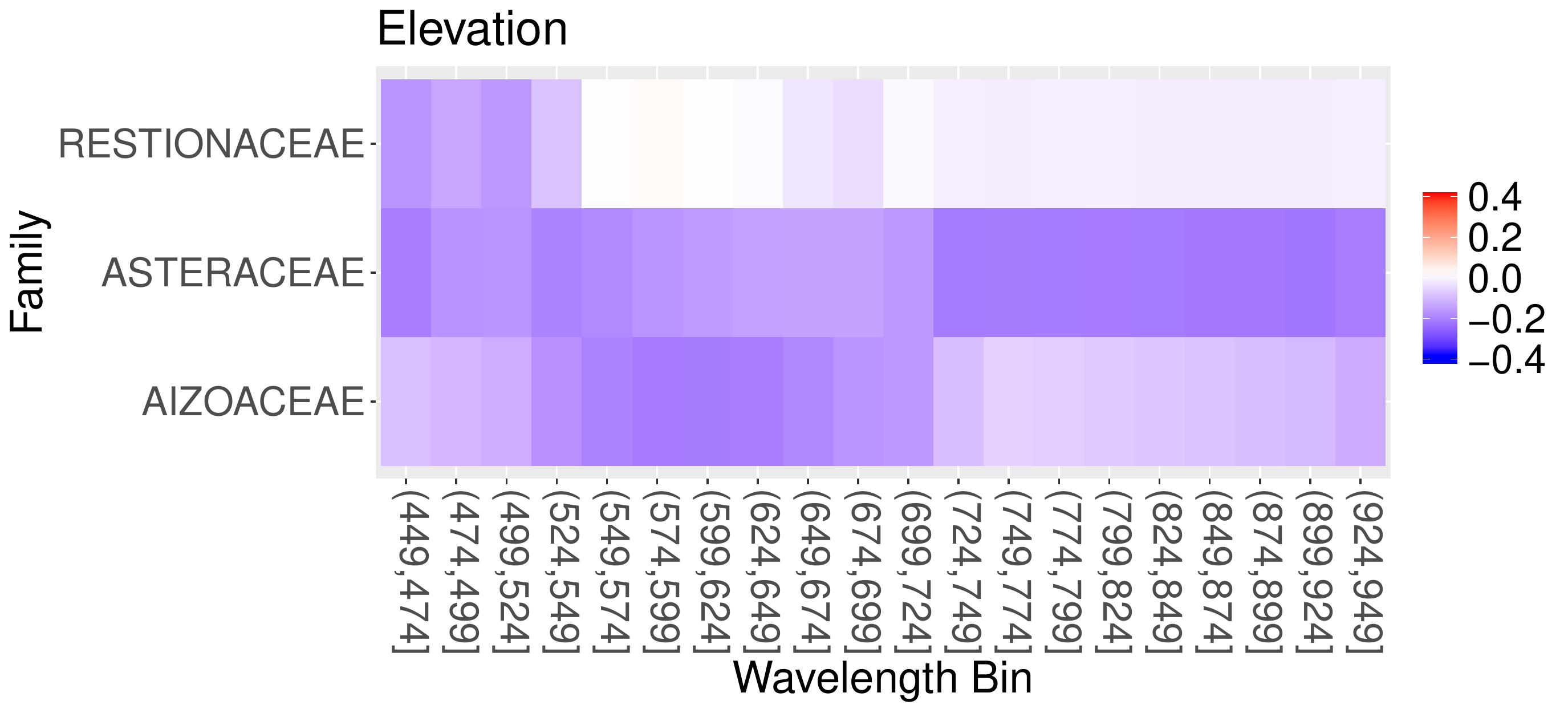}
 \includegraphics[width = 0.48\textwidth]{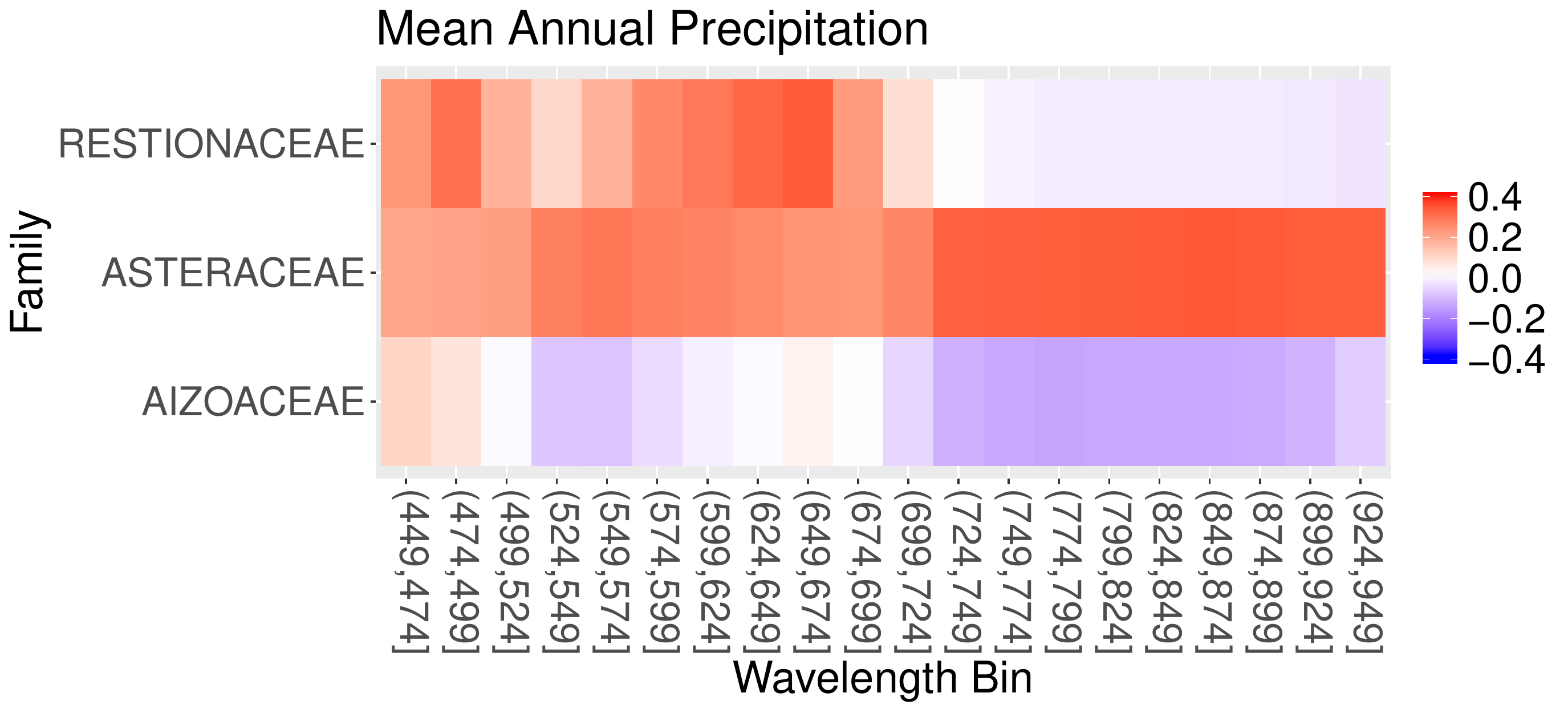}
  \includegraphics[width = 0.48\textwidth]{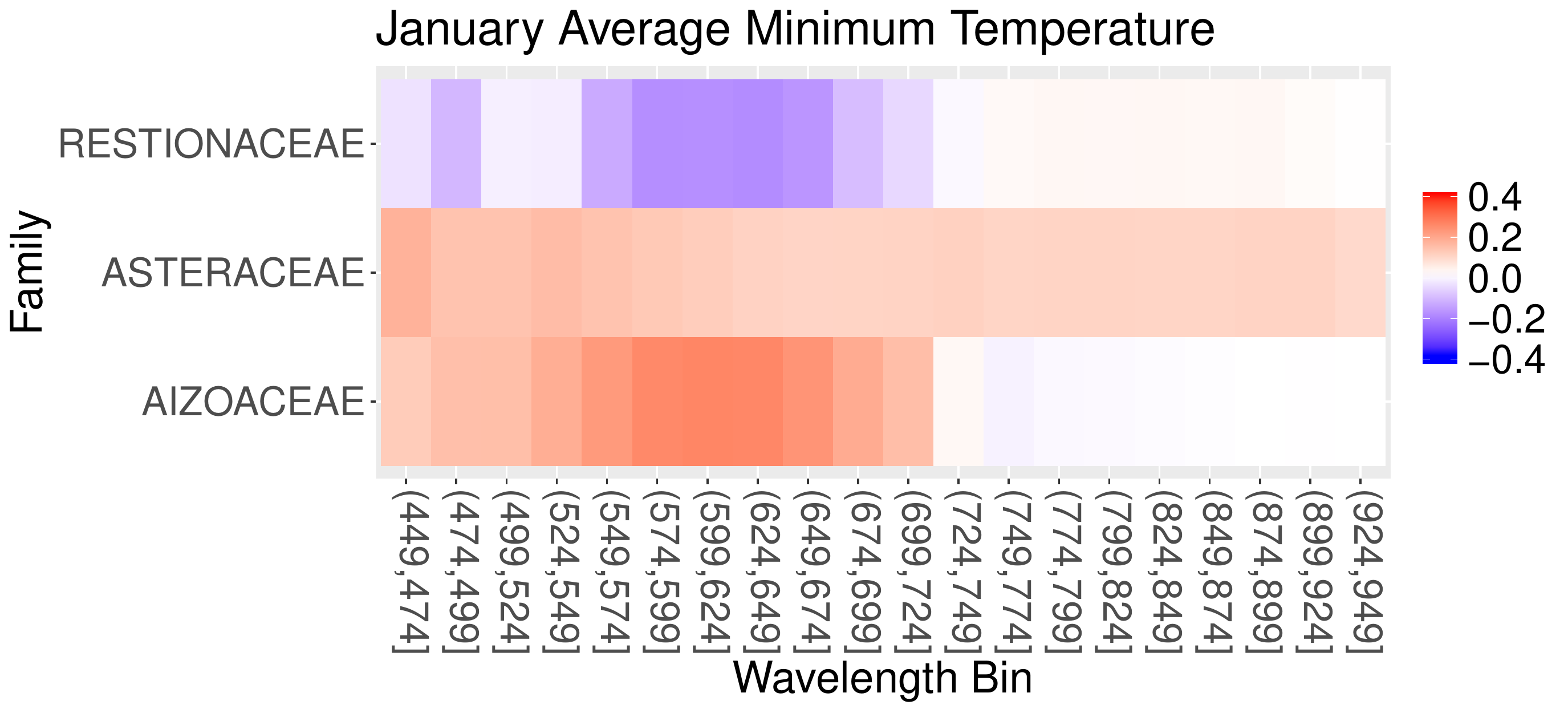}
  \includegraphics[width = 0.48\textwidth]{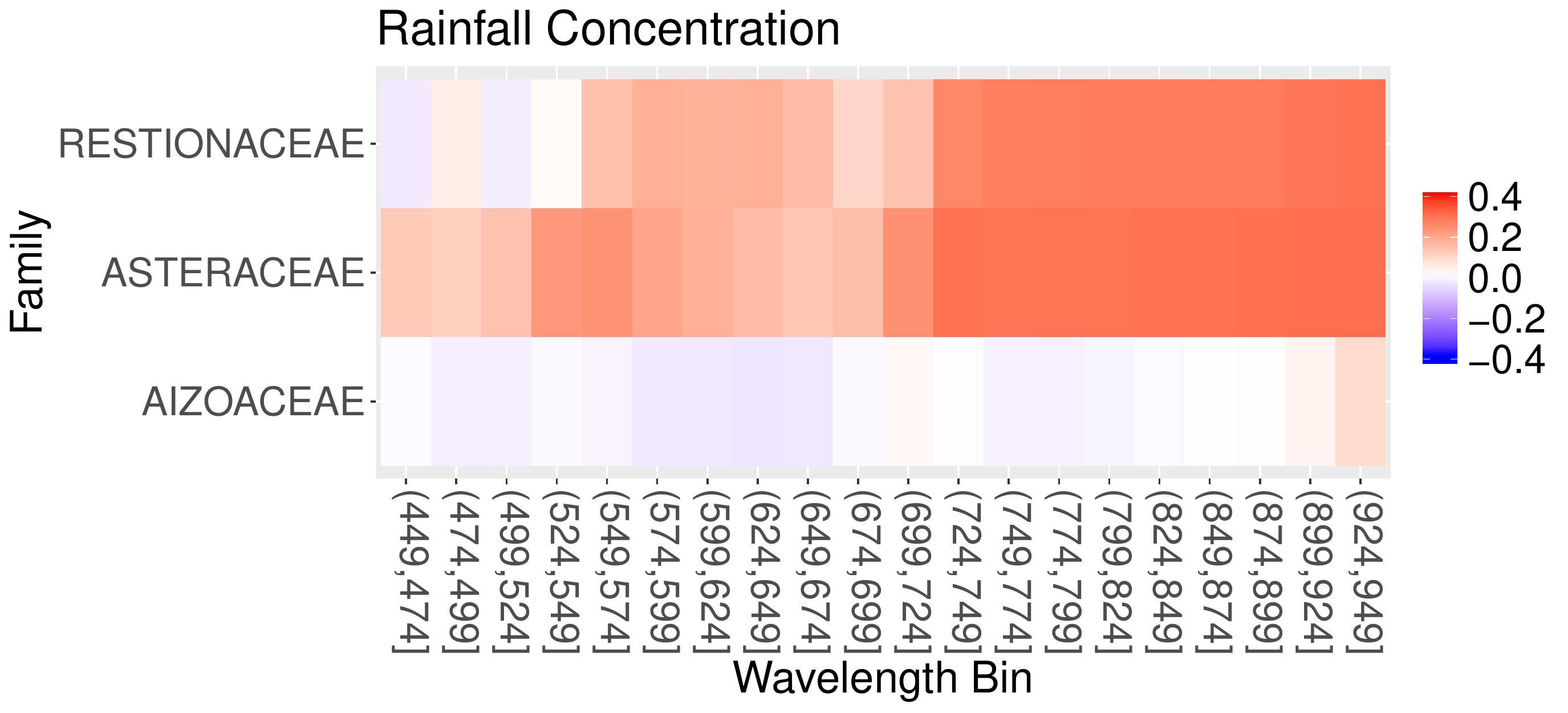}
 \end{center}
 
  \vspace{-4mm}

 \caption{(Top-Left to Bottom-Right) elevation, mean annual precipitation, average minimum temperature in January, rainfall concentration.}\label{fig:cor_mat}
 \end{figure}

For each family, we calculate the correlations between the environmental variables (see Supplemental Material) and the observed log-reflectances, using wavelength bins (See Figure \ref{fig:cor_mat}), to assess whether this relationship changes with wavelength. We find consequential changes in correlation as a function of wavelength. The strongest correlations are of magnitude 0.3 to 0.4.


\subsection{Environmental Variables and Reflectance Spectra}

\section{Spatial Wavelength Modeling}\label{sec:mod}

Functional data modeling for our spatial reflectance spectra was motivated by the foregoing exploratory analyses.  Families are modeled separately, at genus level, treating species within genus as replicates. 
We utilize the following environmental predictors: elevation, annual precipitation, rainfall concentration, and minimum January temperature. We introduce wavelength dependent variances to account for evident heterogeneity.  Model choice focuses on four issues: (i) Do we need wavelength dependent regression coefficients? (ii) Do we need genus specific wavelength random effects? (iii)  Do we need genus specific spatial random effects? (iv) How do we specify space-wavelength interaction?

In Sections \ref{sec:mode_dev} and \ref{sec:spec}, we elaborate the models, while Section \ref{sec:mod_comp} takes up model comparison yielding the model for which results are presented.

\subsection{Model development}\label{sec:mode_dev}

For a given family, let $i$ denote genera within the particular family, let $j$ denote replicates/species within genus.  
Let $\bs$ denote spatial location and $t$ denote wavelength. There is severe imbalance in the data.  The genera observed vary across locations and the number of replicates observed within a genus varies considerably across the locations.
Altogether, our most general model for \emph{log} reflectance takes the form:
\begin{equation}\label{eq:gen_mod}
Y_{ij}(\bs, t) = \mu_i(\bs,t) + \gamma_{i}(t) + \alpha_{i}(\bs) + \eta(\bs,t) + \epsilon_{ij}(\bs,t)
\end{equation}
Specifically, with regard to the site level covariates, $\bX(\bs)$, we write the mean $\mu_i(\bs,t) = \alpha_i +  \bX^{T}(\bs)\bbeta(t)$, where, hierarchically, $\alpha_i|\alpha \sim \mathcal{N}(\alpha,\sigma^2_\alpha)$. 
We have family level regression coefficients, $\bbeta(t)$, which vary with wavelength. So, a first model choice clarification is whether constant coefficients are adequate or whether wavelength varying coefficients are needed. Our EDA (Figure \ref{fig:cor_mat}) suggests the latter, and there is also supporting evidence/suggestion in the literature  \citep{jacquemoud_ustin_2019a}.  We do not consider these coefficients at genus level; with the very irregular observation (including absence) of genera across locations, we cannot learn about coefficients at genus scale.  However, we can learn about genus specific intercepts, the $\alpha_i$.  

Further, we introduce genus level spatial ($\alpha_{i}(\bs)$) and wavelength ($\gamma_i(t)$) random effects but family level space-wavelength interaction effects, $\eta(\bs,t)$. 
In the Supplemental Material, we note different spatial patterns for different wavelength bins, as well as residual dependence by genus and wavelength. Thus, an additive model (removing $\eta(\bs,t)$) seems inadequate; the $\eta$'s allow the functional model for the reflectances to vary more adaptively over space.  However, $\eta(\bs,t)$ is not genus specific.  While we have enough data to examine additivity in wavelength and spatial random effects at genus scale, we are unable to find genus level explanation for the interaction. Then, two model choice comparisons are whether the $\gamma$'s and whether the $\alpha$'s should still be genus specific?



As is customary, heterogeneity in the variance arises through the $\epsilon_{ij}(\bs,t)$ terms where we would have $\text{var}(\epsilon_{ij}(\bs, t)) = \sigma^{2}(t)$. We can accommodate this using a log GP for $\sigma^2(t)$, or perhaps just binned variances over suitable wavelength bins. For simplicity and flexibility, we specify $\log(\sigma^2(t))$ to be piecewise linear with knots every 20 nm from 440 - 960 nm. For all knot selections, we use boundary knots slightly beyond the wavelength range.

\subsection{Explicit Specifications}\label{sec:spec}

The specification for each $\alpha_{i}(\bs)$ is a genus-level mean $0$ Gaussian process with mean of $0$ and exponential covariance function. The GPs are conditionally independent across genera given a shared decay and shared scale parameter. We specify $\gamma_i(t)$ using process convolution of normal random variables \citep{higdon1998process,higdon2002space}. We adopt process convolutions because of their simple connection to GPs; the kernels of the process convolution connect the low-rank process to the GP covariance \citep{higdon1998process}. We adopt wavelength knots $t^{\gamma}_1,...,t^{\gamma}_{J_\gamma} $, spaced every 25 nm from 437.5-962.5 nm (22 in total). 

Specifically, we let $\gamma_i(t) = \sum^{J_\gamma}_{j = 1} k_{t^\gamma_j}(t - t^\gamma_j ;\theta^{(\gamma)}_{t^\gamma_j}) \gamma_i^*(t^\gamma_j)$,
where $\gamma_i^*(t^\gamma_j)$ are independent, normally distributed, and centered on a common $\gamma^*(t^\gamma_j)$. We use Gaussian kernels for  $k_{t_j}(\cdot;\theta_{t_j}^{(\gamma)})$ with bandwidths $\theta_{t_j}^{(\gamma)}$  (standard deviation of the Gaussian pdf) varying over wavelength. We assume that the log-bandwidths follow a multivariate normal distribution with global log-bandwidth and \\ Cov$\left[ \log \left(\theta_{t_j}^{(\gamma)}\right) ,\log\left(\theta_{t_{j'}}^{(\gamma)}\right) \right] = v_{\gamma}^2 \exp \left(- |t_{j} - t_{j'}|/ \phi_\gamma \right)$, yielding a non-stationary process because of the heterogeneous bandwidth. We found that this nonstationary specification outperformed a full-rank stationary GP with squared-exponential covariance (See Supplemental Material).

We specify $\bbeta(t)$ using kernel convolutions, where $\bbeta(t) = B K_\beta(t)$. With $p$ covariates, $B$ supplies a $p \times q$ matrix representation of the $p$ regression coefficient functions $\bbeta(t)$. Here, the kernel convolution has knots every 25 nm from 437.5 - 962.5 nm. As with $\gamma_i(t)$, we use Gaussian kernels to specify $K_\beta(t)$; however, unlike $\gamma_i(t)$, we 
assume common bandwidths for all kernels, for all wavelengths, and for each coefficient function.

Turning to $\eta(\bs,t)$, 
we use wavelength kernel convolutions of spatially-varying variables. That is, we consider low-rank but heterogeneous and nonstationary (in the wavelength domain) specifications. We select a set of wavelength knots $t^\eta_1,...,t^\eta_{J_\eta}$, spaced every 25 nm from 437.5-962.5 nm (22, in total). We define the space-wavelength function as
\begin{equation}\label{eq:extended}
\begin{aligned}
\eta(\bs,t) =  \bK(t)^T \bz(\bs) = \sum_{j = 1}^{J_\eta} k_{t^\eta_j}(t - t^\eta_j;\theta^{(\eta)}) z_{t^\eta_j}(\bs),
\end{aligned}
\end{equation}
where $z_{t^\eta_j}(\bs)$ are spatially-varying random variables associated with Gaussian wavelength kernels $k_{t^\eta_j}(\cdot;\theta_{t^\eta_j})$. Unlike the kernel structure for $\gamma(t)$, we use a common bandwidth $\theta^{(\eta)}$ for all knots. The construction in \eqref{eq:extended} allows heterogeneity and nonstationarity in wavelength space, where the heterogeneity is introduced through $z_{t^\eta_j}(\bs)$ \citep[See][for a similar construction in the context of spatial monotone regression]{white2020hierarchical}.


As an aside, we remark on choosing the form $\eta(\bs, t) = \bK^{T}(t)\bz(\bs)$ vs. $\eta(\bs, t) = \bK^{T}(\bs)\bz(t)$.  With $n$ sites, the former introduces $J_{\eta}n$ random effects, the latter $500n$ random effects.  With $J_{\eta}$ relatively small, the former is preferred computationally. More importantly, it yields much better fits to the data (see the Supplemental Material).  


While we may want dependence between components in $\bz(\bs)$ at $\bs$, that dependence should have nothing to do with the $t^\eta_j$'s. We are capturing association with regard to the distances between wavelength knots through the $K$'s and our objective for the $z$'s is to obtain perhaps nonseparable and nonstationary covariance structure for $\eta(\bs,t)$.  So, we write $\bz(\bs) = A\bw(\bs)$ where $A$ is $J_\eta \times r$ and the components of $\bw(\bs)$ are independent mean $0$ GP's with variance $1$ and correlation functions, $\rho_{r}(\bs - \bs')$. 

When $r = J_\eta$, we have the familiar linear model of coregionalization \citep{wackernagel1998}. We consider using $A=I$ and $A_{p \times r}$, for various $r$, as well as a separable specification for $\bz(\bs)$, where, with $\bV$ a positive definite matrix, $\text{Cov}\left(\bz(\bs), \bz(\bs')\right) = \exp\left(-\phi_{z} \| \bs - \bs' \|\right) \bV$. With $A_{p \times r}$, we constrain the decay parameters of the $(w_1(\bs),...,w_r(\bs))^T$ to be increasing \citep[see][]{white2020multivariate}, so that the latent GPs have different spatial decays ($\phi_z$). The resulting processes for $\bz(\bs)$ are very flexible. We compare the various choices through out-of-sample prediction in Section \ref{sec:mod_comp}.



Under the general form $\eta(\bs, t) = \bK^{T}(t)(\bs)A\bw(\bs)$, 
$\text{cov}(\eta(\bs, t), \eta(\bs',t')) = \bK^{T}(t)A\Sigma_{\bw(\bs), \bw(\bs')}A^{T}\bK(t').$
If $A=I$, we have $\Sigma_{\bw(\bs), \bw(\bs')} = D(\bs-\bs')$, a $J_{\eta} \times J_{\eta}$ diagonal matrix with entry $d_{jj}= \rho_{j}(\bs-\bs')$.  Thus, $\text{cov}(\eta(\bs, t), \eta(\bs',t')) = \bK^{T}(t)D(\bs-\bs')\bK(t')= \sum_{j} k_{t^\eta_{j}}(t-t_{j})k_{t^\eta_{j}}(t'-t^\eta_{j})\rho_{j}(\bs-\bs')$.  
The covariance is always nonseparable and, if $A$ is unconstrained it is nonstationary.


As an illustration, if we take $A$ to be $J_{\eta} \times 2$, we have $\Sigma_{\bw(\bs), \bw(\bs')} = \left(
                                                                                                   \begin{array}{cc}
                                                                                                     \rho_{1}(\bs-\bs') & 0 \\
                                                                                                     0 & \rho_{2}(\bs - \bs') \\
                                                                                                   \end{array}
                                                                                                 \right)$.
Now, with $\ba_1$ and $\ba_2$ the two columns of $A$, $\text{cov}(\eta(\bs, t), \eta(\bs',t'))= (\bK^{T}(t)\ba_{1})(\bK^{T}(t')\ba_{1})\rho_{1}(\bs-\bs')+ (\bK^{T}(t)\ba_{2})(\bK^{T}(t')\ba_{2}) \\ \rho_{2}(\bs-\bs')$.  We achieve both dimension reduction and space-wavelength interaction. Further, we have nonseparability and nonstationarity (in the wavelengths) if there are different bandwidths for the different $t_{j}$. If we set $r=1$, we have separability but still nonstationarity in the wavelengths. 

\subsection{Model Comparison}\label{sec:mod_comp}

We carry out model comparison for Asteraceae, the most abundant family, using 10-fold cross-validation (described below). In the Supplemental Material, we present cross-validation results examining various specifications of the spatial process in $\eta(\bs,t)$. When comparing models with different specifications of $\eta(\bs,t)$, all models include spatially-varying genus-specific intercepts $\alpha_i + \alpha_i(\bs)$, a global (not genus-specific) wavelength random effect $\gamma(t)$, and functional regression coefficients $\bbeta(t)$. For $\eta(\bs,t)$, we compare separable, independent, and latent factor models. We find that the latent factor specification of $\eta(\bs,t)$ with $r = 10$ has the best out-of-sample predictive performance and use this for $\eta(\bs,t)$ in the remainder of the manuscript. For this specification of $\eta(\bs,t)$, we focus our model comparison on eight special cases of \eqref{eq:gen_mod} arising by (i) including or excluding $\alpha_i(\bs)$, (ii) using $\gamma_i(t)$ or only $\gamma(t)$, and (iii) having functional coefficients $\bbeta(t)$ or scalar coefficients $\bbeta$.

We hold out reflectances imagining the setting where researchers visited a site but failed to measure reflectances for some genus at that site. So, at random, we leave out spectra that have (i) at least one other observed reflectance spectrum at the same site and (ii) at least one other observed spectrum of the same genus located elsewhere. For Asteraceae, this yields 117 candidates out of the 185 in total. Holding out a subset, we fit the model using Markov chain Monte Carlo, and, with each posterior sample, we predict the hold-out reflectance spectra. We compare models by averaging across the wavelengths to obtain the predicted mean squared error (MSE), mean absolute error (MAE), and the mean continuous ranked probability score (MRCPS), see \cite{gneiting2007strictly}. The results are summarized in Tables \ref{tab:mod_comp1} and in the Supplement.

\begin{table}[ht]
\centering
\begin{tabular}{lllrrrr}
  \hline
$\alpha_i$/$\alpha_i(\bs)$ & $\gamma(t)$/$\gamma_i(t)$ & $\bbeta$/$\bbeta(t)$ & MSE & MAE & MCRPS & Relative MCRPS \\
  \hline
 $\alpha_i$ & $\gamma(t)$ & $\bbeta$ & 0.148 & 0.301 & 0.241 & 1.229 \\
 $\alpha_i$ & $\gamma(t)$ & $\bbeta(t)$ & 0.141 & 0.293 & 0.234 & 1.196 \\
$\alpha_i$ & $\gamma_i(t)$ & $\bbeta$ & 0.175 & 0.320 & 0.265 & 1.355 \\
 $\alpha_i$ & $\gamma_i(t)$ & $\bbeta(t)$  & 0.169 & 0.318 & 0.262 & 1.340 \\
 $\alpha_i(\bs)$ & $\gamma(t)$ & $\bbeta$ & 0.102 & 0.244 & 0.200 & 1.023 \\
 $\alpha_i(\bs)$ & $\gamma(t)$ & $\bbeta(t)$ & 0.097 & 0.237 & 0.196 & 1.000 \\
  $\alpha_i(\bs)$ & $\gamma_i(t)$ & $\bbeta$ & 0.348 & 0.435 & 0.393 & 2.009 \\
  $\alpha_i(\bs)$ & $\gamma_i(t)$ & $\bbeta(t)$ & 0.290 & 0.420 & 0.380 & 1.940 \\
   \hline
\end{tabular}

\caption{Out-of-sample predictive performance model comparison. Models vary by including or excluding genus-specific terms, as well as comparing scalar and functional coefficients. All models use $r = 10$ spatial factors to construct $\eta(\bs,t)$.}\label{tab:mod_comp1}
\end{table}

Following the results in Table \ref{tab:mod_comp1} and the Supplemental Material, we adopt a model with (1) a global wavelength random effect, (2) a spatially-varying genus-specific intercept, (3) functional regression coefficients, and (4) a space-wavelength random effect specified through the wavelength kernel convolution of a multivariate spatial process with 10 latent spatial GPs having different decay parameters. We use this model to analyze the CFR data.

For the sensitivity of model fit to change in other specifications (e.g., knot spacing and GP/process convolution), we use average deviance, the deviance information criterion, and estimated model complexity \citep{spiegelhalter2002}, 
as supplied in the Supplemental Material.  To summarize, we employ a heterogeneous process convolution specification of $\gamma(t)$ because it gave a better fit than a full-rank homogeneous GP with squared-exponential covariance and a process convolution with a common bandwidth for all wavelength knots. We also specify $\bbeta(t)$ using kernel convolutions where $\bbeta(t) = B K_\beta(t)$, where we space knots every 25 nm from 437.5 - 962.5 nm. We also find that the Gaussian kernel, which corresponds to the Gaussian covariance function, was preferred to using double-exponential kernels for $\gamma(t)$, $\bbeta(t)$, and $\eta(\bs,t)$. For $\gamma(t)$, the model fit was improved when bandwidths $\theta^{(\gamma)}_{t_j}$ varied over wavelength; however, a common bandwidth for the kernels was prefered for $\eta(\bs,t)$. The knot spacing, discussed in Section \ref{sec:spec}, was also determined through sensitivity analysis.

\subsection{Confounding and Orthogonalization}\label{sec:ortho}

The flexibility of the residual specification in our best performing model results in annihilation of the significance of the spatial regressors. This is a well-documented problem in the literature \citep[see, e.g.,][]{hodges2010adding,khan2020restricted}. A solution in the literature is orthogonalization; that is, projection of the random effects (the spatial residuals) onto the orthogonal complement of the manifold spanned by the spatial covariates.  This yields revised regression coefficients with direct interpretation in the presence of the random effects. The coefficients are more aligned with those that arise from model fitting ignoring spatial random effects.

We propose a similar orthogonalization approach here but our setting is more demanding because we have both space and wavelengths in our residuals. We have to introduce orthogonalization with regard to the manifold spanned by the spatial covariates as well as with regard to the manifold spanned through the use of kernel functions with knots.  We present the details below for the simpler case where we have no replicates at locations.  However, in our application, we have replicates associated with the spatial locations and also with different genera.  So, formally, the orthogonalization requires us to introduce a location by genus matrix to align the number of observed sites with the number of observed reflectances. We present the more detailed argument in the Supplemental Material.

With $n$ sites and $500$ wavelengths, we can express \eqref{eq:gen_mod} in matrix form as
\begin{equation}
Y = \alpha \bone  + XBK_\beta^{T} + \eta^{*} + \epsilon
\label{eq:matrix}
\end{equation}
where $Y$ is the $n \times 500$ matrix of log-reflectance spectra data by sites, $\bone$ is a $n \times 500$ matrix of ones, $\alpha$ is the global mean,  $X$ is the $n \times p$ spatial design matrix (with $p$ covariates), $B$ is $p \times J_\beta$ with $J_\beta$ knots, $K_\beta$ is the $500 \times J_\beta$ kernel design matrix with $J_\beta$ knots.  $\eta^*$ is also $n \times 500$ summing the corresponding matrix forms for the mean-zero random effects ($\gamma(t)$, $\alpha_i(\bs)$, $\alpha_i - \alpha$, and $\eta(\bs,t)$).
Then, using standard results, we can vectorize \eqref{eq:matrix} to
\begin{equation}
vec(Y) = \alpha \bone + (X \otimes K_\beta) vec(B) + vec(\eta*) + vec(\epsilon)
\label{eq:vec}
\end{equation}
where $vec(Y)$ is an $n500 \times 1$ vector with $X \otimes K_\beta$ an $n500 \times pJ$ matrix.

Now, define the joint projection matrix,
\begin{equation}
\begin{aligned}
P_{XK_\beta} &\equiv (X \otimes K_\beta)((X \otimes K_\beta)^{T}(X \otimes K_\beta))^{-1}(X \otimes K_\beta)^{T} \\ &= (X (X^T X)^{-1} X^T) \otimes (K_\beta (K_\beta^T K_\beta)^{-1} K_\beta^T) = P_X \otimes P_{K_\beta},
\end{aligned}
\end{equation}
and write $vec(\eta^{*}) = P_{XK_\beta}vec(\eta^{*}) + (I-P_{XK_\beta})vec(\eta^{*})$.  Then, we can write $vec(Y) = (X \otimes K_\beta)vec(B^{*}) + (I-P_{XK_\beta})vec(\eta^{*}) + vec(\epsilon)$, where the updated \emph{unconfounded} coefficients are (in vec and block form)
\begin{equation}
\begin{aligned}
vec(B^{*}) &=  vec(B) + ((X \otimes K_\beta)^{T}(X \otimes K_\beta))^{-1}(X \otimes K_\beta)^{T}vec(\eta^{*}), \\
B^{*} &= B + (X^T X)^{-1} X^T \eta^{*} K_\beta (K_\beta^T K_\beta)^{-1}.
\end{aligned}
\label{eq:Bstar}
\end{equation}
Here, $vec(B^{*})$ and $B^{*}$ provide the vector and matrix of regression coefficients, respectively, under the orthogonalization. The model is fitted using \eqref{eq:gen_mod}.  Then, with the posterior samples of the $\beta$'s, $\gamma$'s, $\alpha$'s, and $\eta$'s along with the $\bX(\bs)$ and $\bK_\beta(t)$, the unconfounded $B^*$'s can be obtained using \eqref{eq:Bstar}.

\section{Analysis of the CFR Reflectance Spectra Data}\label{sec:analysis}

We focus discussion on a comparison between families but give specific attention to the results on Asteraceae, the most abundant family. 
We compare and discuss results from the orthogonalized coefficients using the approach in Section \ref{sec:ortho}. In addition, we summarize covariate importance on log-reflectance. Again using the orthogonalized random effects and unconfounded regression functions, we discuss the proportion of variance explained by each model term.



The confounding between random effects (genus, wavelength, and spatial) and covariates pushes $\bbeta(t)$ to zero, obliterating any significant inference with regard to the effect of environmental variables on log-reflectance. For each MCMC posterior sample, we calculate the proportion of the variance in each random effect ($\alpha_i(\bs)$, $\gamma(t)$, $\eta(\bs,t)$) explained by $\bX$ and $\bK_\beta$.
We orthogonalize our random effects with respect to $\bX$ and $\bK_\beta$ as described in Section \ref{sec:ortho} to remove the diminishing of the effect of the regressors.

\begin{figure}[ht]
\begin{center}
\includegraphics[width=.44\textwidth]{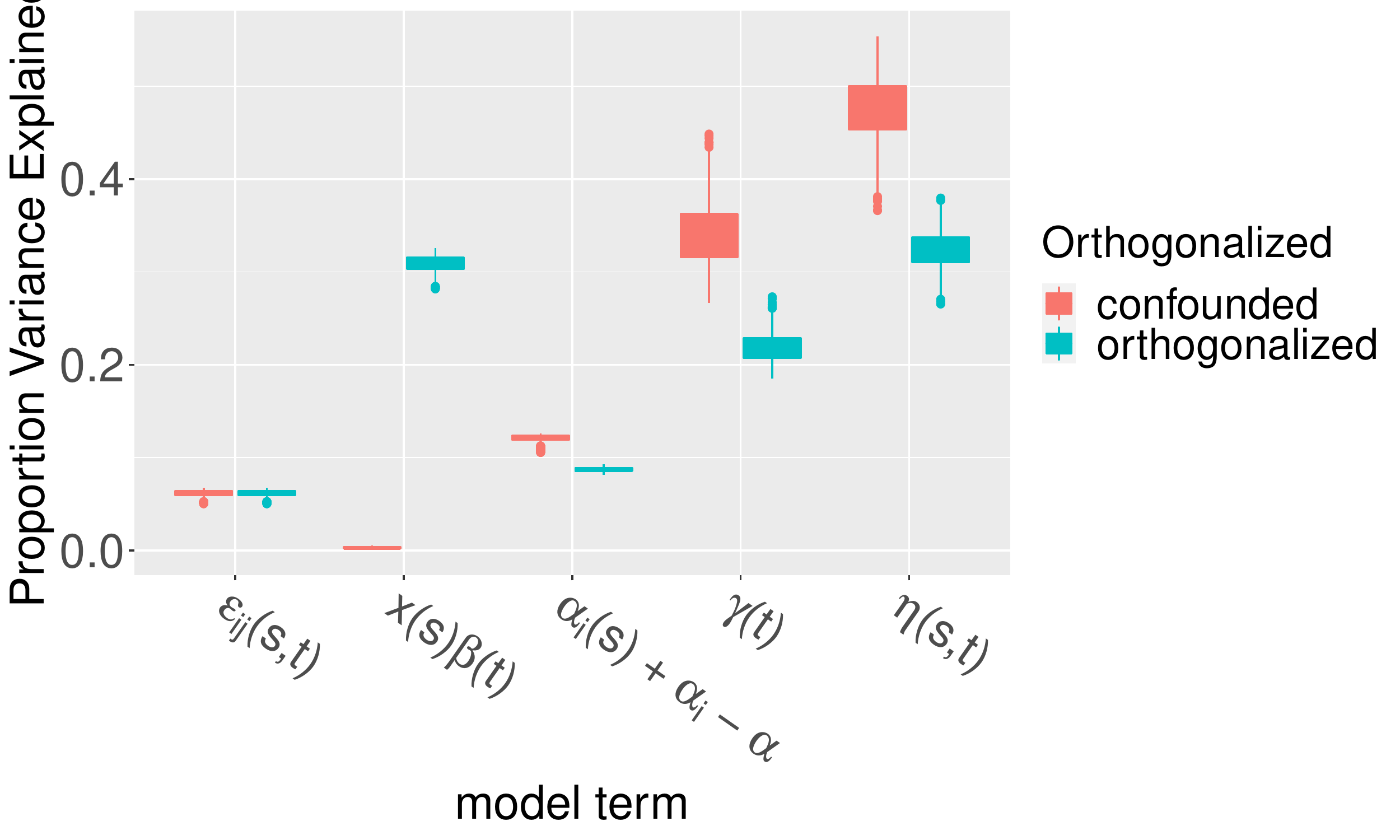}
\includegraphics[width=.54\textwidth]{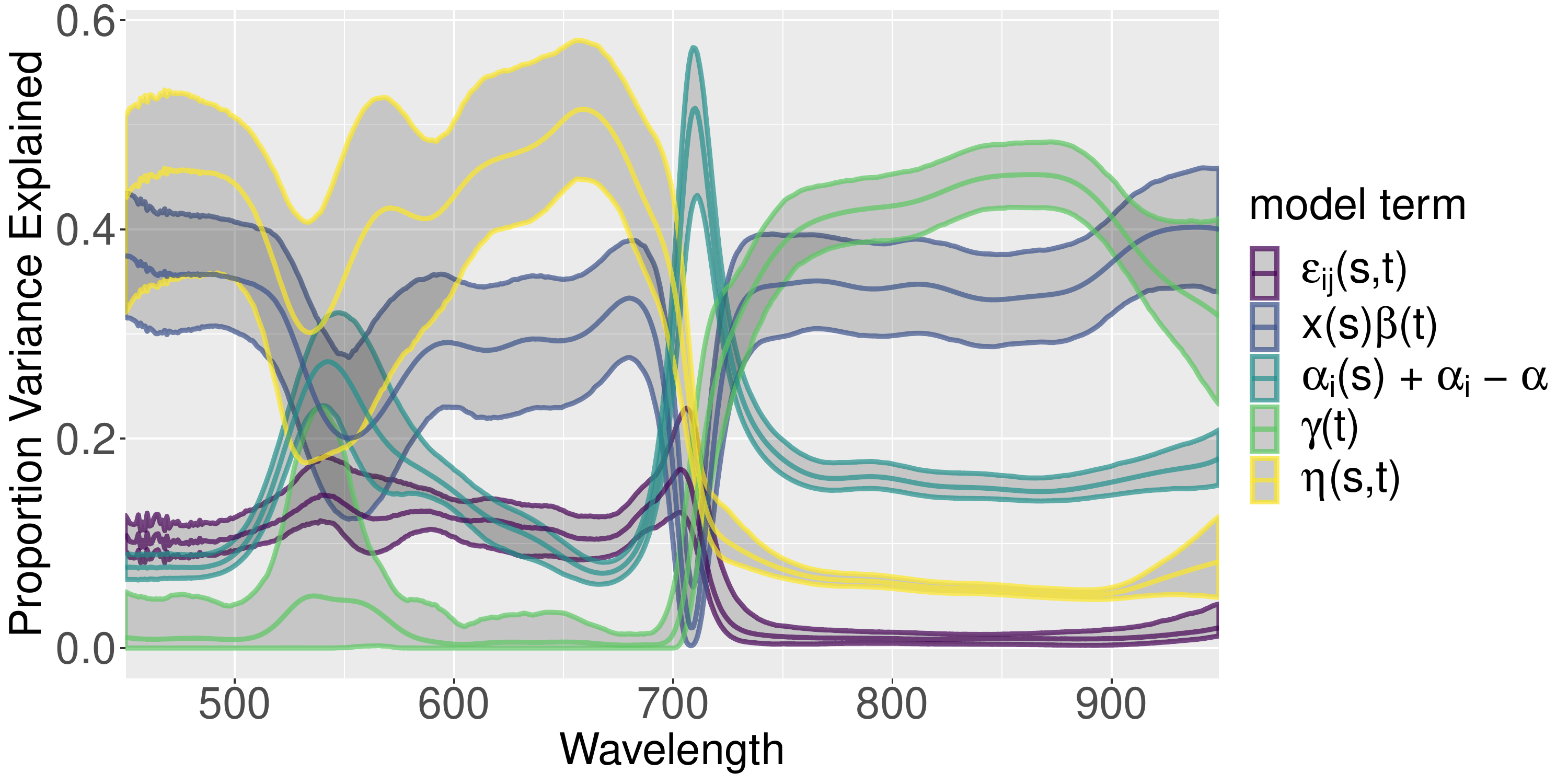}
\end{center}

 \vspace{-4mm}

\caption{(Left) Proportion of variance explained by each model term and (Right) Proportion of between-site variance explained by model terms.}\label{fig:var_explain}
\end{figure}

For the Asteraceae family, we explore the proportion of the variance explained by each of the mean-zero model terms. For every posterior sample, we calculate the empirical variance of all nonorthogonalized and orthogonalized terms (See Figure \ref{fig:var_explain} to the 95\% credible regions): $\epsilon_{ij}(\bs,t)$, $\bx(s)^T \bbeta(t)$, $\alpha_i(\bs) + \alpha_i - \alpha$, $\gamma(t)$, and $\eta(\bs,t)$. We take $\alpha_i(\bs) + \alpha_i$ to capture both genus-specific terms and subtract $ \alpha $ to make $\alpha_i(\bs) + \alpha_i - \alpha$ a mean-zero random effect. For orthogonalized terms, $\gamma(t)$ explains slightly under 25\% of the variability of the data, while both $\eta(\bs,t)$ and $\bx(s)^T \bbeta(t)$ explain over 30\% of the total variance. Without orthogonalization of the random effects, the environmental regression explains almost no variance. The genus-specific spatially-varying intercept $(\alpha_i(\bs) + \alpha_i) - \alpha$ explains over 10\% of the total variance while $\epsilon_{ij}(\bs,t)$ accounts for about 5\% of variance in the data.

In Figure \ref{fig:var_explain}, we plot the proportion of between-spectrum variability explained by all orthogonalized mean-zero terms as a function of wavelength (posterior mean and 95\% credible interval). Even though $\gamma(t)$ is common to all spectra, after orthogonalization, it is no longer a constant term for all spectra. For wavelengths less than 700 nm, we find that unconfounded environmental regression and  space-wavelength random effects are most important in explaining between-spectrum variance. For higher wavelengths ($>750$ nm), where there is little variation in the wavelength functions; the orthogonalized global wavelength random effects $\gamma(t)$ and the unconfounded environmental regression explain the most between-spectrum variance. The spatially-varying genus-specific offset, $(\alpha_i(\bs) + \alpha_i) - \alpha$ explains between 10-20\% of between-spectrum variance for most wavelengths but appears particularly influential for wavelengths between (675-725 nm). The $\epsilon_{ij}(\bs,t)$ account for the 0 to 10\% of unexplained between-spectrum variance in log-reflectance, depending on wavelength.

After updating $\bbeta(t)$ in the presence of orthogonalization, we present inference on covariates for all families. Figure \ref{fig:family_coefs} shows the posterior mean, 95\% credible interval for each element of $\bbeta(t)$. 
All coefficient functions are significantly non-zero for most wavelengths (around 99\% for all wavelengths). With covariates centered and scaled, (i) we can interpret effects as the expected change in log-reflectance for a one standard deviation change in the covariate, holding the other covariates constant, and (ii) we can compare the scales of the coefficient functions among covariates.

The four covariates have positive effects for some wavelengths, negative effects for others, with a transition around 700 nm, a threshold/boundary between visible (450-700nm) and near-infrared regions (NIR, 700-1400nm) of the spectrum. The visible region is most strongly affected by differences in plant pigment composition/concentration while the NIR is most affected by structural properties related to the cell wall, to air interface within the leaf \citep{jacquemoud_ustin_2019a}.  Traits can exhibit uniform effects across multiple parts of the spectrum (e.g., often in water content) or can cause increased reflectance in parts of the spectrum and decreased reflectance in others \citep{feng2008monitoring, jacquemoud_ustin_2019b}. Different sets of traits acting in concert in response to environment likely drive the positive and negative shifts across the 700 nm threshold in Figure \ref{fig:family_coefs}. 

For Asteraceae, we estimate that higher elevations are associated with lower reflectance levels at wavelengths less than 700 nm but higher reflectance at wavelengths above 700 nm. The relationships of precipitation and temperature with reflectance are similar. On the other hand, rainfall concentration is positively correlated with reflectance at low wavelengths and becomes negatively correlated with reflectance as wavelength increases. We note that rainfall concentration, the environmental feature that reflectance responds differently to, is largely longitudinally driven in comparison to the other features. Specifically, the extreme western and to some extent the extreme eastern sample sites have significantly higher rainfall concentrations than more central locations. Because there is between-covariate correlation, the coefficient functions must be interpreted as partial slopes, i.e., holding all other covariates constant.

To compare covariate importance, we calculate the mean integrated absolute coefficient  over the wavelength domain, $\overline{|\beta_j|}=\frac{1}{500}\int_{450}^{950} | \beta_j(t)| dt \approx \frac{1}{500}\sum_{i =1}^{500} |\beta_j(t_i)|,$ for each covariate. This metric weights the contribution of the coefficient equally regardless of sign or wavelength. 
We calculate $\overline{|\beta_j|}$ for every posterior sample and plot these in Figure \ref{fig:family_import}. In terms of $\overline{|\beta_j|}$, elevation and temperature are more influential on reflectance than precipitation and rainfall concentration. 

\subsection{Comparison across families}\label{sec:family_comparison}

\begin{figure}[ht]
\begin{center}
\includegraphics[width=.4\textwidth]{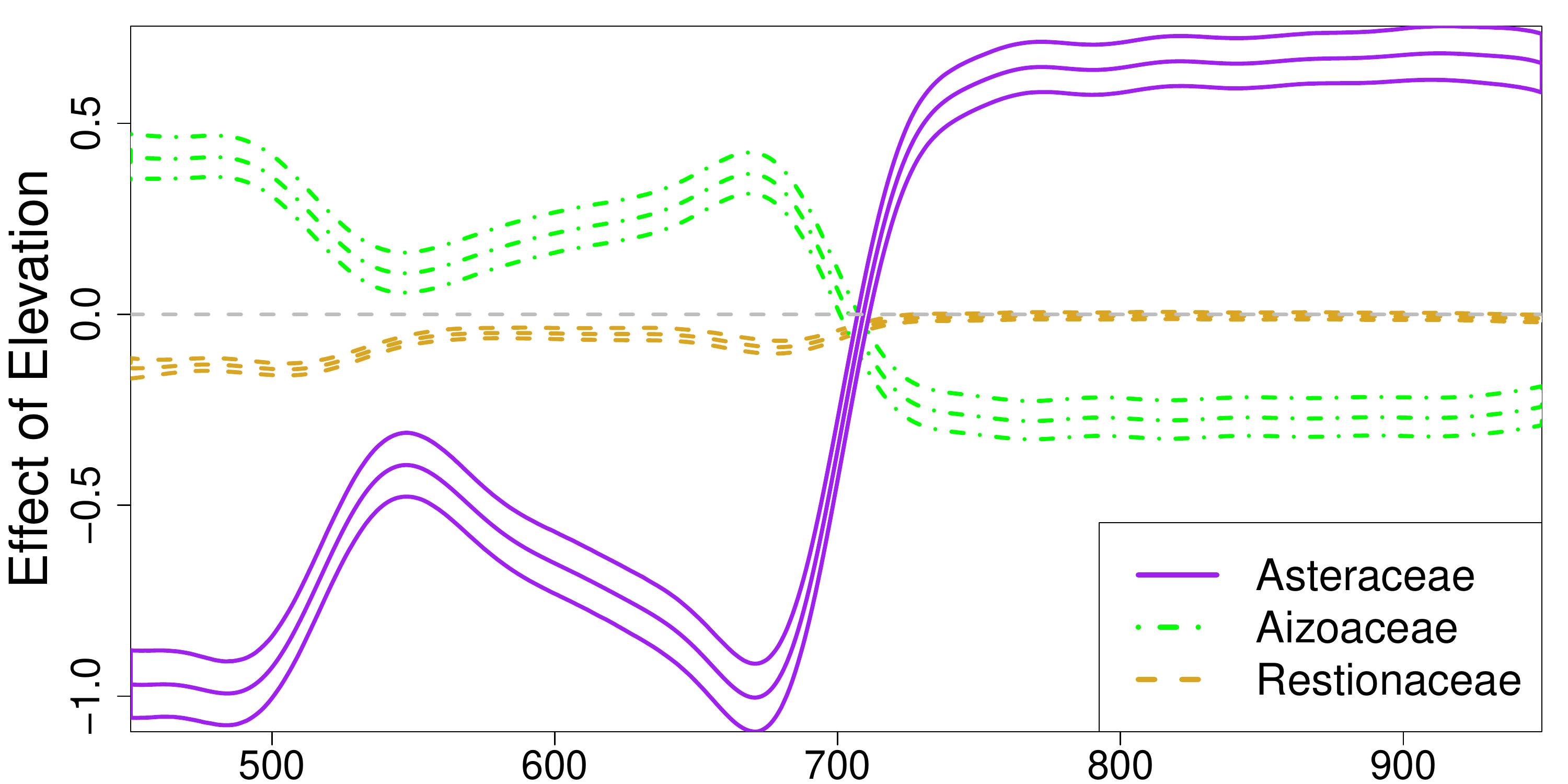}
\includegraphics[width=.4\textwidth]{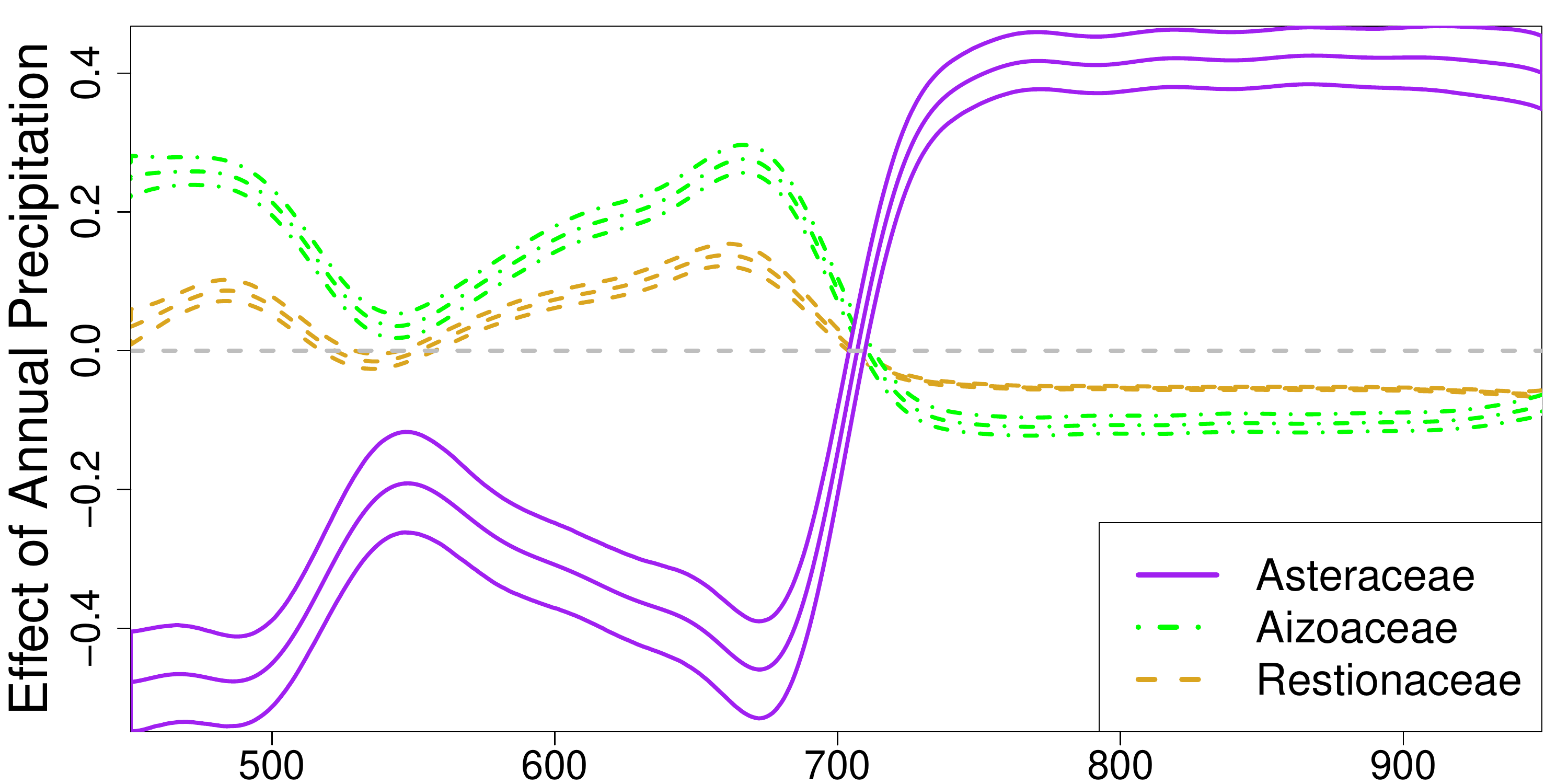}

\includegraphics[width=.4\textwidth]{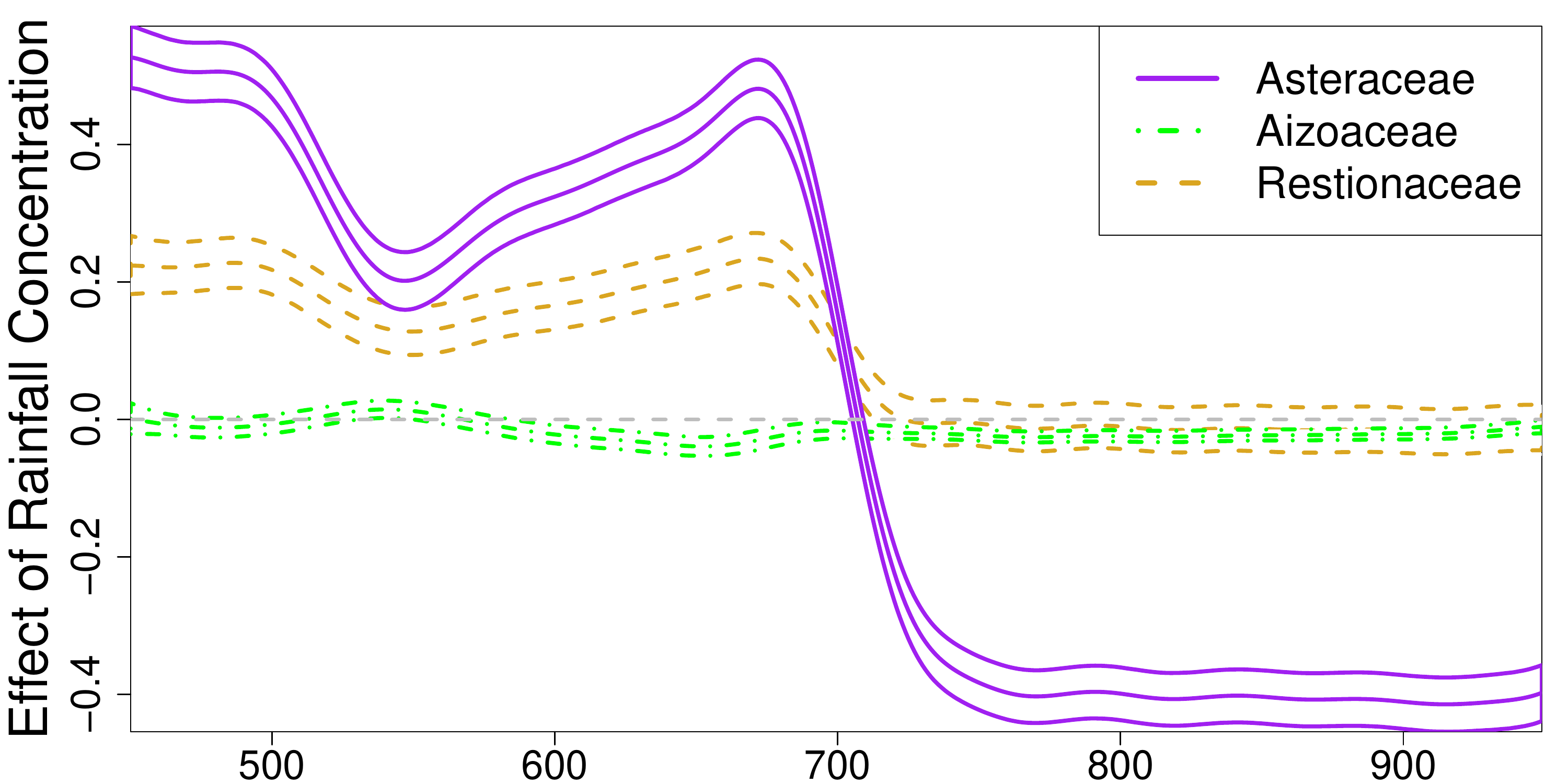}
\includegraphics[width=.4\textwidth]{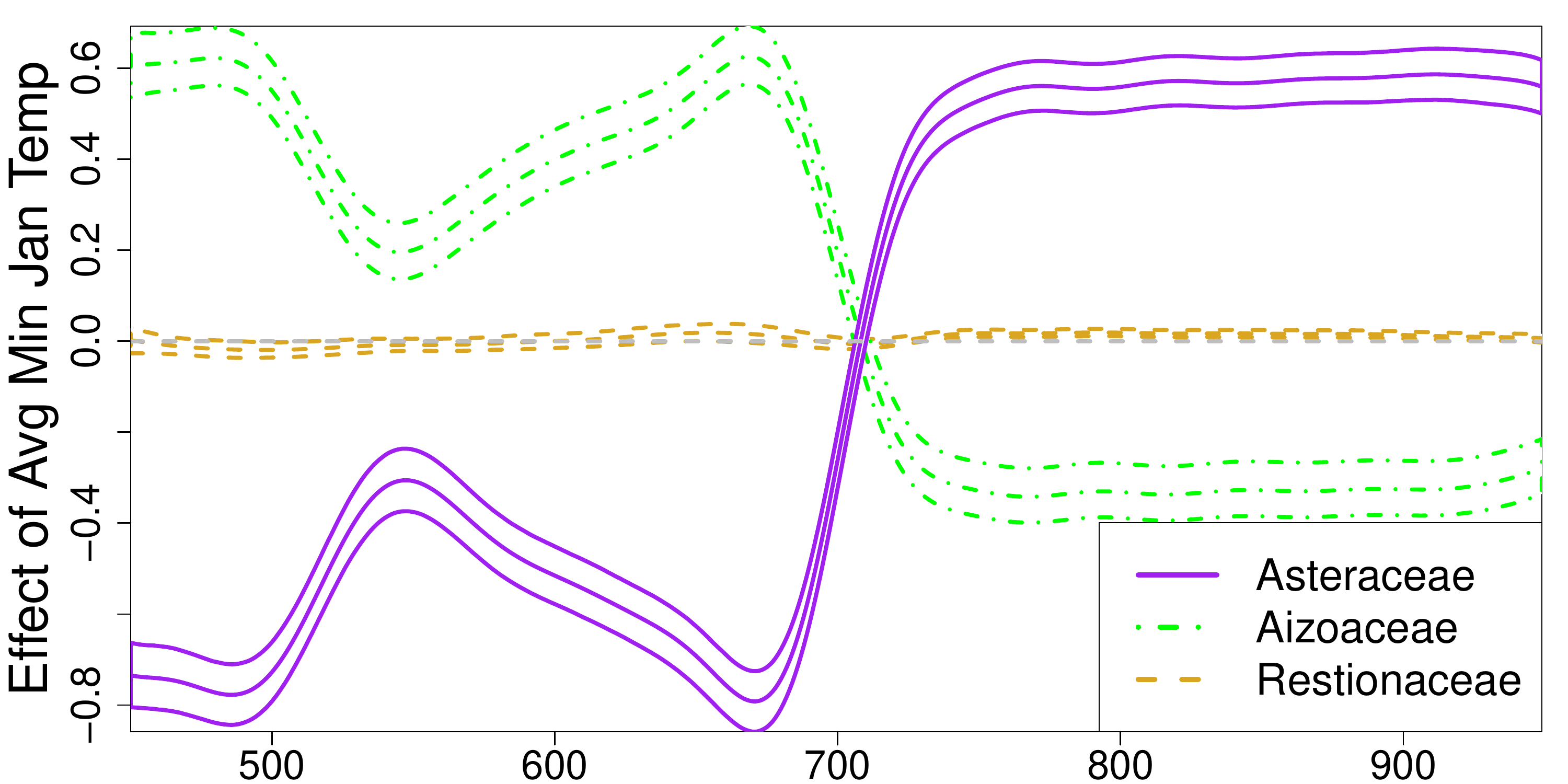}

\includegraphics[width=0.5\textwidth]{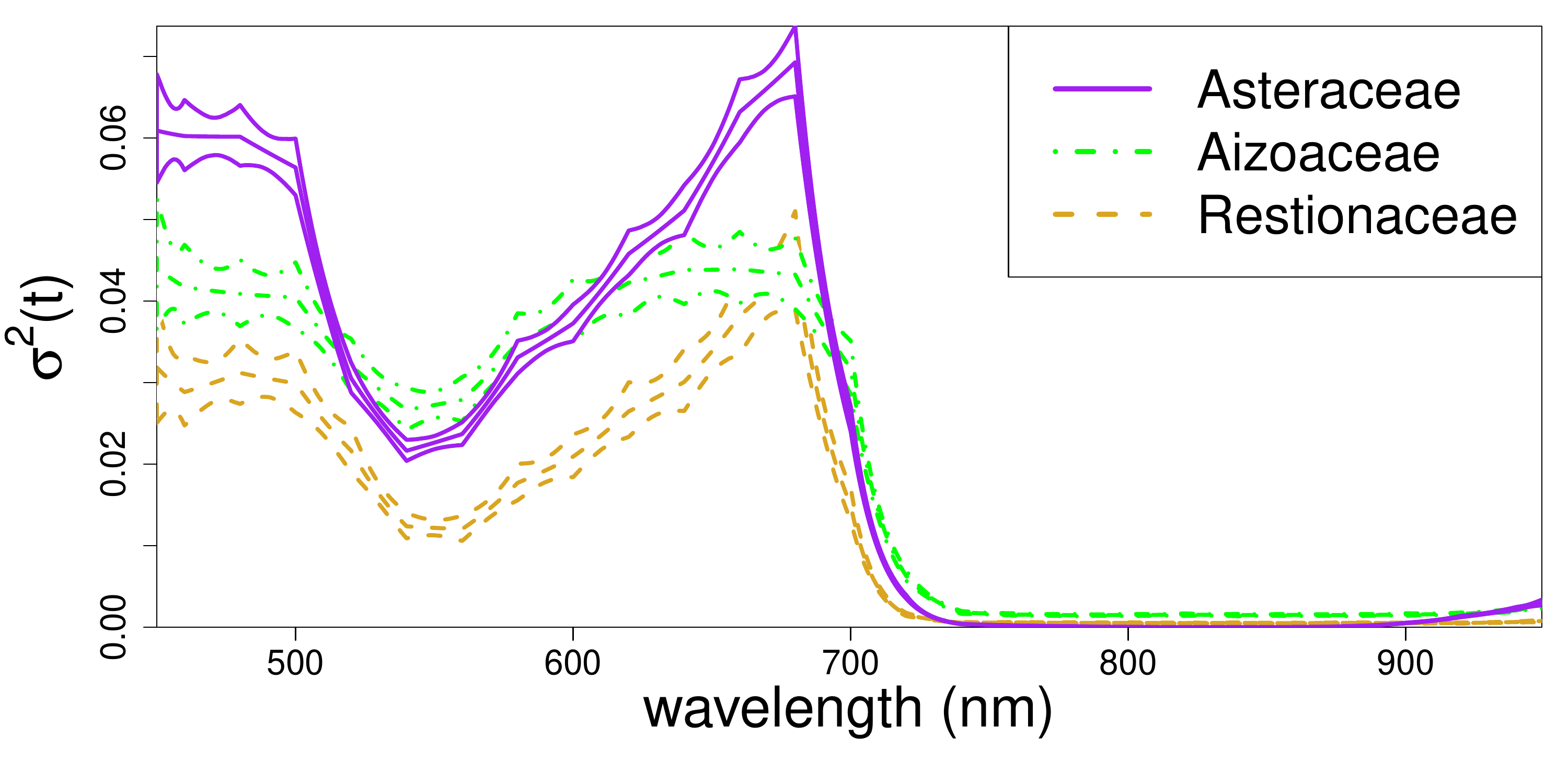}

\end{center}

\vspace{-4mm}

\caption{Between family comparison of (Top-Left to Middle-Right) environmental regression coefficient functions (Elevation, Annual Precipitation, Rainfall Concentration, and Average Minimum January Temperature) and (Bottom) wavelength-varying variance $\sigma^2(t)$.}\label{fig:family_coefs}
\end{figure}

We compare the regression coefficient functions for the three families in this study (posterior mean and 95\% credible interval): Aizoaceae, Asteraceae, and Restionaceae (See Figure \ref{fig:family_coefs}). The regression coefficient functions are clearly distinct across the families. However, between-covariate correlation or different spatial sites covered by each family may account for some of these differences.

The estimated effects of elevation, annual precipitation, and temperature are opposite in direction for all wavelengths between Asteraceae and Aizoaceae. For these covariates, we see positive effects on Aizoaceae log-Reflectance for wavelengths $<700$ nm and negative effects for wavelengths $>700$ nm, with opposite patterns for Asteraceae. For Asteraceae, the estimated effects of rainfall concentration are positive for lower wavelengths and negative for higher wavelengths, while they are nearly zero for Aizoaceae.
Restionaceae has very small estimated temperature effects. For elevation and rainfall concentration, Restionaceae shows significant effects on log-reflectance for wavelengths $< 700$ nm, but essentially no effect for higher wavelengths. The estimated effect of precipitation for Restionaceae is similar to Aizoaceae in pattern but is smaller in magnitude.

In Figure \ref{fig:family_coefs}, we also plot the variance function for $\epsilon_{ij}(\bs,t)$ for each family (posterior mean and 95\% credible interval). Asteraceae has the highest estimated variance for most low wavelengths (450 - 700 nm), a trend that matches the between spectrum variance patterns in Figure \ref{fig:within_between}. Restionaceae has the lowest estimated variance for (450 - 700 nm). All families have very low estimated variance for most high wavelengths (700 - 950 nm).

The differing responses in visible and near-infrared reflectance to environment between Aizoaceae and Asteraceae \ref{fig:family_coefs} likely indicate that genera within the two families employ different adaptive strategies in response to their local environments across the landscape. The Aizoaceae family consists of small succulent stemmed and leafed plants while the Asteraceae family largely consists of non-succulent leafed herbs and shrubs. Both plant families adapt via other traits tied to aridity tolerance (e.g., water storage for periods of drought) and avoidance (e.g., leaf hairs, wax, and anthocyanin pigmentation that block UV radiation). The adaptive traits in the respective "evolutionary toolboxes" of Aizoaceae and Asteraceae are constrained by their phylogenetic ancestry, resulting in differing strategic responses to environment in their traits and thus, reflectances. In contrast, the Restionaceae consist of grass-like plants with tough fibrous photosynthetic stems that vary less than the other two families in adaptation to drought.

\begin{figure}[H]
\centering
\begin{center}
\includegraphics[width=.8\textwidth]{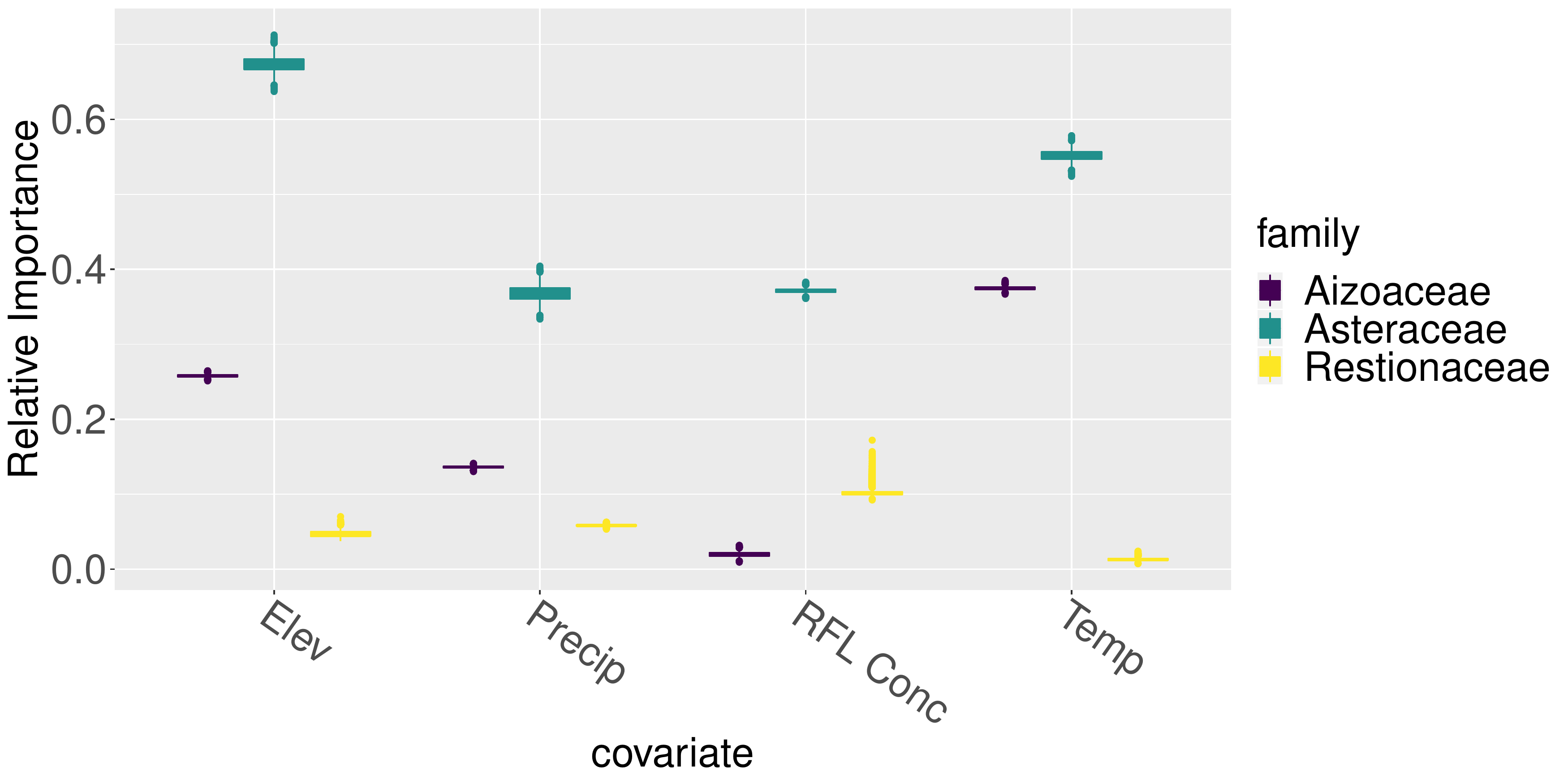}
\end{center}

 \vspace{-4mm}

\caption{Covariate importance $\overline{|\beta|}$ for all families.}\label{fig:family_import}
\end{figure}

We show the posterior distribution (box plots) for $\overline{|\beta_j|}$ across all covariates and families (See Figure \ref{fig:family_import}). Since $\overline{|\beta_j|}$ represents the relative importance of covariates for log-reflectance, we see that the covariates are more important in describing log-reflectances for Asteraceae than Aizoaceae and more important for Aizoaceae than for Restionaceae, except for rainfall concentration. Perhaps the relative importance $\overline{|\beta_j|}$ may be higher for Aizoaceae and Asteraceae because these have more expansive spatial distributions and thus experience higher variability in environmental variables.

Despite the differences in spatial ranges, the families differ in terms of which environmental variables have the highest relative importance to their reflectance signals. The most important variable for Asteraceae is elevation, likely a proxy for several environmental factors; prominent among them is the biome shift from the higher elevation Fynbos biome within the Cederberg mountains to the lower elevated Succulent Karoo biome. These biomes differ widely in their environments, the Fynbos biome having nutrient-poor soils and a regular fire cycle while the Succulent Karoo is largely arid with low levels of rainfall. Asteraceae is the only family to fully span both biomes in large numbers and these biomes feature a wide difference in environments.  The most important variable for Aizoaceae is the minimum average temperate in January (the peak austral summer month), a strong indicator of the maximum temperature a plant can tolerate. This suggests that the major driver of Aizoaceae reflectances are underlying adaptations related to heat tolerance/avoidance. While more limited in its spatial extent, the Restionaceae reflectance spectra responded most to rainfall concentration. Under the notion that higher concentrations of rainfall in fewer months out of the year would lead to more dramatic periods without water, much of the differences in Restionanceae reflectance may be in response to underlying traits managing water during times of drought.


\section{Summary and Future Work}\label{sec:conc}

We have offered plant reflectance modeling to capture variation over space between reflectance across genera within a family. We incorporate wavelength heterogeneity, spatial dependence, and also wavelength - covariate interaction as well as space - wavelength interaction.  We have fitted these models to reflectances from the Cape Floristic Region in South Africa, demonstrating successful model performance and revealing a range of novel inference as well as successful spatial prediction.

This work has several future applications and opportunities for further development. Our current data only included the visible and near-infrared reflectance spectra of leaves. These data could be expanded to include the reflectance of plant canopies across a broader spectral range to make predictions relevant to the reflectance spectra collected by broader band sensors aboard aerial and satellite remote sensing platforms. Our spatially explicit predictions of plant reflectance would be highly relevant for spectral unmixing analyses which seek to predict the abundances of spectral end members, i.e., individual species, in a canopy of vegetation. Future modeling efforts include exploring reflectance signatures following evolutionary history, explicitly taking into account phylogeny among different groups of plants.

Our space-wavelength model could also be adapted for space-time applications. 
For suitable spatiotemporal settings, it may be useful to construct spatial kernel convolutions of wavelength/temporal GPs. Also, our approach to spatial orthogonalization for functional regression coefficients could be applied to dynamic regression in spatiotemporal settings.

\section*{Acknowledgement}

We thank Matthew Aiello-Lammens, Douglas Euston-Brown, Hayley Kilroy Mollmann, Cory Merow, Jasper Slingsby, Helga van der Merwe, and Adam Wilson for their contributions in the data collection and curation. Special thanks to Cape Nature and the Northern Cape Department of Environment and Nature Conservation for permission to collection leaf spectra and traits. Data collection efforts were made possible by funding from National Science Foundation grant DEB-1046328 to J.A. Silander. Additional support was provided by NASA FINESST grant award 19-EARTH20-0266 to H.A. Frye and J.A. Silander.

\section*{Supplementary Material}
Extended data analysis, residual analysis, orthogonalization, and results. (LINK ADDED LATER)

\bibliographystyle{imsart-nameyear}
\bibliography{ref}

\begin{thebibliography}{41}

\bibitem[\protect\citeauthoryear{Asner and
  Martin}{2016}]{asner_spectranomics_2016}
\begin{barticle}[author]
\bauthor{\bsnm{Asner},~\bfnm{Gregory~P.}\binits{G.~P.}} \AND
  \bauthor{\bsnm{Martin},~\bfnm{Roberta~E.}\binits{R.~E.}}
(\byear{2016}).
\btitle{Spectranomics: {Emerging} science and conservation opportunities at the
  interface of biodiversity and remote sensing}.
\bjournal{Global Ecology and Conservation}
\bvolume{8}
\bpages{212--219}.
\bdoi{10.1016/j.gecco.2016.09.010}
\end{barticle}
\endbibitem

\bibitem[\protect\citeauthoryear{Asner et~al.}{2017}]{asner2017airborne}
\begin{barticle}[author]
\bauthor{\bsnm{Asner},~\bfnm{Gregory~P}\binits{G.~P.}},
  \bauthor{\bsnm{Martin},~\bfnm{Rainer~E}\binits{R.~E.}},
  \bauthor{\bsnm{Knapp},~\bfnm{DE}\binits{D.}},
  \bauthor{\bsnm{Tupayachi},~\bfnm{R}\binits{R.}},
  \bauthor{\bsnm{Anderson},~\bfnm{CB}\binits{C.}},
  \bauthor{\bsnm{Sinca},~\bfnm{F}\binits{F.}},
  \bauthor{\bsnm{Vaughn},~\bfnm{NR}\binits{N.}} \AND
  \bauthor{\bsnm{Llactayo},~\bfnm{W}\binits{W.}}
(\byear{2017}).
\btitle{Airborne laser-guided imaging spectroscopy to map forest trait
  diversity and guide conservation}.
\bjournal{Science}
\bvolume{355}
\bpages{385--389}.
\end{barticle}
\endbibitem

\bibitem[\protect\citeauthoryear{Besse, Cardot and
  Stephenson}{2000}]{besse2000autoregressive}
\begin{barticle}[author]
\bauthor{\bsnm{Besse},~\bfnm{Philippe~C}\binits{P.~C.}},
  \bauthor{\bsnm{Cardot},~\bfnm{Herv{\'e}}\binits{H.}} \AND
  \bauthor{\bsnm{Stephenson},~\bfnm{David~B}\binits{D.~B.}}
(\byear{2000}).
\btitle{Autoregressive forecasting of some functional climatic variations}.
\bjournal{Scandinavian Journal of Statistics}
\bvolume{27}
\bpages{673--687}.
\end{barticle}
\endbibitem

\bibitem[\protect\citeauthoryear{Cavender-Bares et~al.}{2016a}]{cavender2016}
\begin{barticle}[author]
\bauthor{\bsnm{Cavender-Bares},~\bfnm{Jeannine}\binits{J.}},
  \bauthor{\bsnm{Meireles},~\bfnm{Jose~Eduardo}\binits{J.~E.}},
  \bauthor{\bsnm{Couture},~\bfnm{John~J}\binits{J.~J.}},
  \bauthor{\bsnm{Kaproth},~\bfnm{Matthew~A}\binits{M.~A.}},
  \bauthor{\bsnm{Kingdon},~\bfnm{Clayton~C}\binits{C.~C.}},
  \bauthor{\bsnm{Singh},~\bfnm{Aditya}\binits{A.}},
  \bauthor{\bsnm{Serbin},~\bfnm{Shawn~P}\binits{S.~P.}},
  \bauthor{\bsnm{Center},~\bfnm{Alyson}\binits{A.}},
  \bauthor{\bsnm{Zuniga},~\bfnm{Esau}\binits{E.}} \AND
  \bauthor{\bsnm{Pilz},~\bfnm{George}\binits{G.}}
(\byear{2016}a).
\btitle{Associations of leaf spectra with genetic and phylogenetic variation in
  oaks: Prospects for remote detection of biodiversity}.
\bjournal{Remote Sensing}
\bvolume{8}
\bpages{221}.
\end{barticle}
\endbibitem

\bibitem[\protect\citeauthoryear{Cavender-Bares
  et~al.}{2016b}]{cavender-bares_associations_2016}
\begin{barticle}[author]
\bauthor{\bsnm{Cavender-Bares},~\bfnm{Jeannine}\binits{J.}},
  \bauthor{\bsnm{Meireles},~\bfnm{Jose}\binits{J.}},
  \bauthor{\bsnm{Couture},~\bfnm{John}\binits{J.}},
  \bauthor{\bsnm{Kaproth},~\bfnm{Matthew}\binits{M.}},
  \bauthor{\bsnm{Kingdon},~\bfnm{Clayton}\binits{C.}},
  \bauthor{\bsnm{Singh},~\bfnm{Aditya}\binits{A.}},
  \bauthor{\bsnm{Serbin},~\bfnm{Shawn}\binits{S.}},
  \bauthor{\bsnm{Center},~\bfnm{Alyson}\binits{A.}},
  \bauthor{\bsnm{Zuniga},~\bfnm{Esau}\binits{E.}},
  \bauthor{\bsnm{Pilz},~\bfnm{George}\binits{G.}} \AND
  \bauthor{\bsnm{Townsend},~\bfnm{Philip}\binits{P.}}
(\byear{2016}b).
\btitle{Associations of {Leaf} {Spectra} with {Genetic} and {Phylogenetic}
  {Variation} in {Oaks}: {Prospects} for {Remote} {Detection} of
  {Biodiversity}}.
\bjournal{Remote Sensing}
\bvolume{8}
\bpages{221}.
\bdoi{10.3390/rs8030221}
\end{barticle}
\endbibitem

\bibitem[\protect\citeauthoryear{Cawse-Nicholson}{2021}]{cawse-nicholson_nasas_2021}
\begin{barticle}[author]
\bauthor{\bsnm{Cawse-Nicholson},~\bfnm{Kerry}\binits{K.}}
(\byear{2021}).
\btitle{{NASA}'s surface biology and geology designated observable: {A}
  perspective on surface imaging algorithms}.
\bjournal{Remote Sensing of Environment}
\bvolume{257}
\bpages{112349}.
\bdoi{10.1016/j.rse.2021.112349}
\end{barticle}
\endbibitem

\bibitem[\protect\citeauthoryear{Clark, Roberts and Clark}{2005}]{clark2005}
\begin{barticle}[author]
\bauthor{\bsnm{Clark},~\bfnm{Matthew~L}\binits{M.~L.}},
  \bauthor{\bsnm{Roberts},~\bfnm{Dar~A}\binits{D.~A.}} \AND
  \bauthor{\bsnm{Clark},~\bfnm{David~B}\binits{D.~B.}}
(\byear{2005}).
\btitle{Hyperspectral discrimination of tropical rain forest tree species at
  leaf to crown scales}.
\bjournal{Remote Sensing of Environment}
\bvolume{96}
\bpages{375--398}.
\end{barticle}
\endbibitem

\bibitem[\protect\citeauthoryear{Cornwell
  et~al.}{2014}]{cornwell2014functional}
\begin{barticle}[author]
\bauthor{\bsnm{Cornwell},~\bfnm{William~K}\binits{W.~K.}},
  \bauthor{\bsnm{Westoby},~\bfnm{Mark}\binits{M.}},
  \bauthor{\bsnm{Falster},~\bfnm{Daniel~S}\binits{D.~S.}},
  \bauthor{\bsnm{FitzJohn},~\bfnm{Richard~G}\binits{R.~G.}},
  \bauthor{\bsnm{O'Meara},~\bfnm{Brian~C}\binits{B.~C.}},
  \bauthor{\bsnm{Pennell},~\bfnm{Matthew~W}\binits{M.~W.}},
  \bauthor{\bsnm{McGlinn},~\bfnm{Daniel~J}\binits{D.~J.}},
  \bauthor{\bsnm{Eastman},~\bfnm{Jonathan~M}\binits{J.~M.}},
  \bauthor{\bsnm{Moles},~\bfnm{Angela~T}\binits{A.~T.}} \AND
  \bauthor{\bsnm{Reich},~\bfnm{Peter~B}\binits{P.~B.}}
(\byear{2014}).
\btitle{Functional distinctiveness of major plant lineages}.
\bjournal{Journal of Ecology}
\bvolume{102}
\bpages{345--356}.
\end{barticle}
\endbibitem

\bibitem[\protect\citeauthoryear{Doughty}{2017}]{doughty_can_2017}
\begin{barticle}[author]
\bauthor{\bsnm{Doughty},~\bfnm{Christopher~E.}\binits{C.~E.}}
(\byear{2017}).
\btitle{Can {Leaf} {Spectroscopy} {Predict} {Leaf} and {Forest} {Traits}
  {Along} a {Peruvian} {Tropical} {Forest} {Elevation} {Gradient}?: {Amazonian}
  leaf spectroscopy and traits}.
\bjournal{Journal of Geophysical Research: Biogeosciences}
\bvolume{122}
\bpages{2952--2965}.
\bdoi{10.1002/2017JG003883}
\end{barticle}
\endbibitem

\bibitem[\protect\citeauthoryear{Feng et~al.}{2008}]{feng2008monitoring}
\begin{barticle}[author]
\bauthor{\bsnm{Feng},~\bfnm{W}\binits{W.}},
  \bauthor{\bsnm{Yao},~\bfnm{X}\binits{X.}},
  \bauthor{\bsnm{Zhu},~\bfnm{Y}\binits{Y.}},
  \bauthor{\bsnm{Tian},~\bfnm{YC}\binits{Y.}} \AND
  \bauthor{\bsnm{Cao},~\bfnm{WX}\binits{W.}}
(\byear{2008}).
\btitle{Monitoring leaf nitrogen status with hyperspectral reflectance in
  wheat}.
\bjournal{European Journal of Agronomy}
\bvolume{28}
\bpages{394--404}.
\end{barticle}
\endbibitem

\bibitem[\protect\citeauthoryear{Féret}{2019}]{feret_estimating_2019}
\begin{barticle}[author]
\bauthor{\bsnm{Féret},~\bfnm{J.~B.}\binits{J.~B.}}
(\byear{2019}).
\btitle{Estimating leaf mass per area and equivalent water thickness based on
  leaf optical properties: {Potential} and limitations of physical modeling and
  machine learning}.
\bjournal{Remote Sensing of Environment}
\bvolume{231}
\bpages{110959}.
\bdoi{10.1016/j.rse.2018.11.002}
\end{barticle}
\endbibitem

\bibitem[\protect\citeauthoryear{Gamon et~al.}{2020}]{gamon_consideration_2020}
\begin{bincollection}[author]
\bauthor{\bsnm{Gamon},~\bfnm{John~A}\binits{J.~A.}},
  \bauthor{\bsnm{Wang},~\bfnm{Ran}\binits{R.}},
  \bauthor{\bsnm{Gholizadeh},~\bfnm{Hamed}\binits{H.}},
  \bauthor{\bsnm{Zutta},~\bfnm{Brian}\binits{B.}},
  \bauthor{\bsnm{Townsend},~\bfnm{Phil~A}\binits{P.~A.}} \AND
  \bauthor{\bsnm{Cavender-Bares},~\bfnm{Jeannine}\binits{J.}}
(\byear{2020}).
\btitle{Consideration of scale in remote sensing of biodiversity}.
In \bbooktitle{Remote {Sensing} of {Plant} {Biodiversity}}
\bpages{425--447}.
\bpublisher{Springer, Cham}.
\end{bincollection}
\endbibitem

\bibitem[\protect\citeauthoryear{Gneiting and
  Raftery}{2007}]{gneiting2007strictly}
\begin{barticle}[author]
\bauthor{\bsnm{Gneiting},~\bfnm{Tilmann}\binits{T.}} \AND
  \bauthor{\bsnm{Raftery},~\bfnm{Adrian~E}\binits{A.~E.}}
(\byear{2007}).
\btitle{Strictly proper scoring rules, prediction, and estimation}.
\bjournal{Journal of the American statistical Association}
\bvolume{102}
\bpages{359--378}.
\end{barticle}
\endbibitem

\bibitem[\protect\citeauthoryear{Higdon}{1998}]{higdon1998process}
\begin{barticle}[author]
\bauthor{\bsnm{Higdon},~\bfnm{David}\binits{D.}}
(\byear{1998}).
\btitle{A process-convolution approach to modelling temperatures in the {N}orth
  {A}tlantic {O}cean}.
\bjournal{Environmental and Ecological Statistics}
\bvolume{5}
\bpages{173--190}.
\end{barticle}
\endbibitem

\bibitem[\protect\citeauthoryear{Higdon}{2002}]{higdon2002space}
\begin{bincollection}[author]
\bauthor{\bsnm{Higdon},~\bfnm{Dave}\binits{D.}}
(\byear{2002}).
\btitle{Space and space-time modeling using process convolutions}.
In \bbooktitle{Quantitative Methods for Current Environmental Issues}
\bpages{37--56}.
\bpublisher{Springer}.
\end{bincollection}
\endbibitem

\bibitem[\protect\citeauthoryear{Hodges and Reich}{2010}]{hodges2010adding}
\begin{barticle}[author]
\bauthor{\bsnm{Hodges},~\bfnm{James~S}\binits{J.~S.}} \AND
  \bauthor{\bsnm{Reich},~\bfnm{Brian~J}\binits{B.~J.}}
(\byear{2010}).
\btitle{Adding spatially-correlated errors can mess up the fixed effect you
  love}.
\bjournal{The American Statistician}
\bvolume{64}
\bpages{325--334}.
\end{barticle}
\endbibitem

\bibitem[\protect\citeauthoryear{Jacquemoud and
  Baret}{1990}]{jacquemoud_prospect:_1990}
\begin{barticle}[author]
\bauthor{\bsnm{Jacquemoud},~\bfnm{S.}\binits{S.}} \AND
  \bauthor{\bsnm{Baret},~\bfnm{F.}\binits{F.}}
(\byear{1990}).
\btitle{{PROSPECT}: {A} model of leaf optical properties spectra}.
\bjournal{Remote Sensing of Environment}
\bvolume{34}
\bpages{75--91}.
\bdoi{10.1016/0034-4257(90)90100-Z}
\end{barticle}
\endbibitem

\bibitem[\protect\citeauthoryear{Jacquemoud and
  Ustin}{2019a}]{jacquemoud_modeling_2019}
\begin{bincollection}[author]
\bauthor{\bsnm{Jacquemoud},~\bfnm{S.}\binits{S.}} \AND
  \bauthor{\bsnm{Ustin},~\bfnm{Susan}\binits{S.}}
(\byear{2019}a).
\btitle{Modeling {Leaf} {Optical} {Properties}: {PROSPECT}}.
In \bbooktitle{Leaf {Optical} {Properties}}
\bpublisher{Cambridge University Press}.
\end{bincollection}
\endbibitem

\bibitem[\protect\citeauthoryear{Jacquemoud and
  Ustin}{2019b}]{jacquemoud_ustin_2019b}
\begin{binbook}[author]
\bauthor{\bsnm{Jacquemoud},~\bfnm{Stéphane}\binits{S.}} \AND
  \bauthor{\bsnm{Ustin},~\bfnm{Susan}\binits{S.}}
(\byear{2019}b).
\btitle{Variation Due to Leaf Structural, Chemical, and Physiological Traits}
In \bbooktitle{Leaf Optical Properties}
\bpages{170?194}.
\bpublisher{Cambridge University Press}.
\bdoi{10.1017/9781108686457.006}
\end{binbook}
\endbibitem

\bibitem[\protect\citeauthoryear{Jacquemoud and
  Ustin}{2019c}]{jacquemoud_ustin_2019a}
\begin{binbook}[author]
\bauthor{\bsnm{Jacquemoud},~\bfnm{Stéphane}\binits{S.}} \AND
  \bauthor{\bsnm{Ustin},~\bfnm{Susan}\binits{S.}}
(\byear{2019}c).
\btitle{Leaf Optical Properties in Different Wavelength Domains}
In \bbooktitle{Leaf Optical Properties}
\bpages{124?169}.
\bpublisher{Cambridge University Press}.
\bdoi{10.1017/9781108686457.005}
\end{binbook}
\endbibitem

\bibitem[\protect\citeauthoryear{Khan and Calder}{2020}]{khan2020restricted}
\begin{barticle}[author]
\bauthor{\bsnm{Khan},~\bfnm{Kori}\binits{K.}} \AND
  \bauthor{\bsnm{Calder},~\bfnm{Catherine~A}\binits{C.~A.}}
(\byear{2020}).
\btitle{Restricted Spatial Regression Methods: Implications for Inference}.
\bjournal{Journal of the American Statistical Association}
\bpages{1--13}.
\end{barticle}
\endbibitem

\bibitem[\protect\citeauthoryear{Kokaly et~al.}{2009}]{kokaly2009}
\begin{barticle}[author]
\bauthor{\bsnm{Kokaly},~\bfnm{Raymond~F}\binits{R.~F.}},
  \bauthor{\bsnm{Asner},~\bfnm{Gregory~P}\binits{G.~P.}},
  \bauthor{\bsnm{Ollinger},~\bfnm{Scott~V}\binits{S.~V.}},
  \bauthor{\bsnm{Martin},~\bfnm{Mary~E}\binits{M.~E.}} \AND
  \bauthor{\bsnm{Wessman},~\bfnm{Carol~A}\binits{C.~A.}}
(\byear{2009}).
\btitle{Characterizing canopy biochemistry from imaging spectroscopy and its
  application to ecosystem studies}.
\bjournal{Remote Sensing of Environment}
\bvolume{113}
\bpages{S78--S91}.
\end{barticle}
\endbibitem

\bibitem[\protect\citeauthoryear{Laukaitis}{2008}]{laukaitis2008functional}
\begin{barticle}[author]
\bauthor{\bsnm{Laukaitis},~\bfnm{Algirdas}\binits{A.}}
(\byear{2008}).
\btitle{Functional data analysis for cash flow and transactions intensity
  continuous-time prediction using {H}ilbert-valued autoregressive processes}.
\bjournal{European Journal of Operational Research}
\bvolume{185}
\bpages{1607--1614}.
\end{barticle}
\endbibitem

\bibitem[\protect\citeauthoryear{Leng and
  M{\"u}ller}{2006}]{leng2006classification}
\begin{barticle}[author]
\bauthor{\bsnm{Leng},~\bfnm{Xiaoyan}\binits{X.}} \AND
  \bauthor{\bsnm{M{\"u}ller},~\bfnm{Hans-Georg}\binits{H.-G.}}
(\byear{2006}).
\btitle{Classification using functional data analysis for temporal gene
  expression data}.
\bjournal{Bioinformatics}
\bvolume{22}
\bpages{68--76}.
\end{barticle}
\endbibitem

\bibitem[\protect\citeauthoryear{Locantore et~al.}{1999}]{locantore1999robust}
\begin{barticle}[author]
\bauthor{\bsnm{Locantore},~\bfnm{N}\binits{N.}},
  \bauthor{\bsnm{Marron},~\bfnm{JS}\binits{J.}},
  \bauthor{\bsnm{Simpson},~\bfnm{DG}\binits{D.}},
  \bauthor{\bsnm{Tripoli},~\bfnm{N}\binits{N.}},
  \bauthor{\bsnm{Zhang},~\bfnm{JT}\binits{J.}},
  \bauthor{\bsnm{Cohen},~\bfnm{KL}\binits{K.}},
  \bauthor{\bsnm{Boente},~\bfnm{Graciela}\binits{G.}},
  \bauthor{\bsnm{Fraiman},~\bfnm{Ricardo}\binits{R.}},
  \bauthor{\bsnm{Brumback},~\bfnm{Babette}\binits{B.}} \AND
  \bauthor{\bsnm{Croux},~\bfnm{Christophe}\binits{C.}}
(\byear{1999}).
\btitle{Robust principal component analysis for functional data}.
\bjournal{Test}
\bvolume{8}
\bpages{1--73}.
\end{barticle}
\endbibitem

\bibitem[\protect\citeauthoryear{Ordo{\~n}ez
  et~al.}{2010}]{ordonez2010functional}
\begin{barticle}[author]
\bauthor{\bsnm{Ordo{\~n}ez},~\bfnm{C}\binits{C.}},
  \bauthor{\bsnm{Mart{\'\i}nez},~\bfnm{Javier}\binits{J.}},
  \bauthor{\bsnm{Mat{\'\i}as},~\bfnm{Jos{\'e}~M}\binits{J.~M.}},
  \bauthor{\bsnm{Reyes},~\bfnm{AN}\binits{A.}} \AND
  \bauthor{\bsnm{Rodr{\'\i}guez-P{\'e}rez},~\bfnm{Jos{\'e}~R}\binits{J.~R.}}
(\byear{2010}).
\btitle{Functional statistical techniques applied to vine leaf water content
  determination}.
\bjournal{Mathematical and Computer Modelling}
\bvolume{52}
\bpages{1116--1122}.
\end{barticle}
\endbibitem

\bibitem[\protect\citeauthoryear{Quintano
  et~al.}{2012}]{quintano_spectral_2012}
\begin{barticle}[author]
\bauthor{\bsnm{Quintano},~\bfnm{Carmen}\binits{C.}},
  \bauthor{\bsnm{Fernández-Manso},~\bfnm{Alfonso}\binits{A.}},
  \bauthor{\bsnm{Shimabukuro},~\bfnm{Yosio~E.}\binits{Y.~E.}} \AND
  \bauthor{\bsnm{Pereira},~\bfnm{Gabriel}\binits{G.}}
(\byear{2012}).
\btitle{Spectral unmixing}.
\bjournal{International Journal of Remote Sensing}
\bvolume{33}
\bpages{5307--5340}.
\bdoi{10.1080/01431161.2012.661095}
\end{barticle}
\endbibitem

\bibitem[\protect\citeauthoryear{Ramsay}{2005}]{ramsay2005}
\begin{barticle}[author]
\bauthor{\bsnm{Ramsay},~\bfnm{James}\binits{J.}}
(\byear{2005}).
\btitle{Functional data analysis}.
\bjournal{Encyclopedia of Statistics in Behavioral Science}.
\end{barticle}
\endbibitem

\bibitem[\protect\citeauthoryear{Ramsay and
  Silverman}{2007}]{ramsay2007applied}
\begin{bbook}[author]
\bauthor{\bsnm{Ramsay},~\bfnm{James~O}\binits{J.~O.}} \AND
  \bauthor{\bsnm{Silverman},~\bfnm{Bernard~W}\binits{B.~W.}}
(\byear{2007}).
\btitle{Applied functional data analysis: Methods and case studies}.
\bpublisher{Springer}.
\end{bbook}
\endbibitem

\bibitem[\protect\citeauthoryear{Reich et~al.}{2003}]{reich2003evolution}
\begin{barticle}[author]
\bauthor{\bsnm{Reich},~\bfnm{Peter~B}\binits{P.~B.}},
  \bauthor{\bsnm{Wright},~\bfnm{Ian~J}\binits{I.~J.}},
  \bauthor{\bsnm{Cavender-Bares},~\bfnm{Jeannine}\binits{J.}},
  \bauthor{\bsnm{Craine},~\bfnm{JM}\binits{J.}},
  \bauthor{\bsnm{Oleksyn},~\bfnm{J}\binits{J.}},
  \bauthor{\bsnm{Westoby},~\bfnm{M}\binits{M.}} \AND
  \bauthor{\bsnm{Walters},~\bfnm{MB}\binits{M.}}
(\byear{2003}).
\btitle{The evolution of plant functional variation: Traits, spectra, and
  strategies}.
\bjournal{International Journal of Plant Sciences}
\bvolume{164}
\bpages{S143--S164}.
\end{barticle}
\endbibitem

\bibitem[\protect\citeauthoryear{Reiss and Ogden}{2007}]{reiss2007functional}
\begin{barticle}[author]
\bauthor{\bsnm{Reiss},~\bfnm{Philip~T}\binits{P.~T.}} \AND
  \bauthor{\bsnm{Ogden},~\bfnm{R~Todd}\binits{R.~T.}}
(\byear{2007}).
\btitle{Functional principal component regression and functional partial least
  squares}.
\bjournal{Journal of the American Statistical Association}
\bvolume{102}
\bpages{984--996}.
\end{barticle}
\endbibitem

\bibitem[\protect\citeauthoryear{Schneider et~al.}{2017}]{schneider2017}
\begin{barticle}[author]
\bauthor{\bsnm{Schneider},~\bfnm{Fabian~D}\binits{F.~D.}},
  \bauthor{\bsnm{Morsdorf},~\bfnm{Felix}\binits{F.}},
  \bauthor{\bsnm{Schmid},~\bfnm{Bernhard}\binits{B.}},
  \bauthor{\bsnm{Petchey},~\bfnm{Owen~L}\binits{O.~L.}},
  \bauthor{\bsnm{Hueni},~\bfnm{Andreas}\binits{A.}},
  \bauthor{\bsnm{Schimel},~\bfnm{David~S}\binits{D.~S.}} \AND
  \bauthor{\bsnm{Schaepman},~\bfnm{Michael~E}\binits{M.~E.}}
(\byear{2017}).
\btitle{Mapping functional diversity from remotely sensed morphological and
  physiological forest traits}.
\bjournal{Nature Communications}
\bvolume{8}
\bpages{1--12}.
\end{barticle}
\endbibitem

\bibitem[\protect\citeauthoryear{Schweiger et~al.}{2018}]{schweiger2018}
\begin{barticle}[author]
\bauthor{\bsnm{Schweiger},~\bfnm{Anna~K}\binits{A.~K.}},
  \bauthor{\bsnm{Cavender-Bares},~\bfnm{Jeannine}\binits{J.}},
  \bauthor{\bsnm{Townsend},~\bfnm{Philip~A}\binits{P.~A.}},
  \bauthor{\bsnm{Hobbie},~\bfnm{Sarah~E}\binits{S.~E.}},
  \bauthor{\bsnm{Madritch},~\bfnm{Michael~D}\binits{M.~D.}},
  \bauthor{\bsnm{Wang},~\bfnm{Ran}\binits{R.}},
  \bauthor{\bsnm{Tilman},~\bfnm{David}\binits{D.}} \AND
  \bauthor{\bsnm{Gamon},~\bfnm{John~A}\binits{J.~A.}}
(\byear{2018}).
\btitle{Plant spectral diversity integrates functional and phylogenetic
  components of biodiversity and predicts ecosystem function}.
\bjournal{Nature Ecology \& Evolution}
\bvolume{2}
\bpages{976--982}.
\end{barticle}
\endbibitem

\bibitem[\protect\citeauthoryear{Shi and Wang}{2014}]{shi_incorporating_2014}
\begin{barticle}[author]
\bauthor{\bsnm{Shi},~\bfnm{Chen}\binits{C.}} \AND
  \bauthor{\bsnm{Wang},~\bfnm{Le}\binits{L.}}
(\byear{2014}).
\btitle{Incorporating spatial information in spectral unmixing: {A} review}.
\bjournal{Remote Sensing of Environment}
\bvolume{149}
\bpages{70--87}.
\bdoi{10.1016/j.rse.2014.03.034}
\end{barticle}
\endbibitem

\bibitem[\protect\citeauthoryear{Slingsby and
  Wistow}{2014}]{slingsby2014functions}
\begin{barticle}[author]
\bauthor{\bsnm{Slingsby},~\bfnm{Christine}\binits{C.}} \AND
  \bauthor{\bsnm{Wistow},~\bfnm{Graeme~J}\binits{G.~J.}}
(\byear{2014}).
\btitle{Functions of crystallins in and out of lens: Roles in elongated and
  post-mitotic cells}.
\bjournal{Progress in biophysics and molecular biology}
\bvolume{115}
\bpages{52--67}.
\end{barticle}
\endbibitem

\bibitem[\protect\citeauthoryear{Spiegelhalter
  et~al.}{2002}]{spiegelhalter2002}
\begin{barticle}[author]
\bauthor{\bsnm{Spiegelhalter},~\bfnm{David~J}\binits{D.~J.}},
  \bauthor{\bsnm{Best},~\bfnm{Nicola~G}\binits{N.~G.}},
  \bauthor{\bsnm{Carlin},~\bfnm{Bradley~P}\binits{B.~P.}} \AND
  \bauthor{\bsnm{Van Der~Linde},~\bfnm{Angelika}\binits{A.}}
(\byear{2002}).
\btitle{Bayesian measures of model complexity and fit}.
\bjournal{Journal of the Royal Statistical Society: Series B (Statistical
  Methodology)}
\bvolume{64}
\bpages{583--639}.
\end{barticle}
\endbibitem

\bibitem[\protect\citeauthoryear{Tian et~al.}{2010}]{tian2010cortical}
\begin{barticle}[author]
\bauthor{\bsnm{Tian},~\bfnm{Peifang}\binits{P.}},
  \bauthor{\bsnm{Teng},~\bfnm{Ivan~C}\binits{I.~C.}},
  \bauthor{\bsnm{May},~\bfnm{Larry~D}\binits{L.~D.}},
  \bauthor{\bsnm{Kurz},~\bfnm{Ronald}\binits{R.}},
  \bauthor{\bsnm{Lu},~\bfnm{Kun}\binits{K.}},
  \bauthor{\bsnm{Scadeng},~\bfnm{Miriam}\binits{M.}},
  \bauthor{\bsnm{Hillman},~\bfnm{Elizabeth~MC}\binits{E.~M.}},
  \bauthor{\bsnm{De~Crespigny},~\bfnm{Alex~J}\binits{A.~J.}},
  \bauthor{\bsnm{D’Arceuil},~\bfnm{Helen~E}\binits{H.~E.}} \AND
  \bauthor{\bsnm{Mandeville},~\bfnm{Joseph~B}\binits{J.~B.}}
(\byear{2010}).
\btitle{Cortical depth-specific microvascular dilation underlies laminar
  differences in blood oxygenation level-dependent functional {MRI} signal}.
\bjournal{Proceedings of the National Academy of Sciences}
\bvolume{107}
\bpages{15246--15251}.
\end{barticle}
\endbibitem

\bibitem[\protect\citeauthoryear{Ullah and Finch}{2013}]{ullah2013}
\begin{barticle}[author]
\bauthor{\bsnm{Ullah},~\bfnm{Shahid}\binits{S.}} \AND
  \bauthor{\bsnm{Finch},~\bfnm{Caroline~F}\binits{C.~F.}}
(\byear{2013}).
\btitle{Applications of functional data analysis: A systematic review}.
\bjournal{BMC Medical Research Methodology}
\bvolume{13}
\bpages{43}.
\end{barticle}
\endbibitem

\bibitem[\protect\citeauthoryear{Wackernagel}{1998}]{wackernagel1998}
\begin{bbook}[author]
\bauthor{\bsnm{Wackernagel},~\bfnm{Hans}\binits{H.}}
(\byear{1998}).
\btitle{Multivariate Geostatistics}.
\bpublisher{Springer}.
\end{bbook}
\endbibitem

\bibitem[\protect\citeauthoryear{White and
  Gelfand}{2020}]{white2020multivariate}
\begin{barticle}[author]
\bauthor{\bsnm{White},~\bfnm{Philip~A}\binits{P.~A.}} \AND
  \bauthor{\bsnm{Gelfand},~\bfnm{Alan~E}\binits{A.~E.}}
(\byear{2020}).
\btitle{Multivariate functional data modeling with time-varying clustering}.
\bjournal{TEST}
\bpages{1--17}.
\end{barticle}
\endbibitem

\bibitem[\protect\citeauthoryear{White, Keeler and
  Rupper}{2021}]{white2020hierarchical}
\begin{barticle}[author]
\bauthor{\bsnm{White},~\bfnm{Philip~A}\binits{P.~A.}},
  \bauthor{\bsnm{Keeler},~\bfnm{Durban~G}\binits{D.~G.}} \AND
  \bauthor{\bsnm{Rupper},~\bfnm{Summer}\binits{S.}}
(\byear{2021}).
\btitle{Hierarchical Integrated Spatial Process Modeling of Monotone West
  {A}ntarctic Snow Density Curves}.
\bjournal{To appear in Annals of Applied Statistics}.
\end{barticle}
\endbibitem

\end{thebibliography}


\begin{thebibliography}{}

\bibitem[Spiegelhalter et~al., 2002]{spiegelhalter2002}
Spiegelhalter, D.~J., Best, N.~G., Carlin, B.~P., and Van Der~Linde, A. (2002).
\newblock Bayesian measures of model complexity and fit.
\newblock {\em Journal of the Royal Statistical Society: Series B (Statistical
  Methodology)}, 64(4):583--639.

\bibitem[Tanner, 1996]{tanner1996}
Tanner, M.~A. (1996).
\newblock {\em Tools for Statistical Inference}.
\newblock Springer-Verlag New York, 3 edition.

\end{thebibliography}

\end{document}


\begin{frontmatter}
\title{Supplement to Spatial Functional Data Modeling of Plant Reflectances}
\runtitle{Spatial Modeling of Reflectance}

\begin{aug}
\author{\fnms{Philip A.} \snm{White}\thanksref{t1,m1}\ead[label=e1]{pwhite@stat.byu.edu}},
\author{\fnms{Henry} \snm{Frye}\thanksref{m2}},
\author{\fnms{Michael F.} \snm{Christensen}\thanksref{m3}},
\author{\fnms{Alan E.} \snm{Gelfand}\thanksref{m3}},
\and
\author{\fnms{John A.} \snm{Silander, Jr.}\thanksref{m2}}

\thankstext{t1}{Corresponding Author}
\runauthor{P. White et al.}

\affiliation{Brigham Young University \thanksmark{m1}, University of Connecticut\thanksmark{m2}, and Duke University\thanksmark{m3}}

%
\end{aug}

\begin{abstract}

In the Supplemental Material, we present extended data exploration and preliminary analysis in Section \ref{sec:data}. Specifically, we explore family differences in occurrence, co-occurrence, and location. We also provide detailed residual analysis to motivate our modeling. In Section \ref{sec:mod}, we present additional model development. In Section \ref{sec:cv_eta}, we present the results of the cross-validation for the Asteraceae used to select the form of $\eta(\bs,t)$. We provide an extended sensitivity analysis in Section \ref{sec:sensitivity}. For the model with the best out-of-sample prediction, we present prior distributions, model fitting, and prediction details in Section \ref{sec:gibbs}. Section \ref{sec:ortho} provides expanded discussion on the orthogonalization approach presented in the manuscript. Lastly, we present additional results about the regression coefficient functions for all three families in Section \ref{sec:results}.

\end{abstract}

\end{frontmatter}

\section{Extended Exploratory Analysis}\label{sec:data}

Here, we extend and elaborate on the exploratory analysis in the manuscript. In particular, we summarize simple aspects of the data in Section \ref{sec:dat_sum}, specifically discussing the number of genera in each family and family/genus duplication and co-occurrence. We also explore family and genus-level occurrence by location, as well as some assessment of family differences. In Section \ref{sec:dat_locations}, we provide visual examinations of the family, genus, and species distributions over the CFR region (the region of analysis discussed in the manuscript). In Section \ref{sec:residual}, we include extended residual analysis to motivate the space-wavelength hierarchical model proposed in the manuscript. 

\subsection{Data Summaries}\label{sec:dat_sum}

\begin{figure}[H]
 \begin{center}
 \includegraphics[width = 0.32\textwidth]{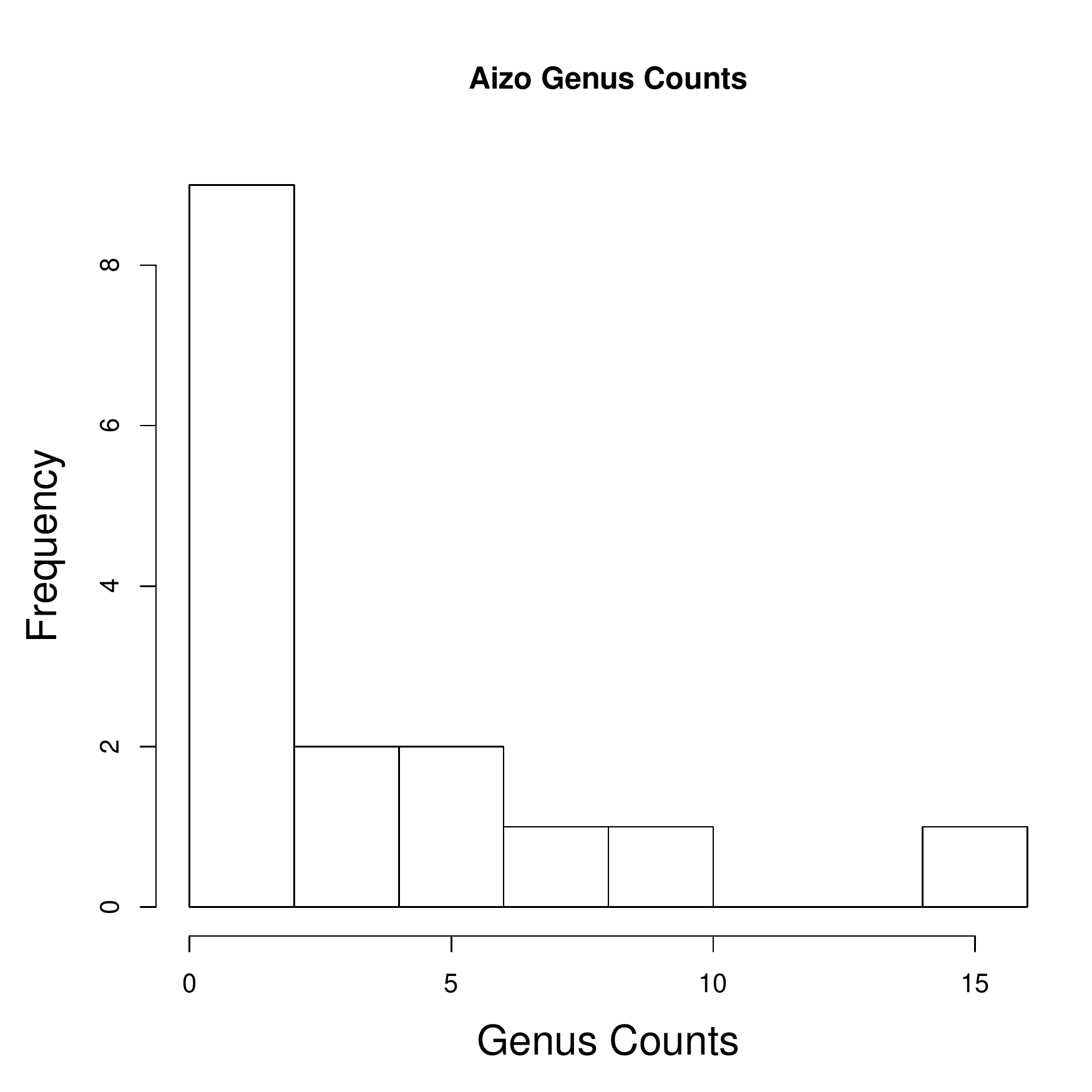}
 \includegraphics[width = 0.32\textwidth]{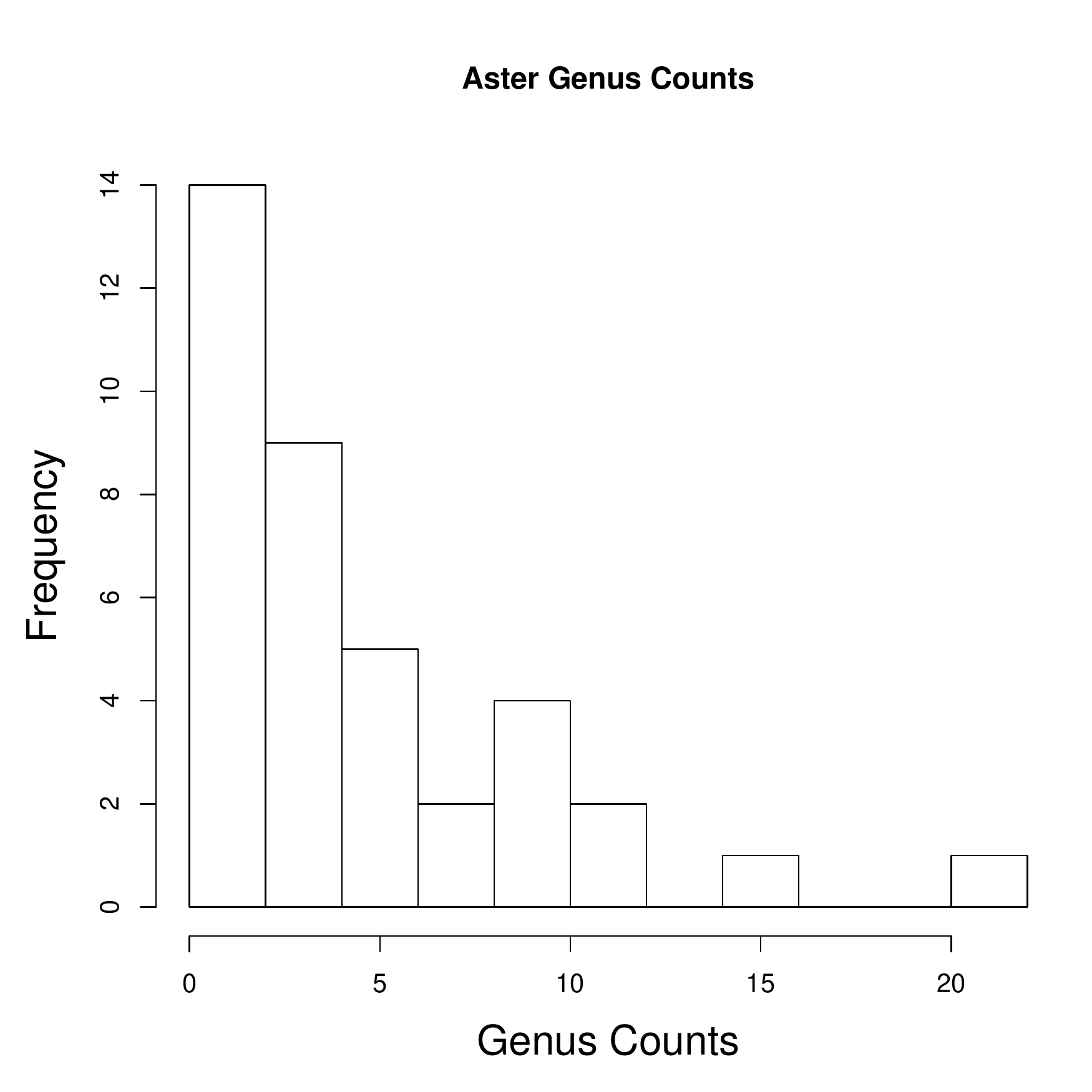}
 \includegraphics[width = 0.32\textwidth]{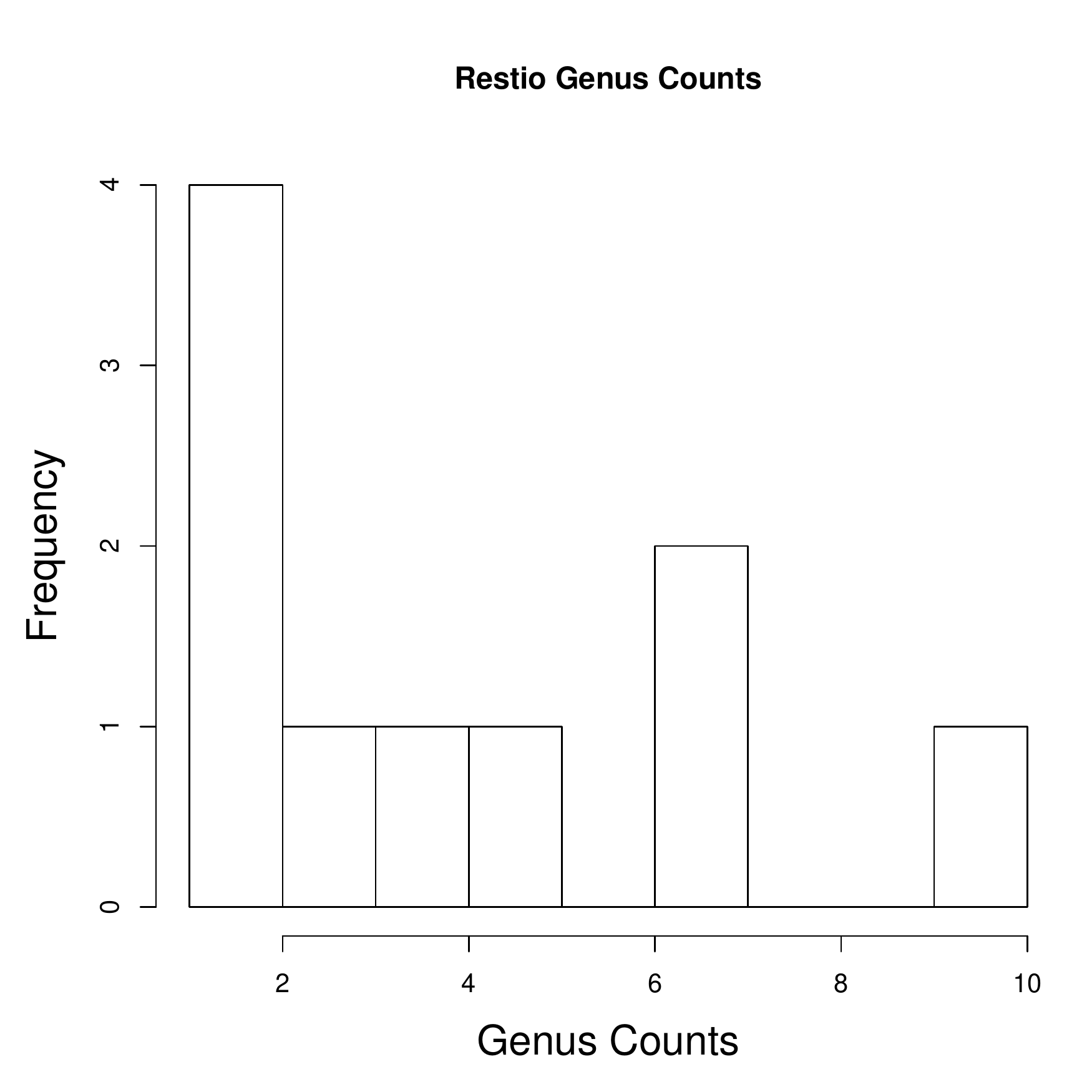}
 \end{center}
 \caption{Number of observed curves for each genus for (Left) Aizoaceae, (Center) Asteraceae, and (Right) Restionaceae.}\label{fig:genus_hist}
 \end{figure}

To better understand the within family diversity present in this dataset, we look at the number of genera with an observed reflectance curve within each family: Aizoaceae - 16, Asteraceae - 38, and Restionaceae - 10. Importantly, many genera are found only once in the data set, and most are observed only a few times. To illustrate this, we plot the histogram of the number of observed reflectance curves by genus for each family in Figure \ref{fig:genus_hist}. We note that for all families most genera are observed fewer than five times. 

In addition, we provide the following summaries: (i) the proportion of sites where each family is present (see Table \ref{tab:site_prop}), (ii) the number of sites with one, two, or three families (see Table \ref{tab:unique_fams}) and (iii) a more detailed breakdown of family co-occurrence (see Table \ref{tab:cooccur}). Asteraceae has the widest spatial spread in the CFR region. Aizoaceae and Restionaceae rarely co-occur, suggesting that joint modeling will not be successful. Importantly, Restionaceae is mostly limited to the Cederberg region with the exception of a couple of HTR observations.


\begin{table}[H]
\centering
\begin{tabular}{rr}
  \hline
 & Proportion of Sites With Family Present \\
  \hline
AIZOACEAE & 0.31 \\
  ASTERACEAE & 0.77 \\
  RESTIONACEAE & 0.21 \\
   \hline
\end{tabular}
\caption{Proportion of sites where each family was present.}\label{tab:site_prop}
\end{table}

\begin{table}[H]
\centering
\begin{tabular}{rr}
  \hline
Number of Families & Number of Sites \\
  \hline
1 &  96 \\
  2 & 35 \\
  3 &  2 \\
   \hline
\end{tabular}
\caption{Number of unique families at each site.}\label{tab:unique_fams}
\end{table}

\begin{figure}[H]
 \begin{center}
 \includegraphics[width = 0.45\textwidth]{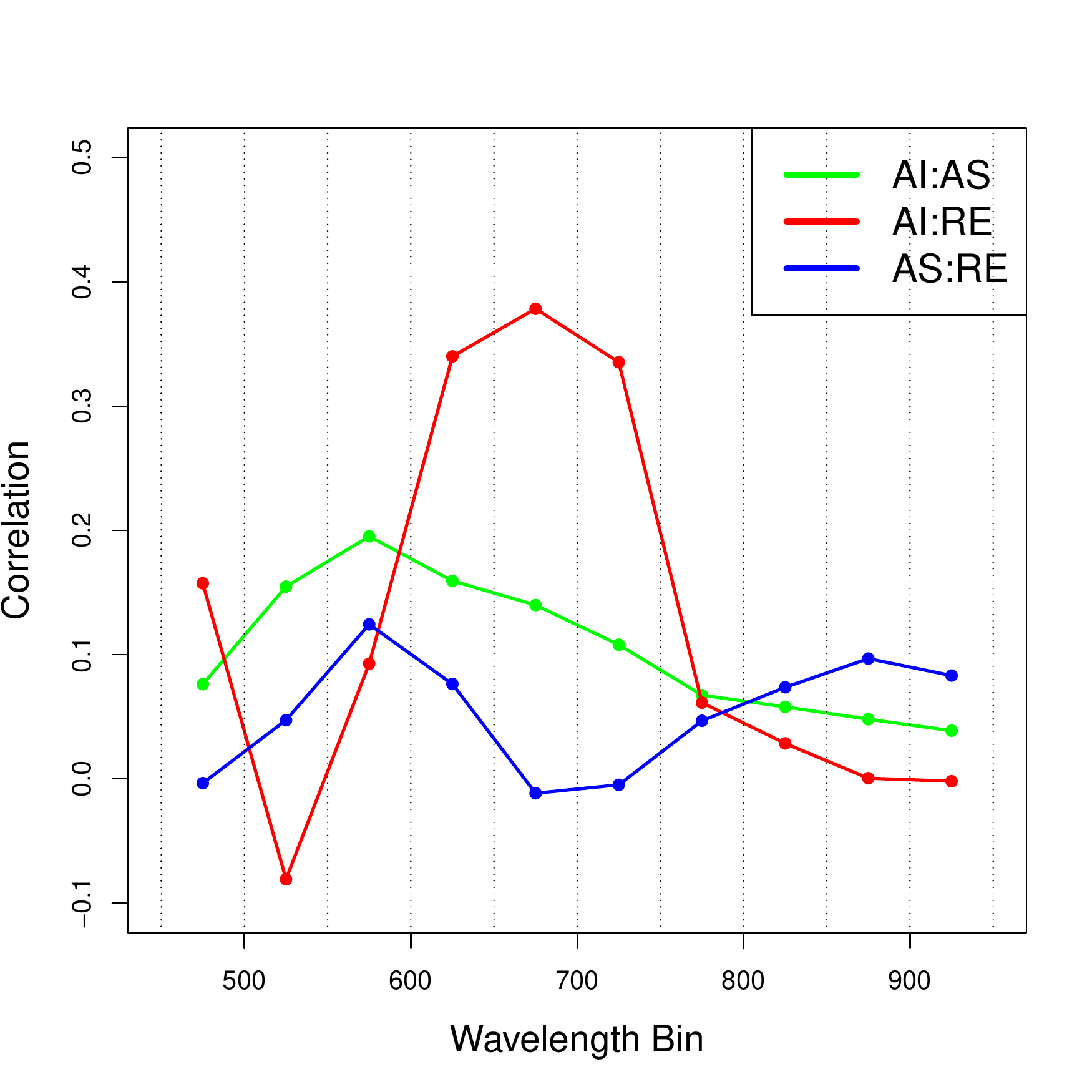}
 \end{center}
 \caption{The between-family correlation for reflectances at the same site, averaged over 10 wavelength bins. In this plot ``AS'' refers to Asteraceae, ``AI'' refers to Aizoaceae, ``RE'' refers to Restionaceae. }\label{fig:between_family}
 \end{figure}

As Table \ref{tab:cooccur} shows, duplication at the family level occurs at approximately 30\% of sites. Moreover, it appears that the between-family relationships at common sites are relatively weak. In Figure \ref{fig:between_family}, we plot the between-family correlation at sites where families co-occur. We calculate the correlations for reflectances over 10 wavelength bins. With the exception of three bins, all correlations are less than 0.2 in magnitude. The combination of low between-family correlations and low family co-occurrence motivate our decision to model the reflectances for each family separately.

In this dataset, it is not common to observe the same genus more than once at the same site (see Table \ref{tab:cooccur_genus}). Duplication is quite rare at the species level; thus, most sites have at most family-level or no duplication. This point is explored in more detail in Section \ref{sec:dat_locations}.

\begin{table}[H]
\centering
\begin{tabular}{rrrrrrrrr}
  \hline
 & $n_{100}$ & $n_{010}$ & $n_{001}$ & $n_{110}$ & $n_{101}$ & $n_{011}$ & $n_{111}$  & Total\\
  \hline
counts $n_{ijk}$ & 14.000 & 69.000 & 13.000 & 22.000 & 3.000 & 10.000 & 2.000 & 133.000 \\
  proportion $n_{ijk}$/$n_s$ & 0.105 & 0.519 & 0.098 & 0.165 & 0.023 & 0.075 & 0.015 & 1.000 \\
   \hline
\end{tabular}
\caption{Letting $i$, $j$, and $k$ be binary, where $i = 1$, $j = 1$, and $k = 1$ indicate the presence of Aizoaceae, Asteraceae, and Restionaceae, respectively. $n_{ijk}$ here represents the number of sites with all possible configurations of sites.}\label{tab:cooccur}
\end{table}

 \begin{table}[H]
\centering
\begin{tabular}{rr}
  \hline
Duplicated genera by site & Counts \\
  \hline
1 & 224 \\
  2 &  31 \\
  3 &   3 \\
   \hline
\end{tabular}
\caption{Level of genus duplication by site.}\label{tab:cooccur_genus}
\end{table}

\begin{figure}[H]
 \begin{center}

 \includegraphics[width = 0.5\textwidth]{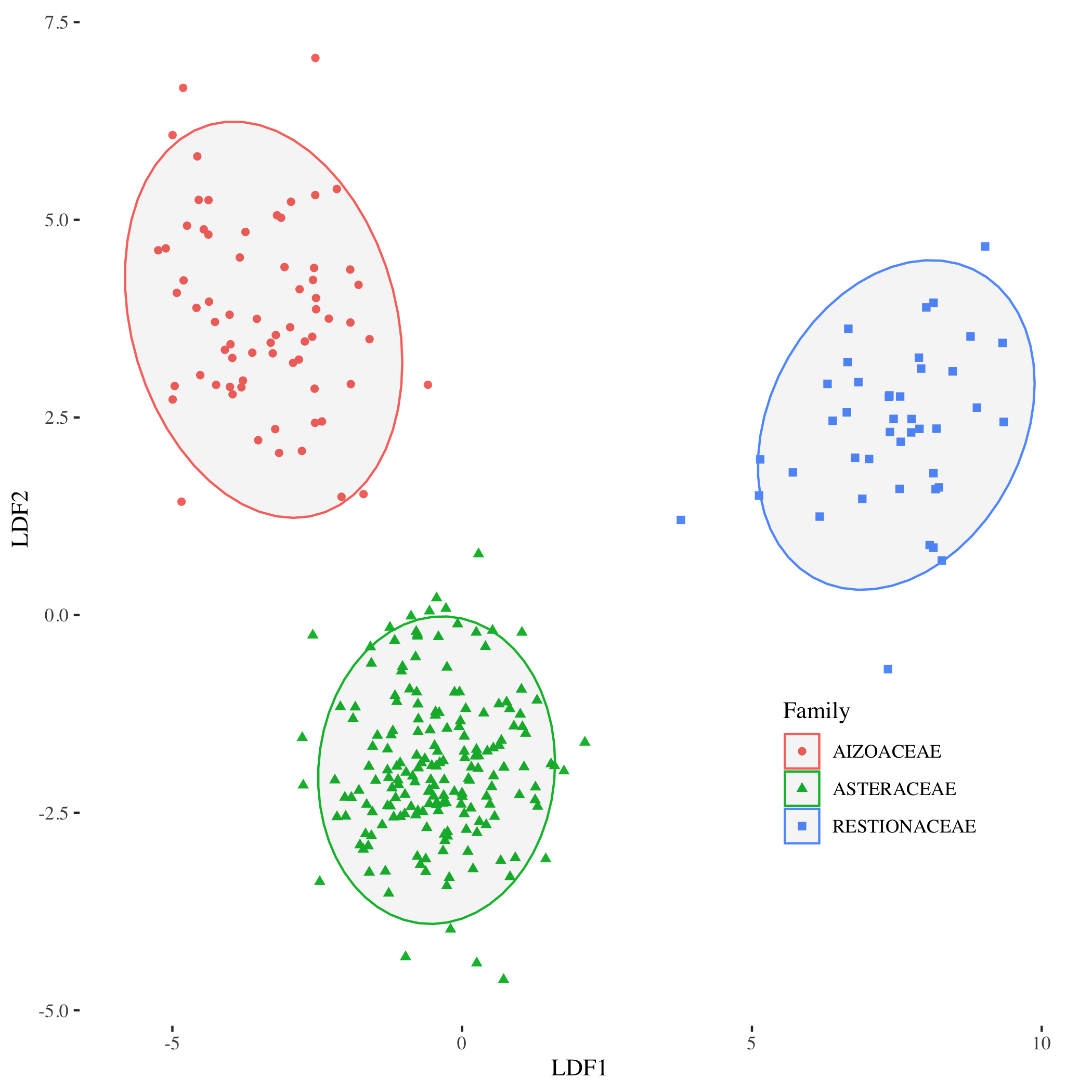}
 \end{center}
 \caption{Linear Discriminant Analysis of family using reflectance curves to predict family.}\label{fig:family_LDA}
 \end{figure}

Although these families have broad overlap in their reflectance curves, as seen in the main manuscript, we find that we can identify families effectively from their reflectance curves using linear discriminant analysis (LDA) (See Figure \ref{fig:family_LDA}). Although this is not the goal of our analysis, the clear separation of the groups indicates that reflectances can be used to effectively predict taxonomic differences. In other words, this effectively demonstrates that reflectance is an ``uber'' trait. We emphasize, however, that this analysis is exploratory to demonstrate that family differences are clearly present in the reflectance curves.

%


\subsection{Data Locations}\label{sec:dat_locations}

\begin{figure}[H]
 \begin{center}

 \includegraphics[width = \textwidth]{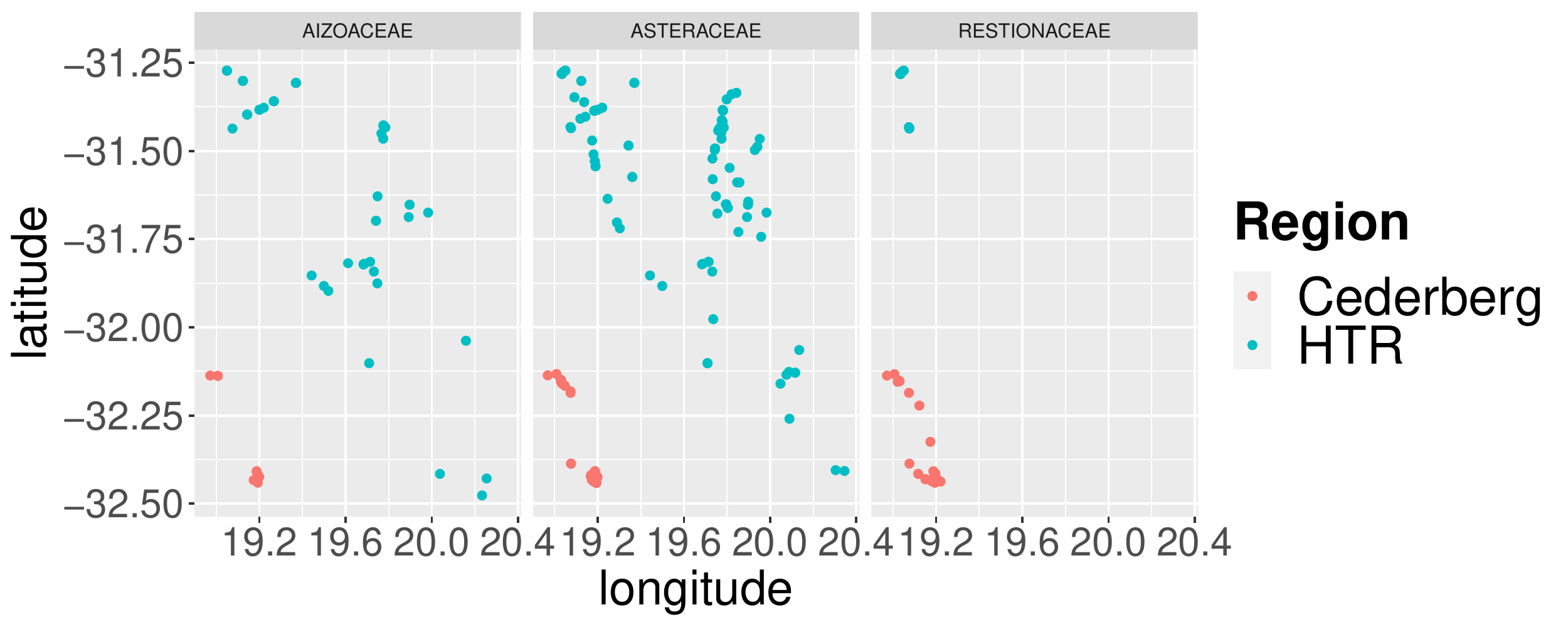}
 \end{center}
 \caption{Family-specific locations colored by region.}\label{fig:family_locs}
 \end{figure}

\begin{figure}[H]
 \begin{center}
 \includegraphics[width = \textwidth]{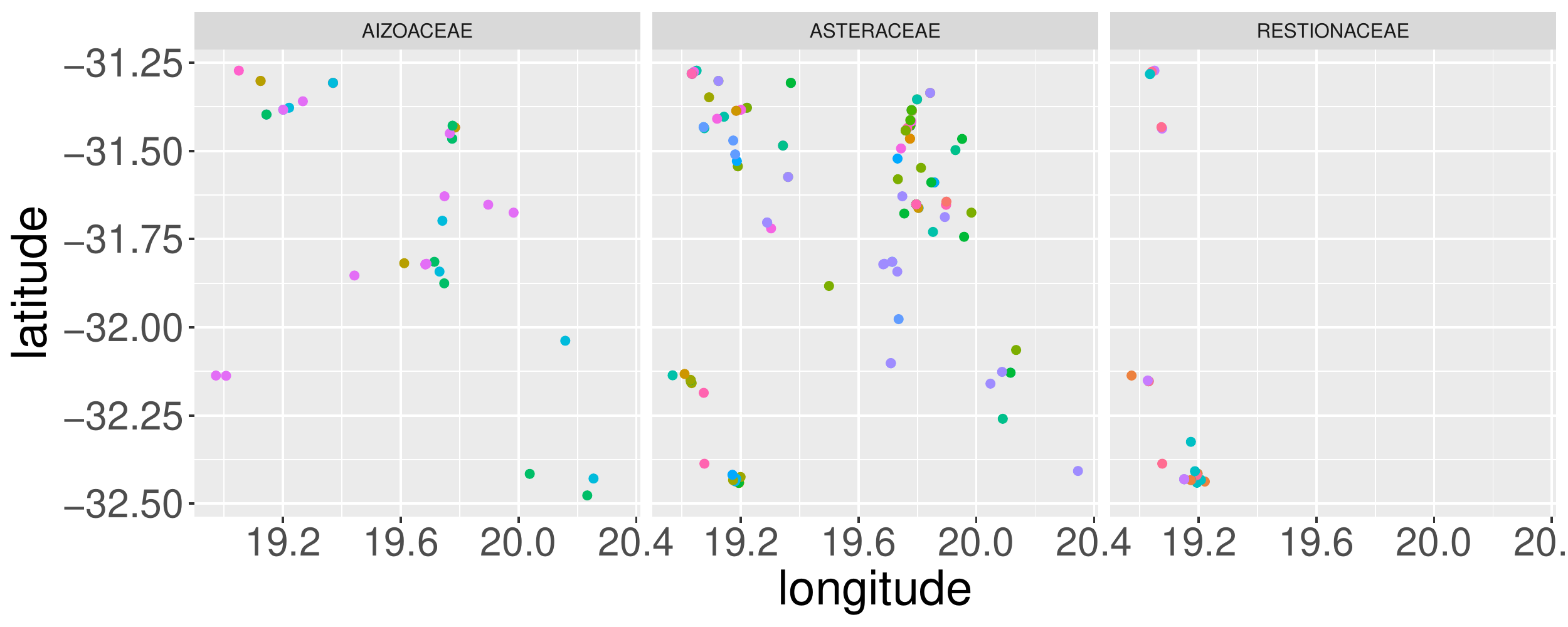}
 \end{center}
 \caption{Locations colored by genus for genera observed at least five times.}\label{fig:genus_locs}
 \end{figure}

In this section, we examine the locations where we have observed plant reflectance curves. This is useful as it motivates decisions about joint modeling and model hierarchy structure. In Figures \ref{fig:family_locs} and \ref{fig:genus_locs}, we show all locations coded by the region they belong to and by the number of families observed at that site.  To visualize the spatial distribution of each family, we plot the locations where each family is observed in Figure \ref{fig:family_locs}; the observed curve locations differ by family. Because of the significant differences between their spatial ranges, joint spatial modeling of the three families does not seem appropriate. Joint modeling may also be discouraged based on the weak between-family correlations presented in Section \ref{sec:dat_sum}.

\begin{figure}[H]
 \begin{center}
 \includegraphics[width = 1.\textwidth]{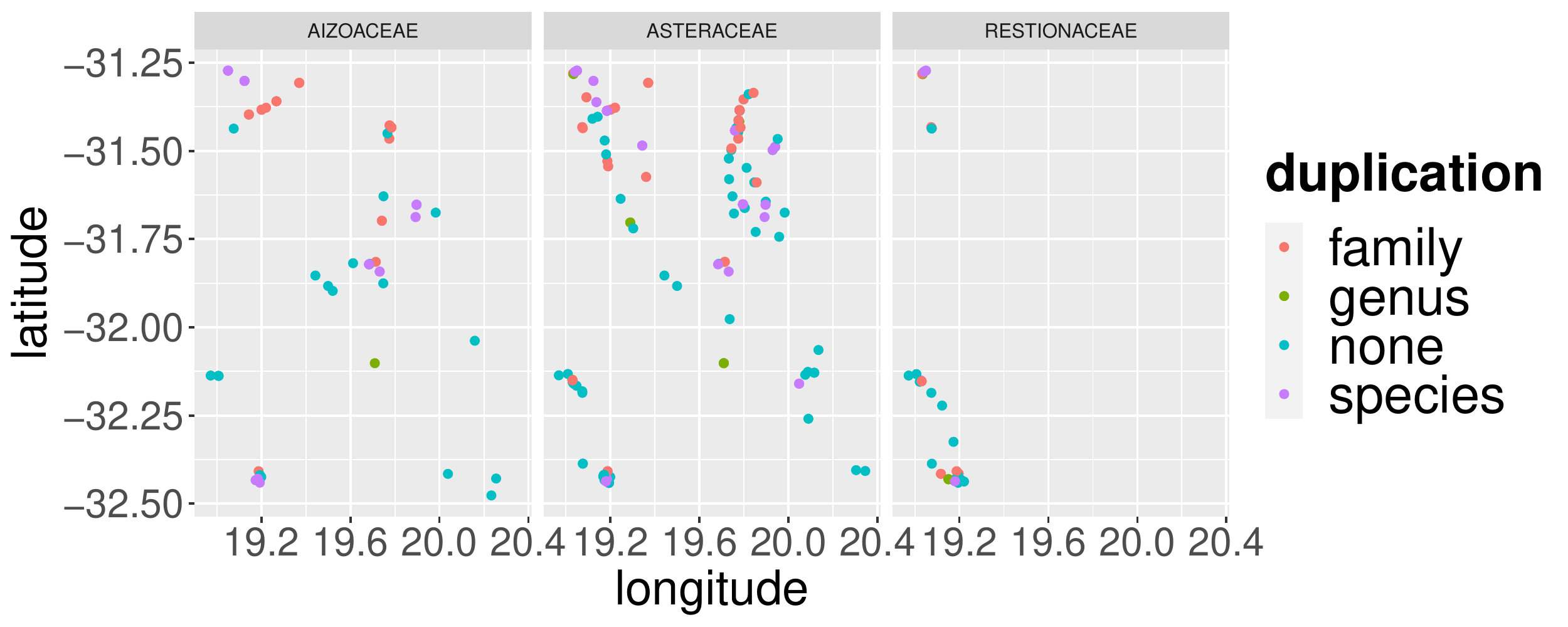}
 \end{center}
 \caption{Presence of family, genus, and species duplication by site. If genus duplication is present, then it is indicated rather than that family duplication that is also present. No species duplication is present at any site. }\label{fig:duplication}
 \end{figure}

Figure \ref{fig:genus_locs} plots the locations of the curves colored by genus for each family. (The plot is limited to those genera that have at least five reflectance curves). Although somewhat difficult to see, this shows that most genera are very concentrated, while others are spread out. This suggests that spatial modeling at a genus-level may be difficult or limited. In our final model presented in the manuscript (selected through cross-validation), we use a spatially-varying genus-specific offset but do not have functional modeling at the genus level.
In the HTR and Cederberg regions, there is little species duplication at these study sites. In Figure \ref{fig:duplication}, for each site, we plot the lowest classification level of duplication observed. For example, if we observe two reflectance curves for the same genus at the same site, we note that as genus duplication.

 \subsection{Variance, Covariance, and Residual Analysis}\label{sec:residual}

%
%
%

To motivate the hierarchical analyses in the manuscript, we fit the functional coefficient model described in the manuscript without any random effects and analyze residual patterns from this simple model. We carry out this analysis for the Asteraceae family; however, similar patterns are also present in the Aizoaceae and Restionaceae families. 
 
\begin{figure}[H]
 \begin{center}

\includegraphics[width=\textwidth]{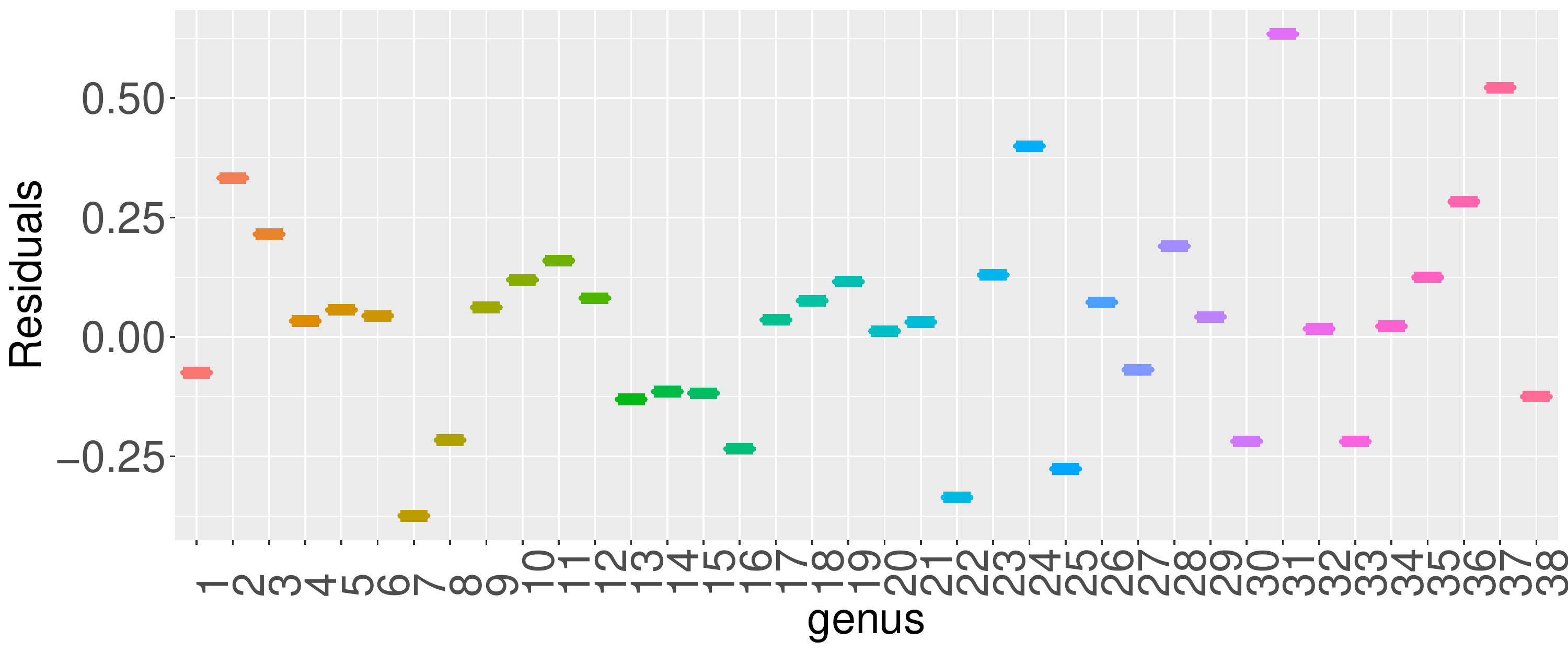}
 \end{center}
 \caption{Mean of residuals by genus for the Asteraceae family. }\label{fig:genus_mean_resid}
 \end{figure}

\begin{figure}[H]
 
 \begin{center}

\includegraphics[width=\textwidth]{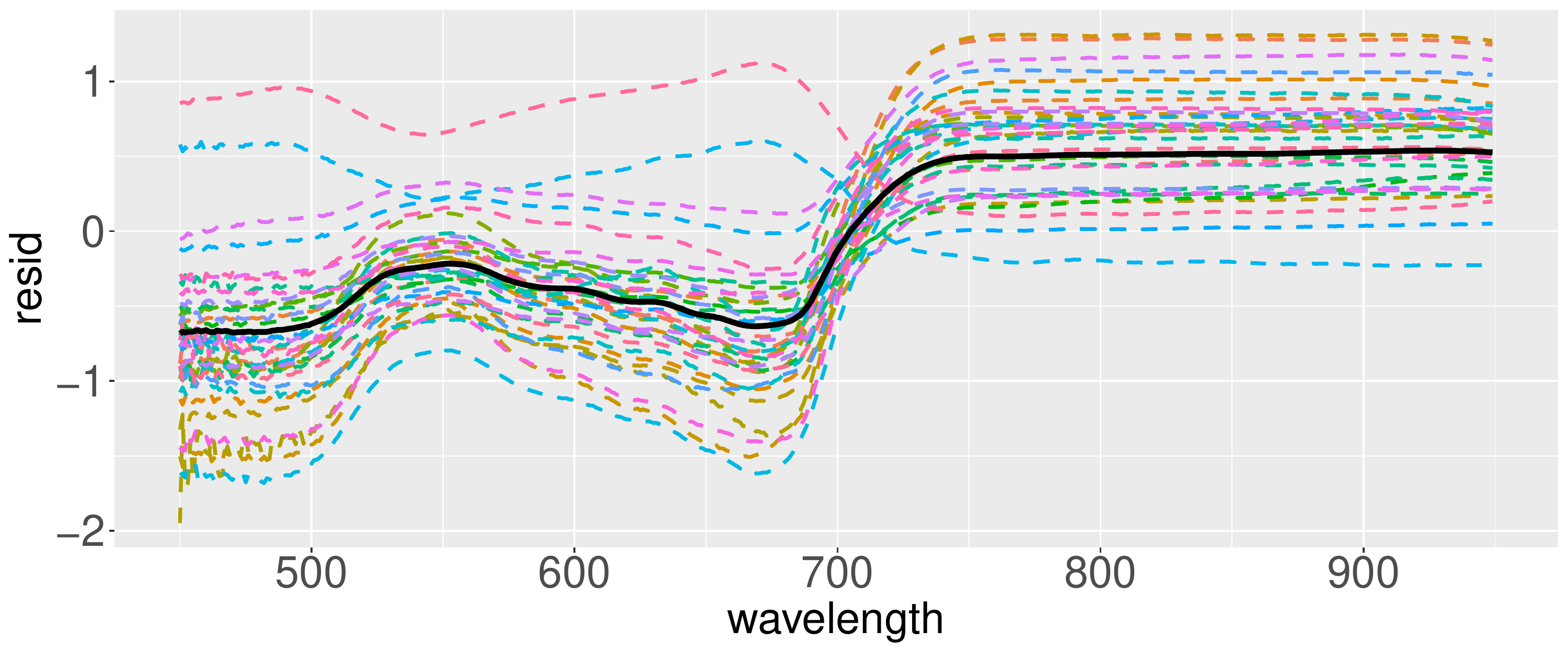}
 \end{center}
  \caption{Mean of residuals by genus and wavelength (in nm) for the Asteraceae family. }\label{fig:genus_wave_resid}
 \end{figure}

\begin{figure}[H]
 \begin{center}

\includegraphics[width=\textwidth]{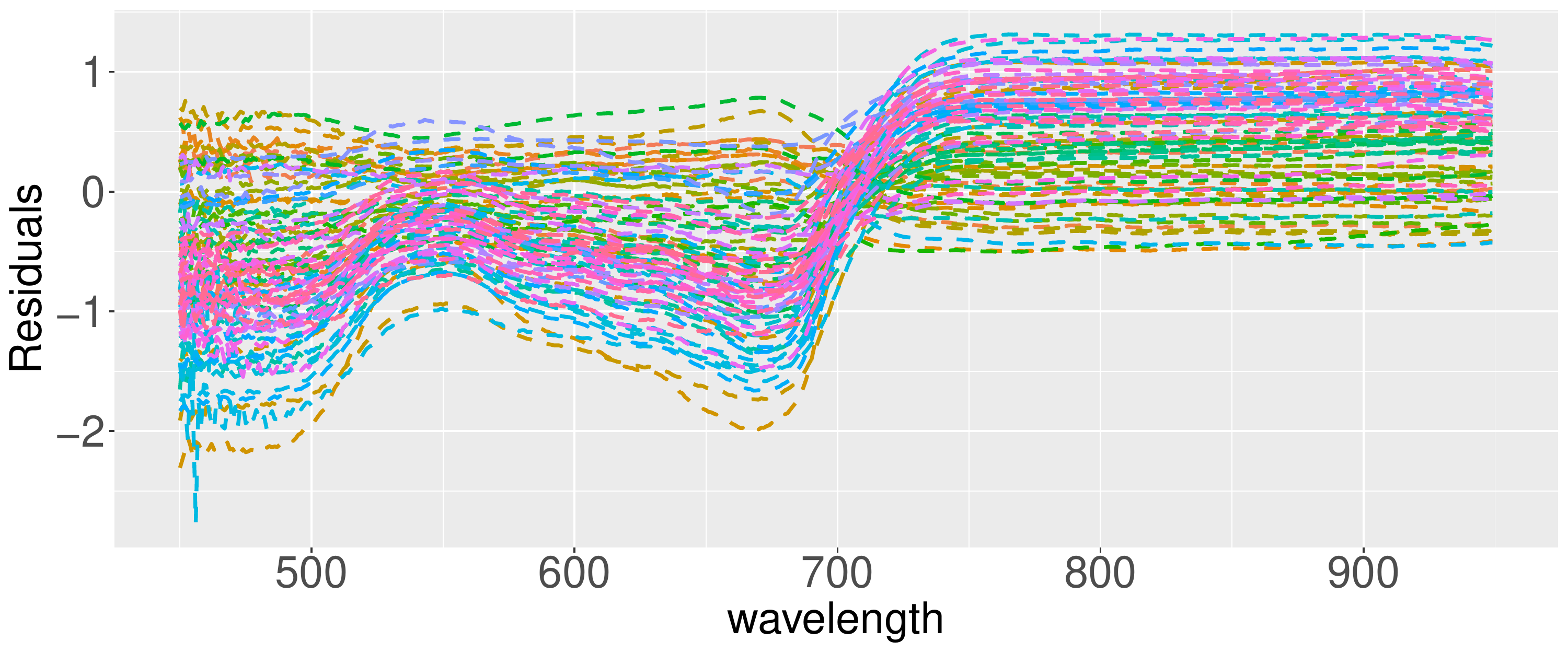}
 \end{center}
  \caption{Mean of residuals by unique location and wavelength (in nm) for the Asteraceae family. }\label{fig:loc_wave_resid}
 \end{figure}

In Figure \ref{fig:genus_mean_resid}, we plot the mean of residuals for each genus. These patterns demonstrate that the overall log-reflectance level is strongly related to the genus.
In Figure \ref{fig:genus_wave_resid} and \ref{fig:loc_wave_resid}, we plot the mean residuals as a function of wavelength by genus and by location, respectively. We group these two residual analyses because the patterns observed are similar and because genus-specific patterns are strongly related to spatial patterns due to the limited spatial range of genera. Although the patterns in Figures \ref{fig:genus_wave_resid} and \ref{fig:loc_wave_resid} are both strong and motivate wavelength functions for genus and location, we find that having genus-specific wavelength functions makes predictive performance worse when a spatial-wavelength random effect is included. Thus, we argue that much of the structure in Figure \ref{fig:genus_wave_resid} is implicit in the patterns seen in Figure \ref{fig:loc_wave_resid}.

\begin{figure}[H]
 \begin{center}
 \includegraphics[width= 0.19\textwidth]{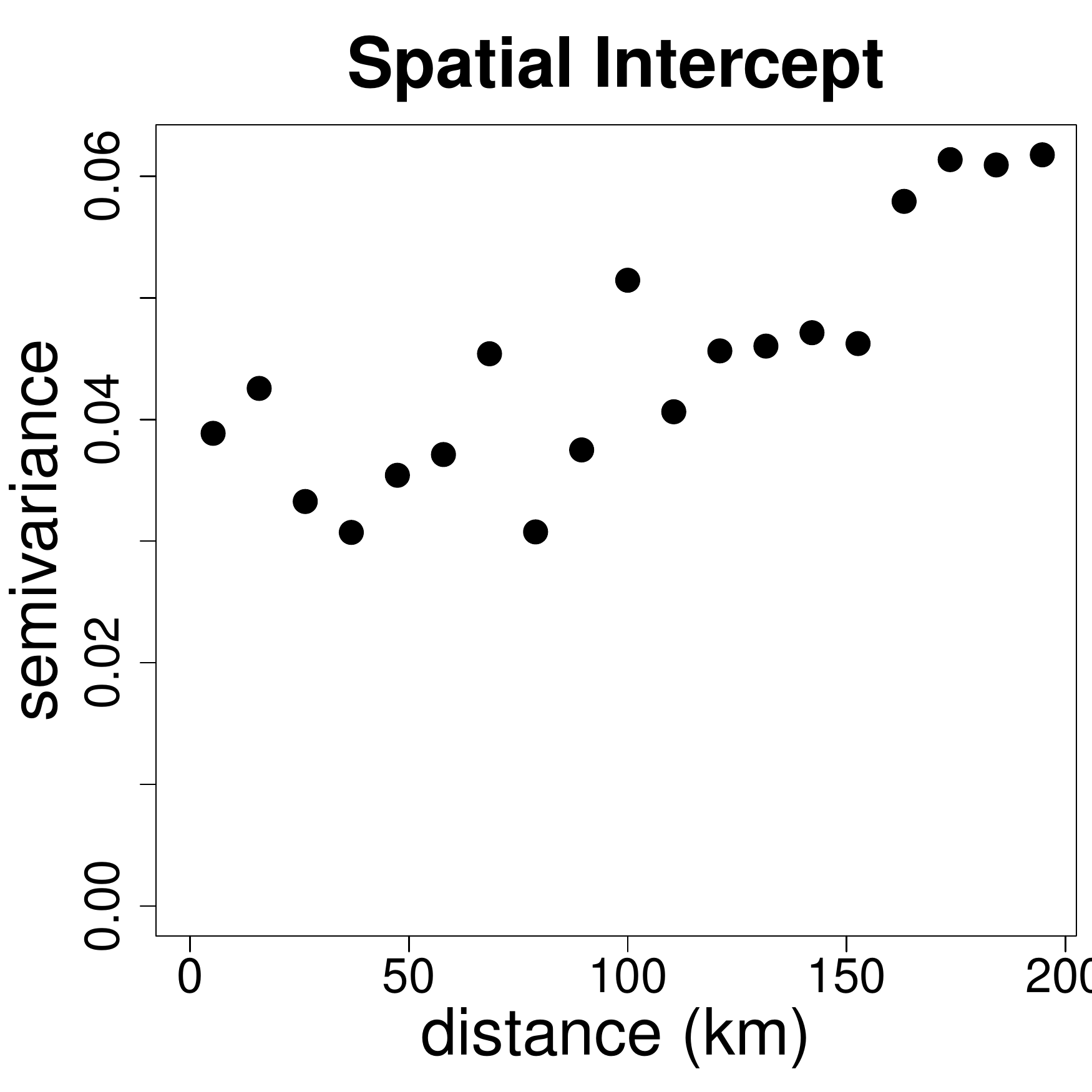}
  \includegraphics[width= 0.19\textwidth]{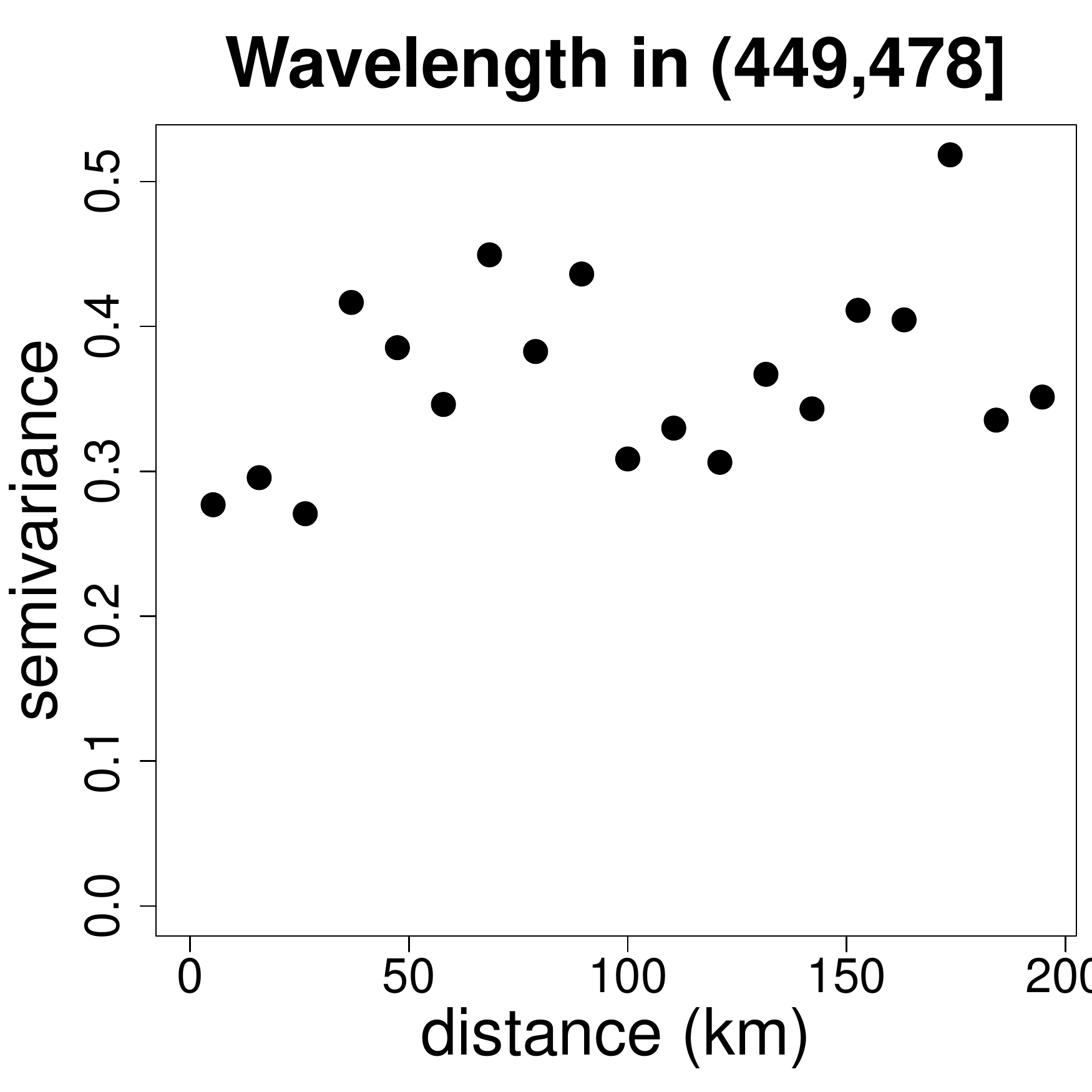}
  \includegraphics[width= 0.19\textwidth]{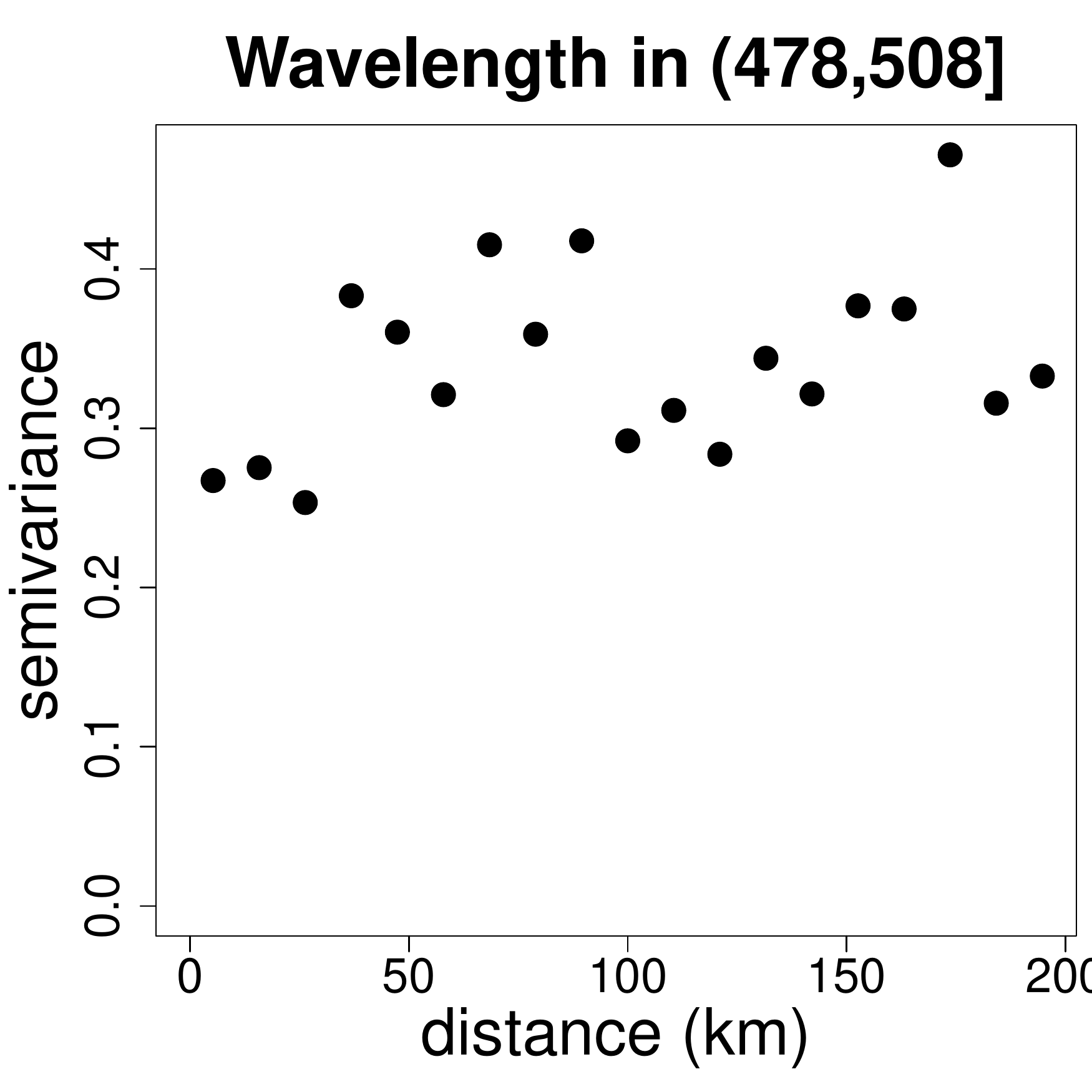}
  \includegraphics[width= 0.19\textwidth]{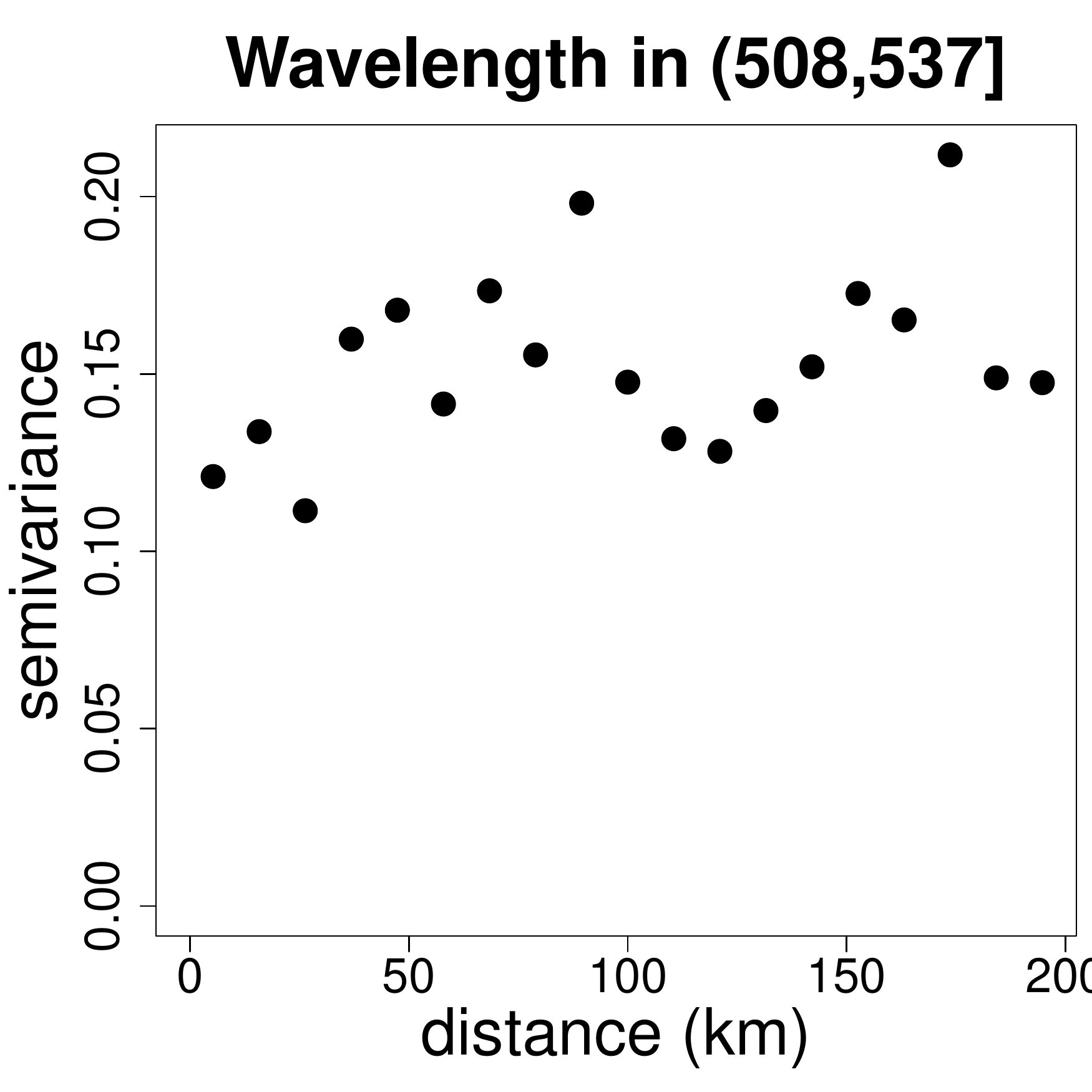}
  \includegraphics[width= 0.19\textwidth]{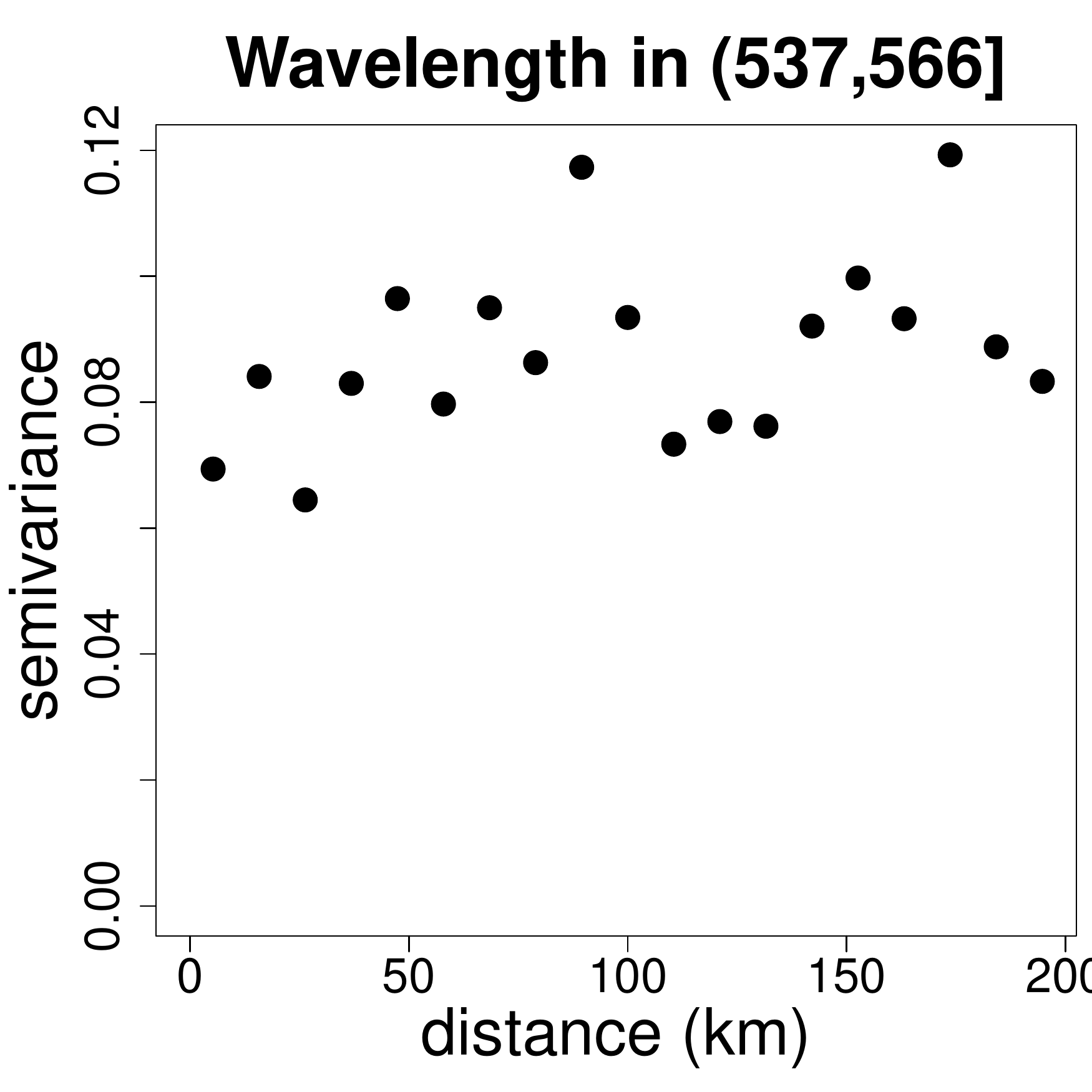}
  \includegraphics[width= 0.19\textwidth]{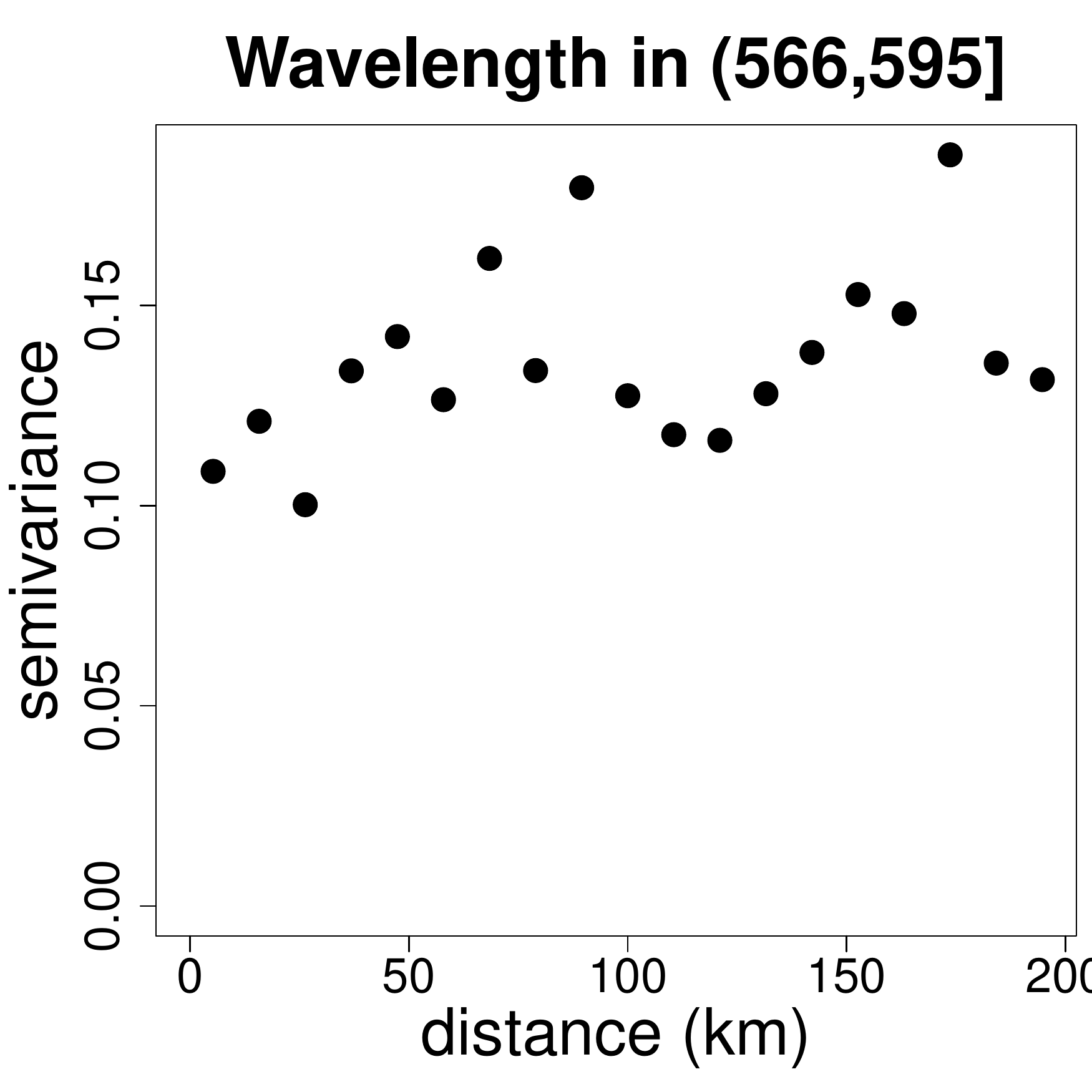}
  \includegraphics[width= 0.19\textwidth]{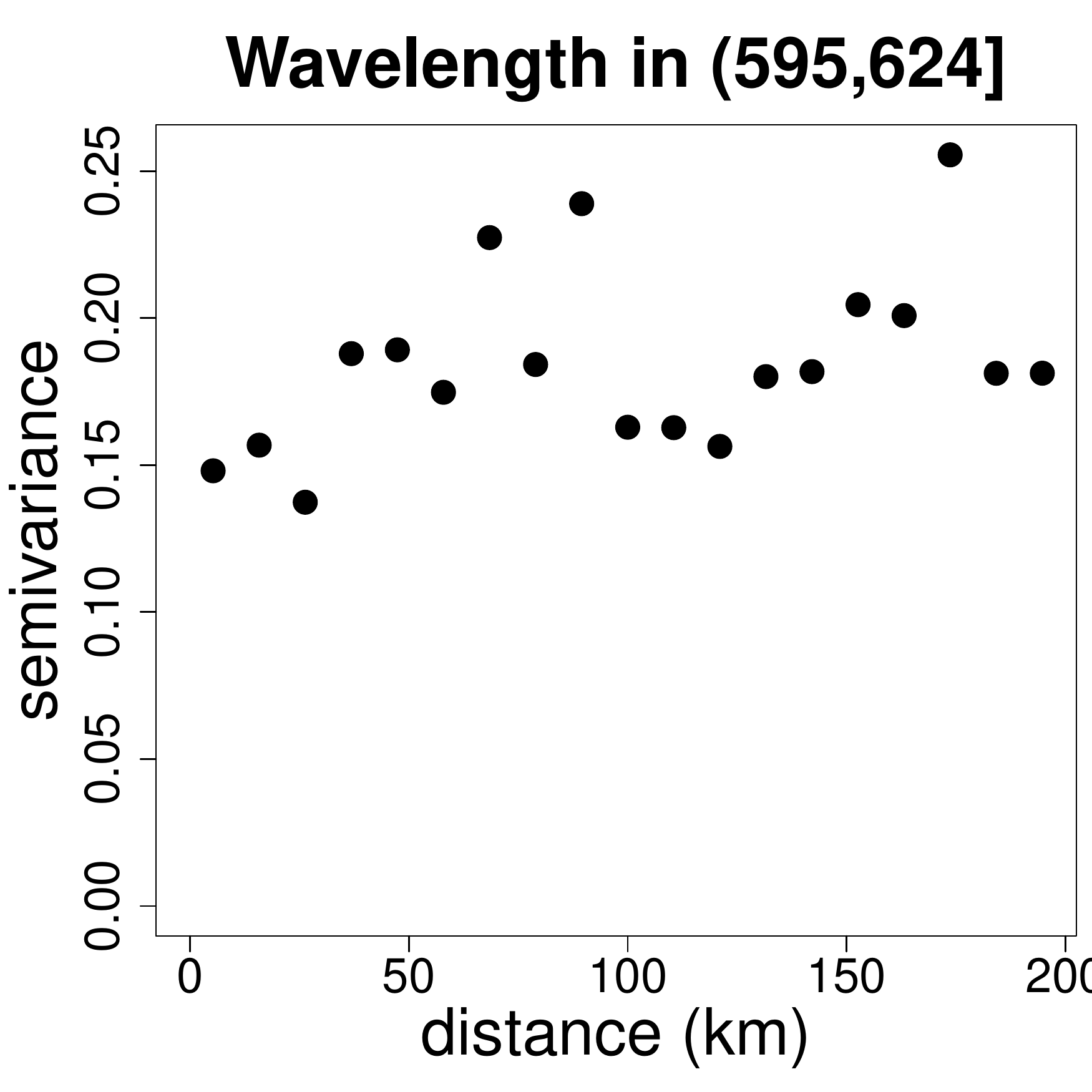}
  \includegraphics[width= 0.19\textwidth]{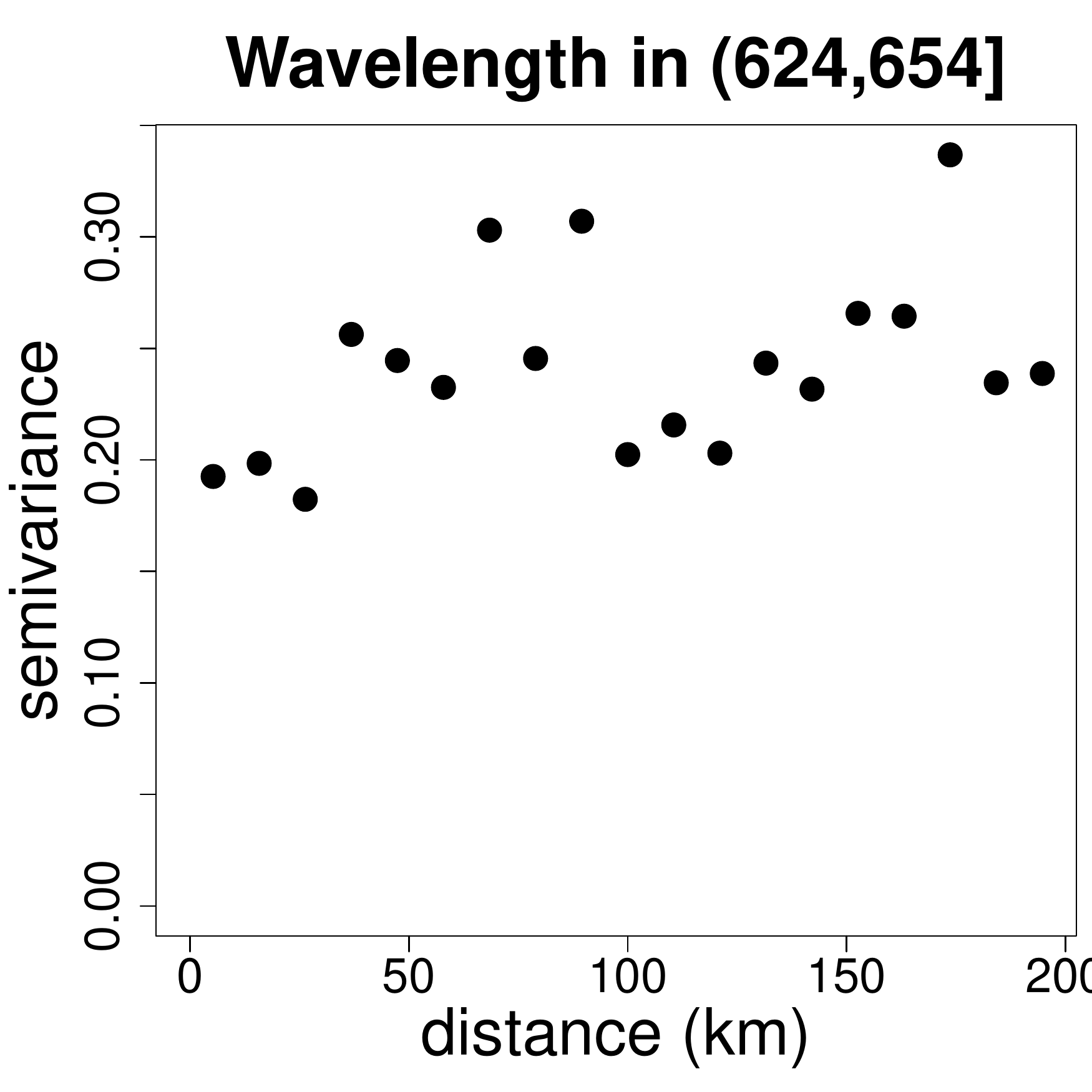}
  \includegraphics[width= 0.19\textwidth]{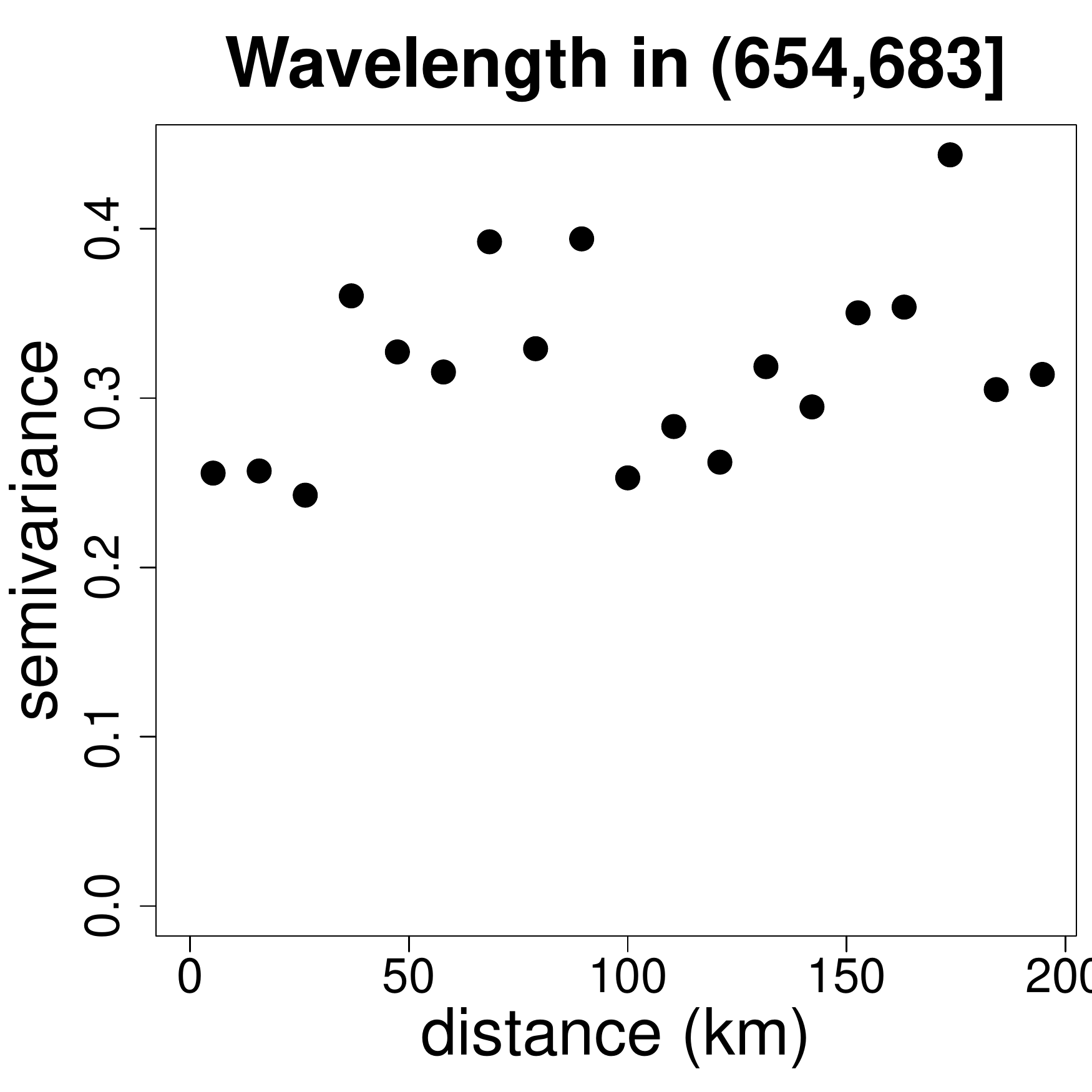}
  \includegraphics[width= 0.19\textwidth]{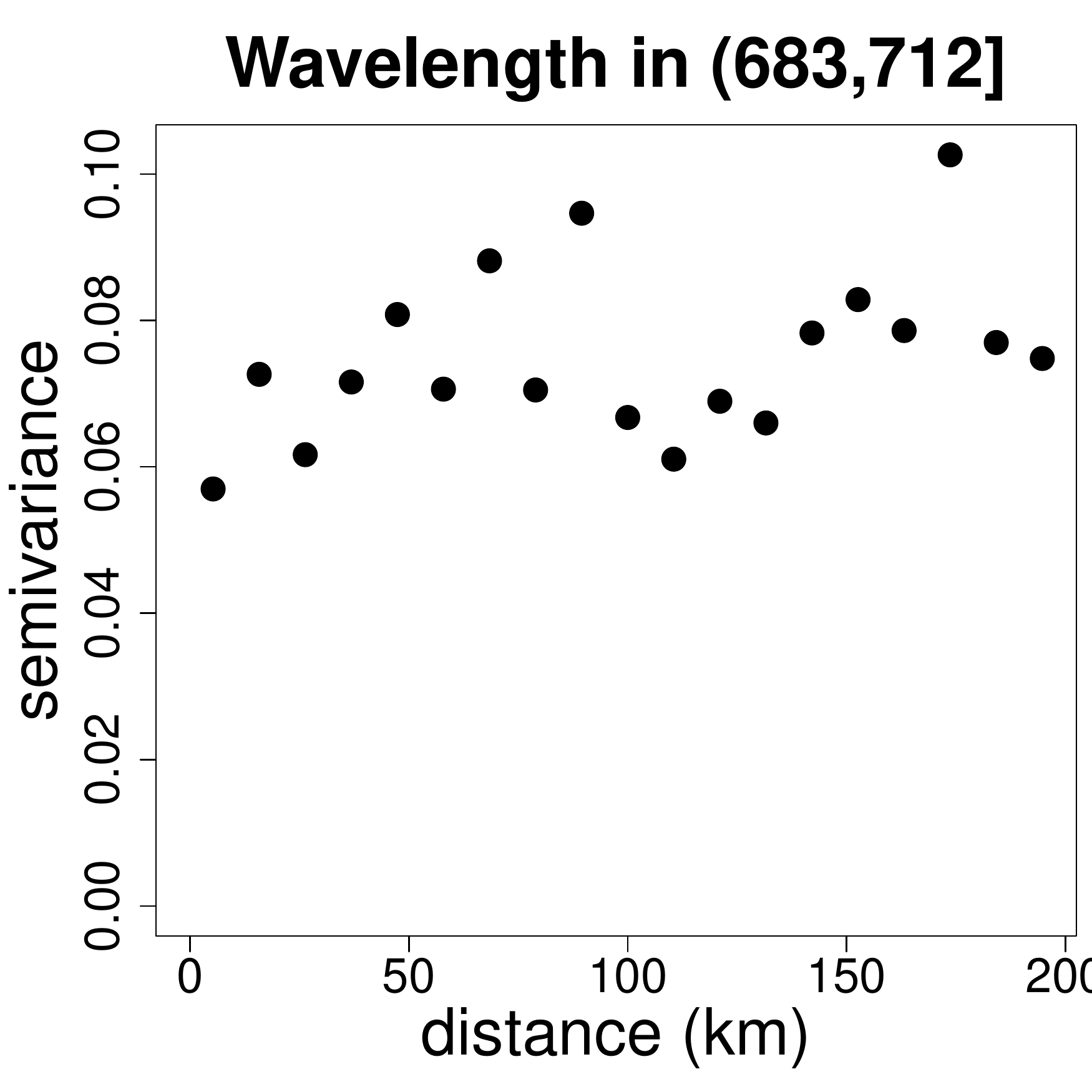}
  \includegraphics[width= 0.19\textwidth]{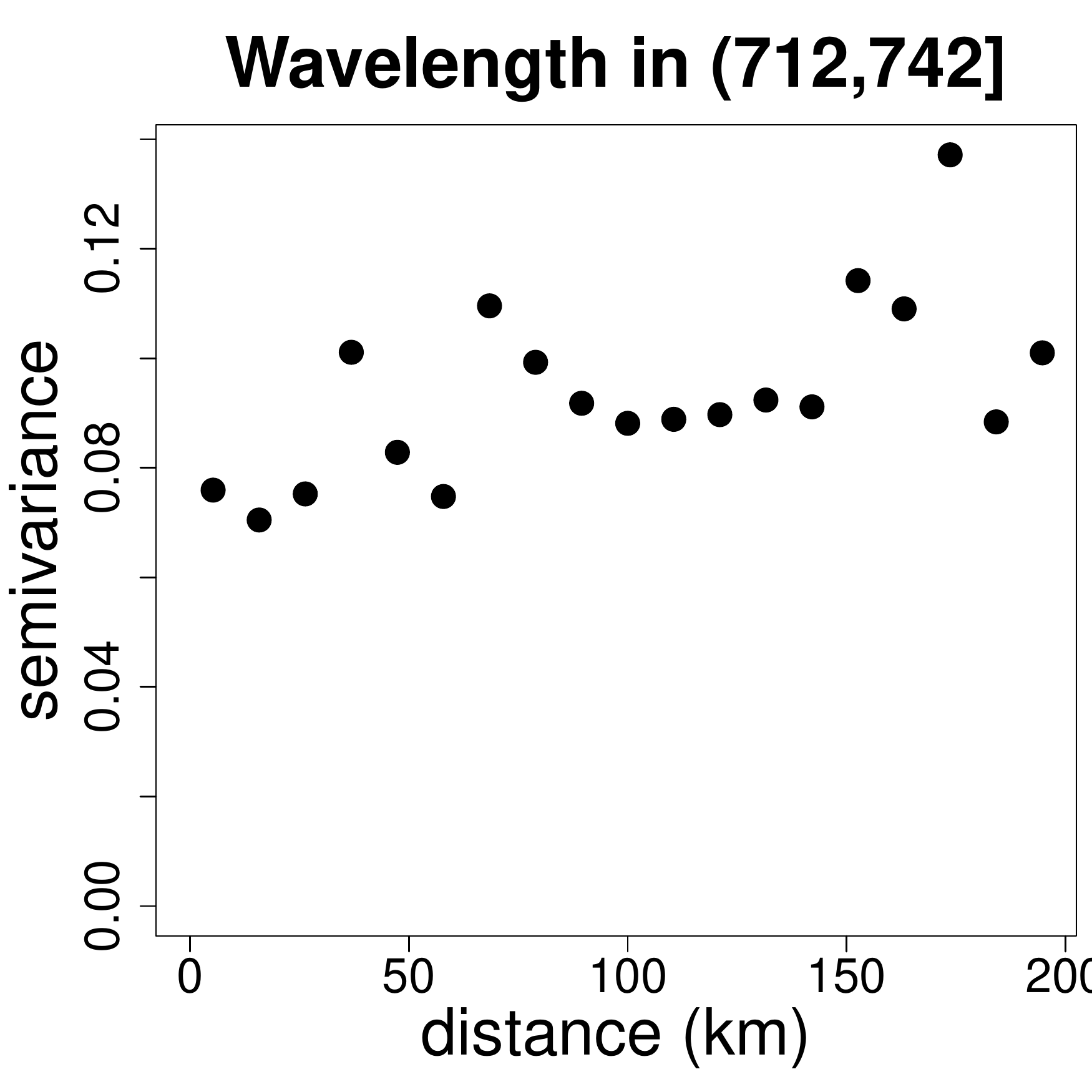}
  \includegraphics[width= 0.19\textwidth]{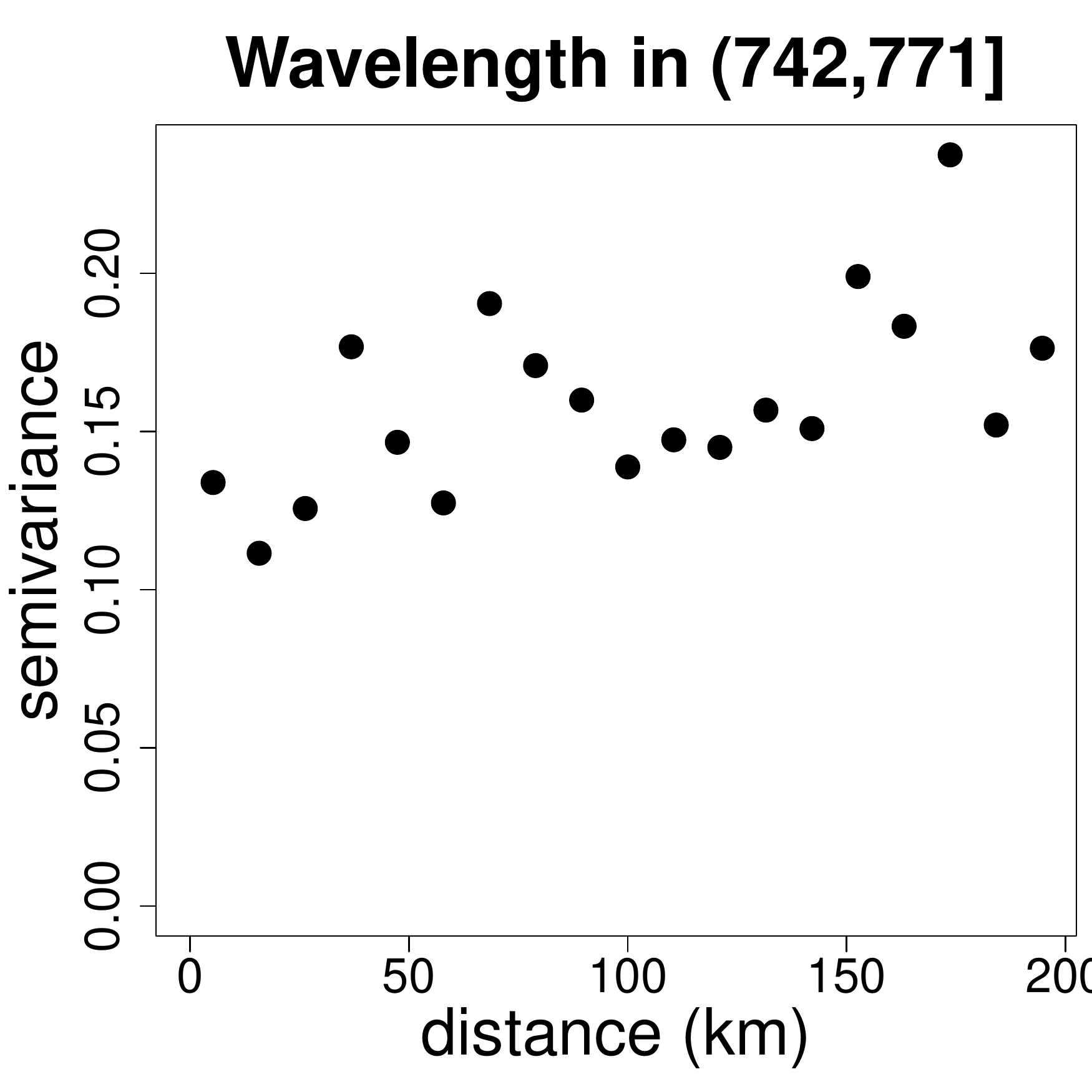}
  \includegraphics[width= 0.19\textwidth]{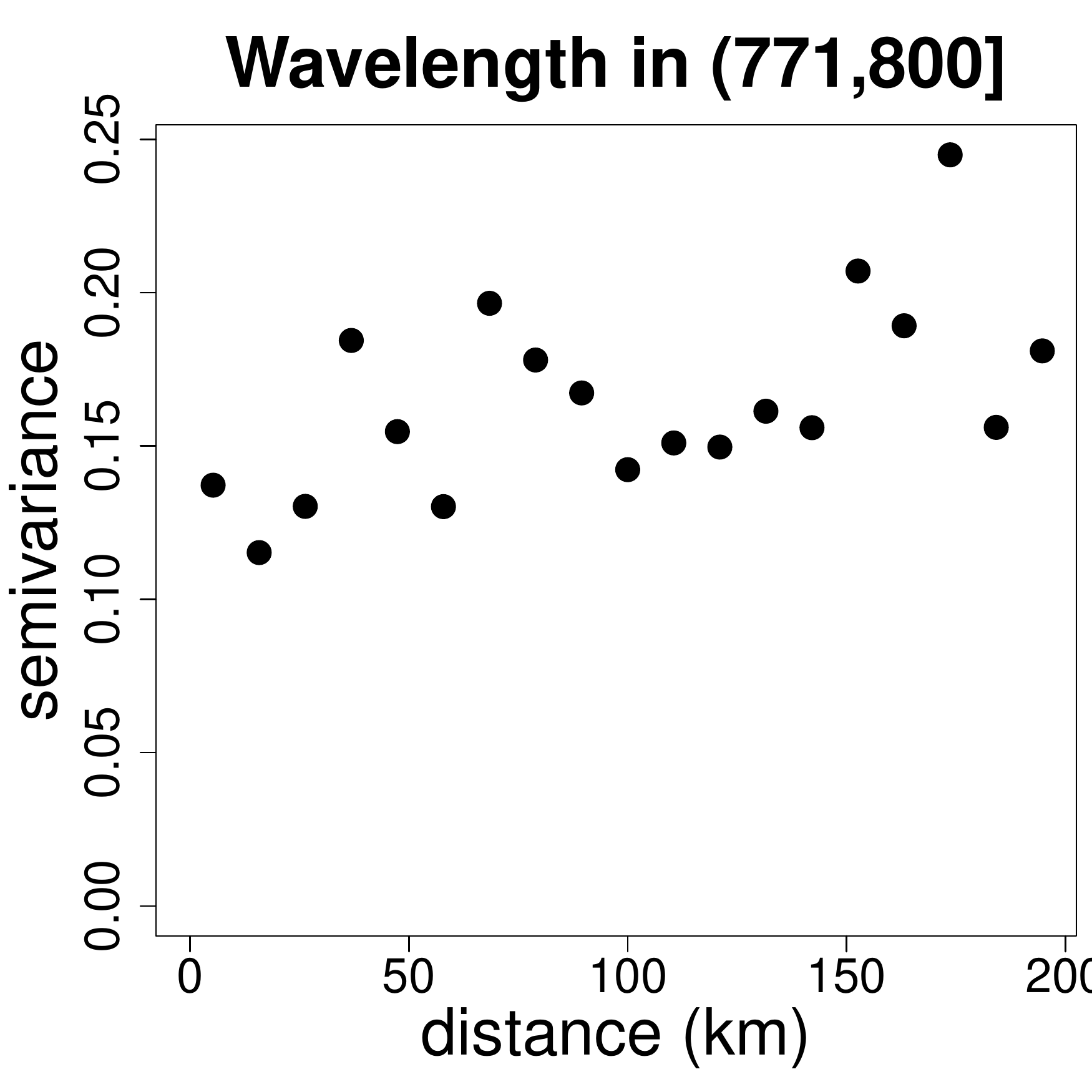}
  \includegraphics[width= 0.19\textwidth]{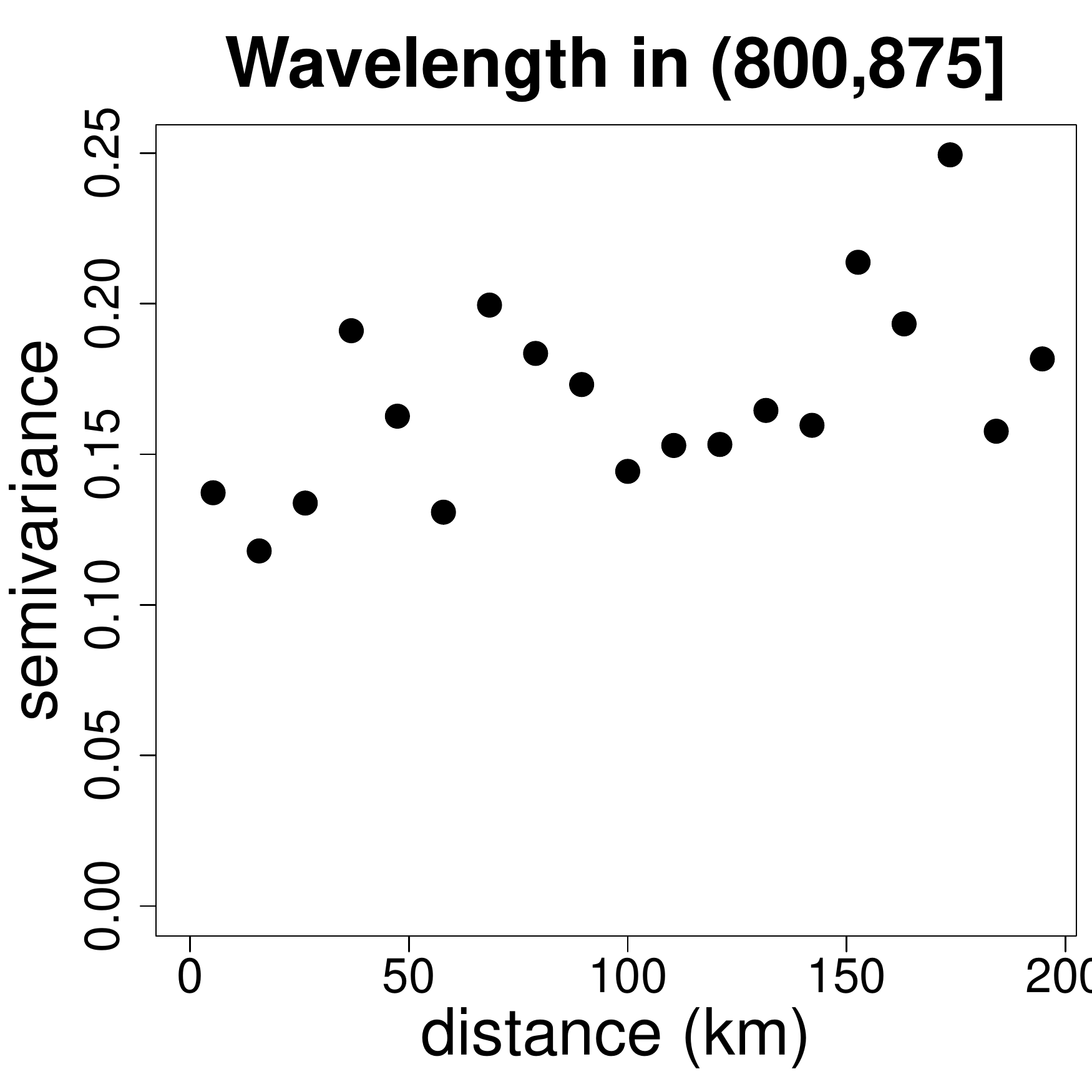}
  \includegraphics[width= 0.19\textwidth]{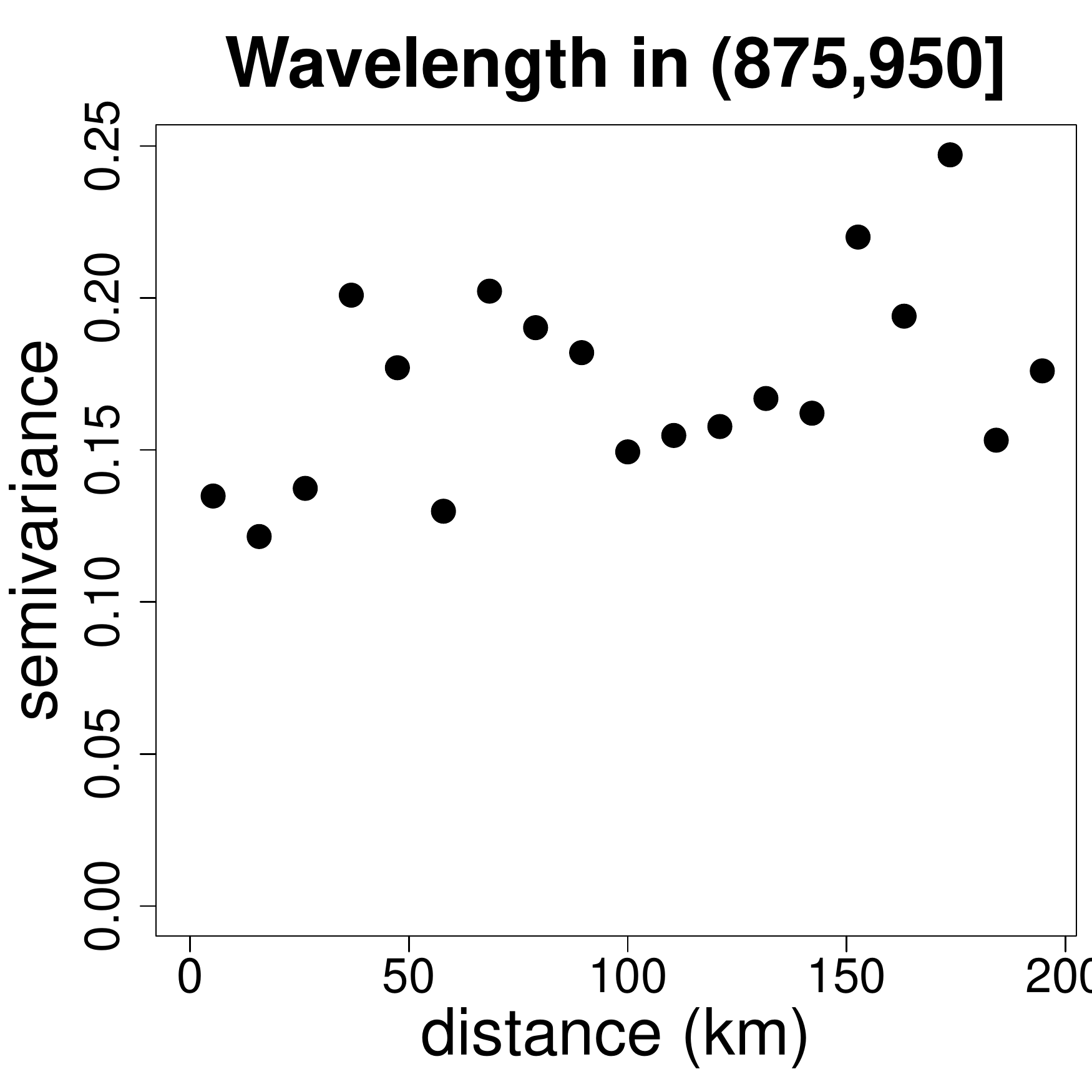}

 \end{center}
   \caption{Empirical binned semivariograms of residuals overall and for 14 wavelength bins (indicated in the plot) for the Asteraceae family. }\label{fig:vario}
 \end{figure}

Lastly, we examine the spatial autocorrelation of residuals overall and over various wavelength bins by computing empirical binned semivariograms. Although our models are explicitly functional and not binned, this is intended to show that these patterns vary over the wavelength domain. These semivariograms are plotted in Figure \ref{fig:vario}. Overall, there is a clear pattern in the semivariance as a function of distance. Additionally, most wavelength bins show some spatial autocorrelation, but most of these spatial patterns are weak relative to the ``nugget'' effect. 
Importantly, the autocorrelation patterns appear to be wavelength (bin) dependent in that the scales of semivariance are wavelength dependent and the spatial ranges appear to vary for different wavelengths. Specifically, lower wavelengths ($<$ 700 nm) seem to have a rapid rise in semivariance of the first 75 km, while, for higher wavelengths, there are increases in semivariance at greater distances.

\begin{figure}[H]
 \begin{center}
 \includegraphics[width = 0.48\textwidth]{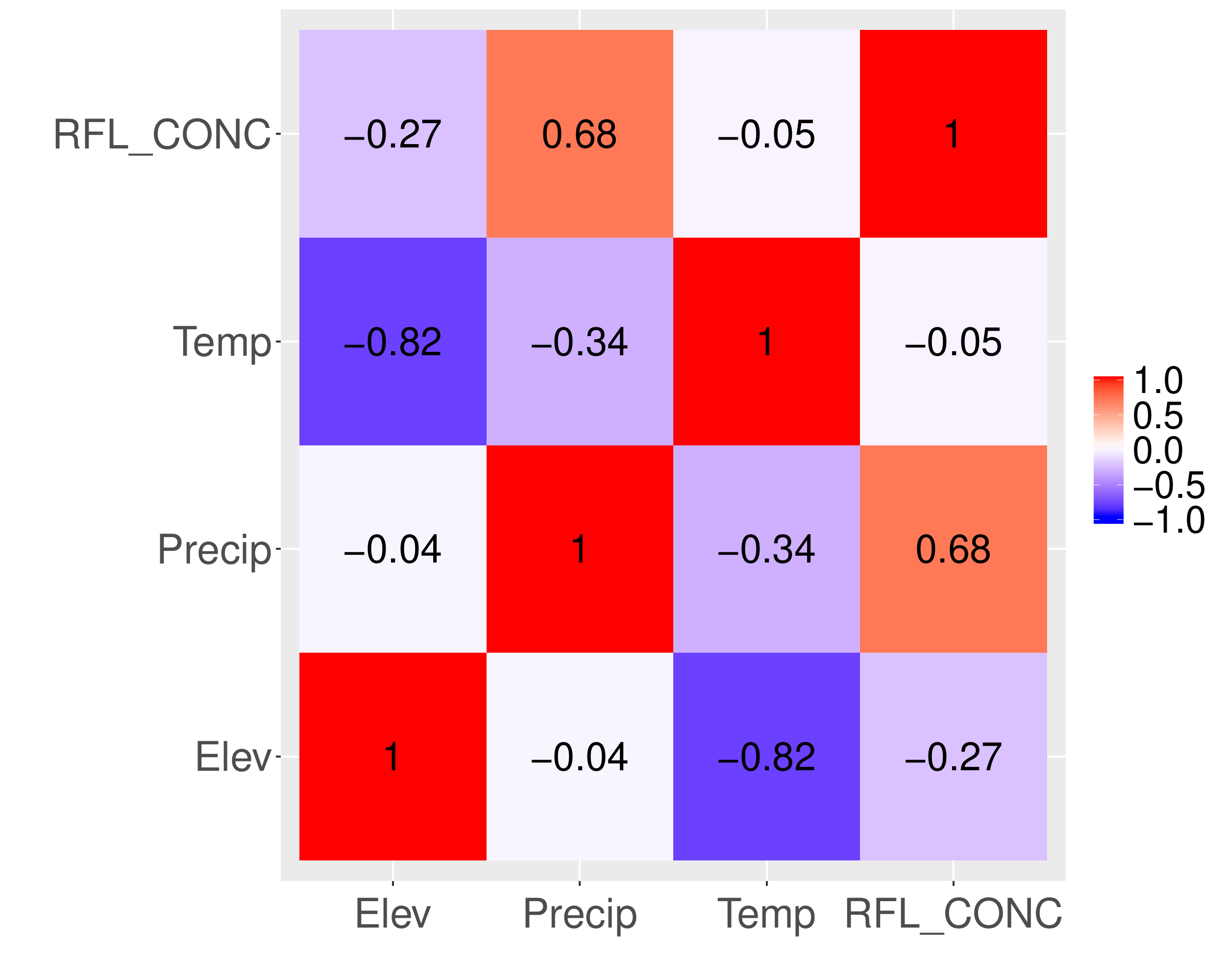}
 \end{center}
 \caption{Between-covariate correlation. }\label{fig:cor_mat_supp}
 \end{figure}

 We also point out that the covariates used in our analysis are collinear. That is, they have some between-covariate correlation (Figure \ref{fig:cor_mat_supp}). Thus, the proposed regression functions must be interpreted with this between-covariate covariance in mind.

 \section{Model Development and Sensitivity Analysis}\label{sec:mod}

 \subsection{Cross-validation for specifying $\eta(\bs,t)$}\label{sec:cv_eta}

As written in the manuscript, we define the space-wavelength function as
\begin{equation}\label{eq:extended}
\begin{aligned}
\eta(\bs,t) =  \bK(t)^T \bz(\bs) = \sum_{j = 1}^{J_\eta} k_{t^\eta_j}(t - t^\eta_j;\theta^{(\eta)}) z_{t^\eta_j}(\bs),
\end{aligned}
\end{equation}
where $z_{t^\eta_j}(\bs)$ are spatially-varying random variables associated with Gaussian wavelength kernels $k_{t^\eta_j}(\cdot;\theta_{t^\eta_j})$. We select the structure of $\bz(\bs)$ through cross-validation as described in the manuscript.  In particular, we compare separable, independent, and various latent factor models, including the linear model of coregionalization (LMC). We present the results of these analyses in Table \ref{tab:mod_comp2}. We find that including $r=10$ factors to construct $\bz(\bs)$ led to the best out-of-sample prediction. The models in Table \ref{tab:mod_comp2} all include spatially-varying genus-specific intercepts $\alpha_i + \alpha_i(\bs)$, a global (not genus-specific) wavelength random effect $\gamma(t)$, and functional regression coefficients $\bbeta(t)$. 

Although this does not consider all possible combinations of mean terms and specifications of $\eta(\bs,t)$, we emphasize the degree to which the models with spatially-varying genus-specific intercepts $\alpha_i + \alpha_i(\bs)$ and a global (not genus-specific) wavelength random effect $\gamma(t)$ outperform other specifications by about 20\%. 

\begin{table}[H]
\centering
\begin{tabular}{llrrrr}
  \hline
$\bz(\bs)$ Specification & Latent Components & MSE & MAE & MCRPS & Relative MCRPS \\
  \hline
Separable & --- & 0.104 & 0.252 & 0.208 & 1.060 \\
Independent & $r = J_\eta $  & 0.125 & 0.271 & 0.225 & 1.147 \\
Factor & $r = 2$ & 0.156 & 0.307 & 0.258 & 1.316 \\
Factor &  $r = 3$ & 0.125 & 0.279 & 0.231 & 1.180 \\
Factor & $r = 5$ & 0.115 & 0.262 & 0.217 & 1.110 \\
Factor & $r = 8$  & 0.105 & 0.249 & 0.208 & 1.064 \\
Factor &  $r = 10$ & 0.097 & 0.237 & 0.196 & 1.000 \\
Factor & $r = 12$ & 0.108 & 0.254 & 0.211 & 1.077 \\
LMC & $r = J_\eta$ & 0.105 & 0.251 & 0.207 & 1.059 \\
   \hline
\end{tabular}
\caption{Out-of-sample predictive performance for various specifications of $\bz(\bs)$. All these models use the best configuration in Table 1 of the Manuscript. As discussed in the manuscript, $J_\eta = 22$.}\label{tab:mod_comp2}

\end{table}

\subsection{Sensitivity Analysis}\label{sec:sensitivity}

We use our MCMC model fitting to obtain $M$ posterior samples. Let $l(\btheta^{(m)})$ be the log-likelihood evaluated at the $m$th posterior sample $\btheta^{(m)}$, $m = 1,..,M$. Additionally, we define the posterior mean to be $\overline{\btheta} = \frac{1}{M}\sum\limits^M_{m=1} \btheta^{(m)}$. For simplicity, we use various quantities related to deviance,
$D(\btheta^{(m)}) = -2 l(\btheta^{(m)}),$
for preliminary model comparison and sensitivity analysis.
Specifically, we use the mean deviance ($\overline{D} = \frac{1}{M}\sum^M_{m=1} D(\btheta^{(m)})$), estimated model complexity ($P_D = \overline{D} - D(\overline{\btheta})$), and deviance information criterion ($DIC = \overline{D} + P_D$) to make some preliminary model selections \citep{spiegelhalter2002}. For these quantities, smaller is better. For each comparison, we look at the relative $\overline{D}$, $P_D$, and $DIC$, where the lowest (best) model is set to 0. For most comparisons, we are interested in selecting the simplest specification that gives approximately the same performance. The exception is for $\bbeta(t)$; in this case, we desire the richest specification that gives approximately the same performance in terms of DIC. For all comparisons, we fix other model specifications on the other ``best'' settings while changing only one component of the model. 

Although not carried out for every model, we fit one model ten times to roughly estimate the standard error of $\overline{D}$ and DIC to get a sense of the scale of differences that may arise by randomness. We found that the standard errors ($\mathbb{SE}$) of $\overline{D}$, $P_D$, and DIC to be about 70, 100, and 120, respectively. Treating $\overline{D}$, $P_D$, and DIC for the models of independent, the standard error of the differences in $\overline{D}$, $P_D$, and DIC is (roughly) $\sqrt{ 2\mathbb{SE}^2} = \sqrt{2}\mathbb{SE}$. We consider differences of roughly two standard errors ``significantly'' different in model performance data (200 in $\overline{D}$, 280 in $P_D$, and 340 in DIC). Again, this is only a sensitivity analysis, so these comparisons are somewhat informal and intended only to discriminate models whose fits are significantly different (better or worse) than others. Primary model questions are answered by out-of-sample prediction in the manuscript and in Section \ref{sec:cv_eta}.

In the first case, we consider variations on the form of $\eta(\bs,t)$ discussed in the manuscript (Table \ref{tab:eta}).  We find that the asymmetric specification of $\eta(\bs,t) = \bk(t)^T \bz(\bs)$ is best here. This is expected based on our exploratory analyses where (1) spatial autocorrelation decays rapidly and (2) the wavelength function is evidently heterogeneous. The spatial convolution model $\bk(\bs)^T \bz(t)$ assumes a low-rank spatial model which fails to capture short-range spatial relationships. We determine the form of $\bz(\bs)$ in the manuscript through out-of-sample predictive performance.
\begin{table}[H]
\centering
\begin{tabular}{rrrr}
  \hline
$\eta(\bs,t)$ & $\overline{D}$ &$P_D$ & DIC \\ 
  \hline
$\bk(t)^T \bz(\bs)$& 0 & 655 & 0 \\
$\bk(\bs)^T \bz(t)$  & 84064 & 1198 & 85752 \\ 
$\alpha(\bs)$ & 47120 & 154 & 47765 \\  
None & 97711 & 0 & 98202 \\ 
   \hline
\end{tabular}
\caption{Comparison of models with different structural forms of $\eta(\bs,t)$.}\label{tab:eta}
\end{table}

%
%

As a follow-up, we also consider four specifications of the decay parameters of the independent GPs $\bw(\bs)$ used to construct $\bz(\bs)$. In the first case, we fixed all $\phi_j$ to be the same value before model fitting. In the second case, we fixed $\phi_j$ to be a sequence of increasing values to allow more flexibility in $\bz(\bs)$. Third, we include a single decay parameter in the model fitting. Lastly, we learn $\phi_j$ and include a prior ordering constraint. Recalling the roughly estimated standard errors, we find that these specifications have almost the same performance (See Table \ref{tab:phi}). We use the second case, a fixed sequence of $\phi_j$, because it excludes $\phi_j$ from model fitting but allows appealing flexibility. 

\begin{table}[H]
\centering
\begin{tabular}{rrrr}
  \hline
$\phi_j$ & $\overline{D}$ &$P_D$ & DIC \\ 
  \hline
Fixed Single & 67 & 156 & 223 \\ 
Fixed Sequence & 73 & 155 & 230 \\ 
Random Single & 69 & 62 & 132 \\ 
Random Sequence & 0 & 0 & 0 \\ 
   \hline
\end{tabular}
\caption{Comparison of models with different specifications of the range parameters of $\bw(\bs)$. }\label{tab:phi}
\end{table}

For $\gamma(t)$, we consider two specifications: one using a process convolution with wavelength-dependent bandwidths and one with a full GP with a single decay parameter. The process convolution specification is nonstationary but low-rank, while the GP is full-rank but stationary. Table \ref{tab:gamma} shows that the nonstationary process convolution is better in terms of all measures.

\begin{table}[H]
\centering
\begin{tabular}{rrrr}
  \hline
$\gamma(t)$ & $\overline{D}$ &$P_D$ & DIC \\ 
  \hline
Non-stationary Process Convolution & 0 & 0 & 0 \\ 
Full (stationary) GP & 283 & 462 & 746 \\ 
   \hline
\end{tabular}
\caption{Comparison of models with different forms for $\gamma(t)$.}\label{tab:gamma}
\end{table}

In Tables \ref{tab:bandwidth} and \ref{tab:kernel_dist}, we consider specifications of the the wavelength kernels. Interestingly, we find that the model with a single kernel bandwidth is competitive with the model with a wavelength-dependent bandwidth. We found that the Gaussian kernel was significantly better than the double exponential kernel. The Gaussian kernel approximates a Gaussian covariance, meaning that the better model is very smooth. Given the smoothness of the observed reflectance curves, this result is not surprising. Ultimately, we use Gaussian kernels with a common bandwidth for all wavelengths.

Because we use low-rank specifications for $\beta(t)$, we wish to examine the sensitivity of the fit to the spacing of knots. We find that the fit is quite robust to the specification of $\beta(t)$; therefore, we use one of the richest specifications that does not make DIC ``significantly'' worse (25 nm). We specify $\log(\sigma^2(t))$ as a piecewise linear function and explore the sensitivity of this to the spacing of the knots (See Table \ref{tab:dx_tau}). The fit appears best when knots for $\log(\sigma^2(t))$ are spaced every 20 nm, but the fit is comparable when spacing is 10 nm or 25 nm. For the spacing of knots for $\eta(\bs,t)$, we find that knots spacing of 25 nm or less yields comparable model fits, so we use 25 nm as the spacing for wavelength knots (See Table \ref{tab:dx_tau}). Lastly, for the spacing of knots for $\gamma(t)$, we find that knot spacing of 25 nm or less yields comparable model fits (See Table \ref{tab:dx_beta}), so we use 25 nm as the spacing for wavelength knots for $\gamma(t)$.


\begin{table}[H]
\centering
\begin{tabular}{rrrr}
  \hline
Kernel & $\overline{D}$ &$P_D$ & DIC \\ 
  \hline
One Bandwidth & 10 & 149 & 160 \\ 
Wavelength-Dependent & 0 & 0 & 0 \\ 
   \hline
\end{tabular}
\caption{Comparison of models with different wavelength bandwidths for $\eta(\bs,t)$.}\label{tab:bandwidth}
\end{table}

\begin{table}[H]
\centering
\begin{tabular}{rrrr}
  \hline
Kernel & $\overline{D}$ &$P_D$ & DIC \\ 
  \hline
Gaussian & 0 & 0 & 0 \\ 
Double-Exponential & 1497 & 69 & 1566 \\ 
   \hline
\end{tabular}
\caption{Comparison of models with different kernel functions for $\gamma(t)$ and $\eta(\bs,t)$.}\label{tab:kernel_dist}
\end{table}


\begin{table}[H]
\centering
\begin{tabular}{lrrr}
  \hline
$\bbeta(t)$ Knot spacing & $\overline{D}$ &$P_D$ & DIC \\ 
  \hline
20 nm  & 0 & 167 & 148 \\ 
25 nm  & 66 & 23 & 69 \\ 
50 nm & 50 & 94 & 124 \\ 
100 nm  & 20 & 0 & 0 \\ 
   \hline
\end{tabular}
\begin{tabular}{rrrr}
  \hline
$\gamma(t)$ - Knot spacing & $\overline{D}$ & $P_D$ & DIC \\ 
  \hline
10 nm  & 0 & 82 & 16 \\ 
20 nm  & 151 & 49 & 135 \\ 
25 nm  & 65 & 0 & 0 \\ 
50 nm  & 277 & 31 & 243 \\ 
   \hline
\end{tabular}
\caption{Knot spacing of process convolution used in $\bbeta(t)$ and $\gamma(t)$.}\label{tab:dx_beta}
\end{table}


\begin{table}[H]
\centering
\begin{tabular}{rrrr}
  \hline
$\log\left[ \sigma^2(t)\right]$ knot spacing & $\overline{D}$ &$P_D$ & DIC \\ 
  \hline
10 nm  & 22 & 91 & 57 \\ 
20 nm  & 0 & 57 & 0 \\ 
25 nm  & 160 & 0 & 103 \\ 
50 nm & 2466 & 33 & 2443 \\ 
100 nm  & 9572 & 0 & 9515 \\ 
   \hline
\end{tabular}
\begin{tabular}{rrrr}
  \hline
$\eta(\bs,t)$ - Knot spacing & $\overline{D}$ & $P_D$ & DIC \\ 
  \hline
10 nm & 0 & 161 & 0 \\ 
20 nm & 129 & 79 & 47 \\ 
25 nm & 167 & 127 & 133 \\ 
50 nm & 3561 & 45 & 3444 \\ 
100 nm & 11701 & 0 & 11540 \\ 
   \hline
\end{tabular}
\caption{Knot spacing of the piecewise linear specification of $\log\left[ \sigma^2(t)\right]$ and $\eta(\bs,t)$. }\label{tab:dx_tau}
\end{table}


 \section{Prior Distributions, Model Fitting, and Prediction}\label{sec:gibbs}

We present the relevant fitting details for the model with the best out-of-sample predictive performance:
$$Y_{ij}(\bs,t) = \alpha_i + \alpha_i(\bs) + \bx(\bs)\bbeta(t) + \gamma(t) + \eta(\bs,t) + \epsilon_{ij}(\bs,t),$$
where the variance of the error is  piecewise linear on the log-scale, $\log(\sigma^2(t)) = K_\sigma(t)^T\bbeta_\sigma, $ as discussed in the manuscript. To present the model completely, we define assumptions and terms mathematically that were described in words in the manuscript and discuss the prior distributions.

We use finite variance inverse gamma prior distributions for variance parameters that, given the scale of our data, are only weakly informative. As discussed, the decay parameters $\phi_{w_j}$ are fixed and ordered to provide flexibility to the latent factor representation of $\bz(\bs)$. $\vec{B}$ and $\bgamma^*$ are the low-rank representations of the regression coefficients and the global wavelength random effect discussed in the manuscript. The correlation matrix $R_\gamma$ is specified through an exponential correlation function with $\phi_\gamma = 1/50$ and imposed correlation between the bandwidths of the nonstationary kernel convolution $\gamma(t)$. 

\begin{equation}
\begin{aligned}
\alpha_i &\overset{ind}{\sim} \mathcal{N}\left(\alpha , \sigma^2_\alpha\right) \\
\alpha &\sim \mathcal{N}\left(0,10^2\right) \\
\alpha_i(\bs)  &\sim \mathcal{GP}\left(0,\sigma^2_{\alpha(\bs)} e^{-\phi_\alpha d(\bs,\bs')}\right) \\
\text{vec}(B) &\sim \mathcal{N}\left(\bzero, \sigma^2_\beta \mathbb{I} \right) \\
\bgamma^* &\sim \mathcal{N}\left(\bzero, \sigma^2_\gamma \mathbb{I} \right) \\
w_j(\bs) &\sim \mathcal{GP}\left(0, e^{-\phi_{w_j} d(\bs,\bs')}\right) \\
\sigma^2_\alpha &\sim \mathcal{IG}\left(3,0.2 \right) \\
\sigma^2_{\alpha(s)} &\sim \mathcal{IG}\left(3,2 \right)  \\
\sigma^2_{\beta} &\sim \mathcal{IG}\left(3,2 \right) 
\end{aligned}
\hspace{5mm}
\begin{aligned}
\phi_\alpha &\sim \text{Unif}\left(1/100,1 \right) \\
\theta^{(\beta)} &\sim \text{Gamma}\left(5,1/10 \right) \\
\log\left(\theta^{(\gamma)}_{t_1^\gamma},...,\theta^{(\gamma)}_{t_J^\gamma} \right) &\sim \mathcal{N}\left(\mu_{\theta_\gamma},\sigma^2_{\theta_\gamma} R_\gamma \right) \\
A_{ik} &\overset{iid}{\sim} \mathcal{N}\left(0,\sigma^2_A \right) \\
\theta^{(\eta)} &\sim \text{Gamma}\left(5,1/10 \right) \\
\bbeta_\sigma &\sim \mathcal{N}\left(0,10^2 \mathbb{I}\right) \\
\mu_{\theta_\gamma} &\sim \mathcal{N}\left(3,3^2\right)\\
\sigma^2_\gamma &\sim \mathcal{IG}\left(3,2 \right) \\
\sigma^2_A &\sim \mathcal{IG}\left(11,10 \right) \\
\sigma^2_{\theta_\gamma} &\sim \mathcal{IG}\left(5,2 \right)
\end{aligned}
\end{equation}

We fit our model using Markov chain Monte Carlo (MCMC) through a Gibbs sampler when closed-form posterior conditional distribution are available and using Metropolis-Hastings within Gibbs when these conditional distributions are not available in closed form. The spatial GP parameter $\phi_\alpha$, bandwidth parameters ($\theta^{(\beta)}$, $\theta^{(\gamma)}_{t_1^\gamma}$,..., $\theta^{(\gamma)}_{t_J^\gamma}$), and variance regression parameters $\bbeta_\sigma$ are all updated using Metropolis-Hastings.
We sample from the posterior distribution 200,000 times. We discard the first 150,000 samples and thin the remaining 50,000 samples to 10,000 posterior samples. We base our inference on these 10,000 posterior samples.

After model fitting, when of interest, we sample from the posterior predictive distribution, 
$$ \int_{\Psi} f(y_{new} | \psi) \pi(\psi | \bY) d \psi,$$
using composition sampling \citep[see][for early reference]{tanner1996}, where $\psi$ denotes all model parameters and $\bY$ represents all data. For model comparison, this predictive approach allows us to compare the entire empirical predictive distribution to hold-out reflectance curves. When interpolating reflectances to a new site, this involves sampling spatially-varying intercepts $\alpha_i(\bs)$ and latent GP parameters $w_j(\bs)$ from the appropriate conditional normal distribution conditional on sampled values.

We use $\cdot | \cdots$ the conditional distribution given all other parameters, as well as the data. For computational speed, we sample all $\bw(\bs)$ for a given site jointly, effectively updating the whole space-wavelength function $\eta(\bs,t)$ site-wise. We also considered updating $w_j(\bs)$ for all sites; however, this did not mix as effectively. For simplicity, we define $\mu_i(\bs,t)_{-\alpha_i}$,  $\mu_i(\bs,t)_{-\alpha_i(\bs)}$, $\mu_i(\bs,t)_{-\beta}$, $\mu_i(\bs,t)_{-\gamma}$, $\mu_i(\bs,t)_{-\eta}$ to the mean term excluding $\alpha_i$, $\alpha_i(\bs)$, $\bx(\bs)\bbeta(t)$, $\gamma(t)$, and $\eta(\bs,t)$, respectively. When an index is excluded, it indicates that we are looking at all terms. For example, $Y_{ij}(\bs)$ represents all log-reflectances for genus $i$, replicate $j$, and at the location $\bs$. The matrix $D_\sigma$ is a diagonal matrix with all $\sigma^2(t)$ on the diagonal.

Let $N_i$ be the number of curves in the dataset for the $i$th genus, $N(\bs)$ be the number of curves at site $\bs$, while $N_i(\bs)$ is the number of curves for the $i$th genus at site $\bs$. We refer to some definitions introduced in Section \ref{sec:ortho}. The full conditional distributions are as follows:
\begin{equation}
\footnotesize
\begin{aligned}
\alpha_i | \cdots &\sim \mathcal{N}\left(v^*_{\alpha_i} m^*_{\alpha_i} , v^*_{\alpha_i}\right) \\
\balpha_i(\bs) | \cdots &\sim \mathcal{N}\left(v^*_{\alpha_i(\bs)}m^*_{\alpha_i(\bs)} , v^*_{\alpha_i(\bs)}\right) \\
\alpha | \cdots &\sim \mathcal{N}\left(v^*_{\alpha}m^*_{\alpha} , v^*_{\alpha}\right) \\
\text{vec}(B) | \cdots  &\sim \mathcal{N}\left(v^*_{\beta}m^*_{\beta} , v^*_{\beta}\right) \\
\gamma^* | \cdots  &\sim \mathcal{N}\left(v^*_{\gamma}m^*_{\gamma} , v^*_{\gamma} \right) \\
\bw(\bs) | \cdots  &\sim \mathcal{N}\left(v^*_{\bw(\bs)}m^*_{\bw(\bs)} , v^*_{\bw(\bs)} \right) \\
\mu_{\theta_\gamma} | \cdots &\sim \mathcal{N}\left(v^*_{\theta_\gamma}m^*_{\theta_\gamma} , v^*_{\theta_\gamma} \right)
\end{aligned}
\begin{aligned}
A_{k} | \cdots&\sim \mathcal{N}\left(v^*_{A_{j}}m^*_{A_{j}} , v^*_{A_{j}}  \right) \\
\sigma^2_\alpha | \cdots &\sim \mathcal{IG}\left(a^*_\alpha,b^*_\alpha  \right) \\
\sigma^2_{\alpha(s)} | \cdots &\sim \mathcal{IG}\left(a^*_{\alpha(bs)},b^*_{\alpha(\bs)} \right) \\
\sigma^2_{\beta} | \cdots  &\sim \mathcal{IG}\left(a^*_{\beta},b^*_{\beta} \right)  \\
\sigma^2_{\gamma} | \cdots  &\sim \mathcal{IG}\left(a^*_{\gamma},b^*_{\gamma} \right)  \\
\sigma^2_{A} | \cdots  &\sim \mathcal{IG}\left(a^*_{A},b^*_{A} \right)  \\
\sigma^2_{\theta_\gamma} | \cdots  &\sim \mathcal{IG}\left(a^*_{\theta_\gamma},b^*_{\theta_\gamma} \right), \qquad \text{where}  \\
\end{aligned}
\end{equation}
\vspace{-2mm}
\begin{equation*}
\tiny
\begin{aligned}
m^*_{\alpha_i} &= \sum_\bs \sum_{j} \sum_t \left( Y_{ij}(\bs,t) - \mu_i(\bs,t)_{-\alpha_i}\right) / \sigma^2(t) + \alpha /\sigma^2_\alpha\\
m^*_{\alpha_i(\bs)} &=  \sum_{j} \sum_t \left( Y_{ij}(\bs,t) - \mu_i(\bs,t)_{-\alpha_i(\bs)}\right) / \sigma^2(t) \\
m^*_{\alpha}&= \frac{1}{\sigma^2_\alpha} \sum_i \alpha_i \\
m^*_{\beta}&= \text{vec}\left[\bK_\beta^T D_\sigma^{-1} \left( \bY - \bmu_{-\beta} \right) \right]\\
m^*_{\gamma}&= \bK_\gamma^T \sum_\bs \sum_i \sum_j D_\sigma^{-1} \left( Y_{ij}(\bs) - \mu_i(\bs)_{-\gamma} \right) \\
m^*_{\bw(\bs)}&= A^T K_\eta^T \sum_i \sum_j D^{-1}_\sigma (Y_{ij}(\bs) - \bmu_i(\bs)_{-\bw(\bs)}) \\ &+ \begin{pmatrix}
 \text{Cov}[w_1(\bs), w_1(-\bs)] \text{Cov}^{-1}[w_1(-\bs)]w_1(-\bs) \\ \vdots \\ \text{Cov}[w_r(\bs), w_r(-\bs)]\text{Cov}^{-1}[w_r(-\bs)]w_r(-\bs)   
\end{pmatrix}\\
m^*_{\theta_\gamma}&= \bone^T R_\gamma^{-1} \log\left(\theta^{(\gamma)}_{t_1^\gamma},...,\theta^{(\gamma)}_{t_J^\gamma}  \right) \bone / \sigma^2_{\theta_\gamma} + 3/9 \\
m^*_{A_k}&=  K_\eta^T  D^{-1}_\sigma \left( \bY - \bmu_{-A_j w_j(\bs)} \right) \bM_s \bw_j \\
a^*_\alpha &= 3 + N_{g}/2 \\
a^*_{\alpha(\bs)} &= 3 + N_i(\bs)/2\\
a^*_\beta &= 3 + J_\beta P / 2 \\
a^*_\gamma &= 3 + J_\gamma/2\\
a^*_A &=11 + rJ_\eta/2 \\
b^*_A &= 10 + \frac{1}{2}\sum_i \sum_k A_{ik}^2  \\
\end{aligned}
\hspace{2mm}
\begin{aligned}
v^*_{\alpha_i} &=  \left(N_i \sum_t 1/ \sigma^2(t) + 1 / \sigma^2_\alpha \right)^{-1} \\
v^*_{\alpha_i(\bs)} &= \left(R_{\bs_i}^{-1}/ \sigma^2_{\alpha(\bs)} +  \mathbb{I} N_i(\bs) \sum_t 1/ \sigma^2(t) \right)^{-1} \\
v^*_{\alpha}&= \left( 1/100 + N_i / \sigma^2_\alpha\right)^{-1} \\
v^*_{\beta}&= \left( \mathbb{I} /\sigma^2_\beta + \bX^T\mathbf{M}_s^T \bM_s \bX \otimes \bK_\beta ^T \bK_\beta  \} \right)^{-1}\\
v^*_{\gamma}&= \left( \mathbb{I} /\sigma^2_\gamma + \bK_\gamma^T \bK_\gamma  \} \right)^{-1} \\
v^*_{\bw(\bs)}&= \left( \text{diag}\left(\text{var}(w_j(\bs) |w_j(-\bs)^{-1}\right) + N(\bs) A^T K_\eta^T D_{\sigma}^{-1} K_\eta A  \right)^{-1}\\
v^*_{\theta_\gamma}&= \left(\bone^T R^{-1}_\gamma \bone/\sigma^2_{\theta_\gamma} + 1/9 \right)^{-1} \\
v^*_{A_k}&= \left( K_\eta^T D_\sigma^{-1} K_\eta +  \mathbb{I}\sigma^2_A \right)^{-1} \\
b^*_\alpha &= 0.2 + \frac{1}{2} \sum_i \left(\alpha_i - \alpha\right)^2  \\
b^*_{\alpha(\bs)} &= 2 + \frac{1}{2}\balpha_i(\bs)^T R_{\bs_i}^{-1} \balpha_i(\bs)\\
b^*_\beta &= 2 + \bone^T \text{vec}(B)/2 \\
b^*_\gamma &= 2 + \bone^T \bgamma^*/2 \\
a^*_{\theta_\gamma} &= 5 + J_\gamma/2\\
b^*_{\theta_\gamma} &= 2 + \\ &\frac{1}{2} \left[\log\left(\theta^{(\gamma)}_{t_1^\gamma},...,\theta^{(\gamma)}_{t_J^\gamma} \right)-\bone \mu_{\theta_\gamma} \right]^{T}R_\gamma^{-1} \left[\log\left(\theta^{(\gamma)}_{t_1^\gamma},...,\theta^{(\gamma)}_{t_J^\gamma} \right) -\bone \mu_{\theta_\gamma} \right]
\end{aligned}
\end{equation*}

\section{Extended Discussion on Orthogonalization}\label{sec:ortho}

Because we have the dual goal of using this model for prediction and estimation, we discuss how to preserve informative inference on the regression coefficient functions. 
In this section, we expand our discussion in the main manuscript to explicitly address the imbalance in our data. Specifically, we do not have a unique location for each reflectance curve; genera are observed at some sites and not others. 

We resolve issues of spatial and functional basis confounding between $\eta(\bs,t) + \gamma(t) + \alpha_i +  \alpha_i(\bs)$ and $\bx(\bs)^T \bbeta(t)$. To illustrate how the functional basis confounding contributes to this problem, we rewrite the model. To do this, we define several terms. Let $N_{wave} = 500$ be the number of log-reflectances measured in each reflectance curve, $N_{rep}$ be the number of observed reflectance curves, and $N_s$ be the number of unique spatial sites, and $N_g$ be the number of unique genera. In addition, we define a variety of other terms, including several index matrices that map random effects to the correct genus, location, or both Our primary goal here is to demonstrate that the form of $\boeta^*$ is complicated because our dataset is not balanced.  We let:
\begin{itemize}
\item $\bY$ be a $N_{wave} \times N_{rep}$ matrix of log-reflectances.
\item $\bM_s$ be a $N_{rep} \times N_s$ matrix that indexes the spatial site of the $i$th specturm. 
\item $\bM_{sg}$ be a $N_{rep} \times (N_{s}N_g)$ matrix that indexes the spatial site/genus combination of the $i$th specturm.
\item $\bX$ be a $N_{s} \times P$ covariate matrix, where we have centered and scaled each variable such that $\bM_{sg} \bX$ has mean of 0 and variance of 1. 
\item $\bK_\beta$ be the $N_{wave} \times N_{\beta,knot}$ matrix of wavelegnth kernel/basis weights associated with the coefficient functions.
\item $\bK_\eta$ be the $N_{wave} \times N_{s,knot}$ matrix of wavelength kernel/basis weights associated with the space-wavelength random effect $\eta(\bs,t)$
\item $\bgamma = (\gamma(t_1),...,\gamma(t_{N_wave})^T$ be the vector of all global wavelength random effects.
\item $\bw = (\bw_1,...,\bw_r)^T$ be a $N_s\times r$ matrix with $r$ independent spatial GPs.
\item $\balpha_s$ be an $N_{s} \times N_g$ matrix of mean-zero genus-specific spatially-varying random intercepts ($\alpha_i + \alpha_i(\bs) - \alpha$). In practice, we only sample terms of this matrix that appear in our dataset. However, given the model we have proposed, the genus-specific terms can be estimated at any location. 
\end{itemize}

With these terms defined, we more carefully specify the block model discussed in the manuscript to explicitly show the form of $\boeta^*$ while accounting for the data imbalance: 
\begin{equation}\label{eq:block_model}
\begin{aligned}
\bY &=\alpha\bone + \bK_\beta B (\bM_s\bX)^T + \boeta^* + \bepsilon \\ 
&= \alpha\bone + \bK_\beta B (\bM_s\bX)^T + \bK_\eta \bA (\bM_s\bw)^T  + \bgamma \bone^T + \bone (\bM_{sg}\balpha_s)^T  + \bepsilon,\\
\text{vec}(\bY) &=  \alpha\bone + \left(\bM_s\bX \otimes \bK_\beta \right) \text{vec}(B) + \left(\bI \otimes \bK_\eta \bA \right)  \text{vec}(\bM_s\bw) + \bone^T \otimes \bgamma + \bM_{sg}\balpha_s \otimes \bone + \text{vec}(\bepsilon).
\end{aligned}
\end{equation}
However, as previously discussed, we do not have the same inference on $B$, the low-rank coefficients, as we do if we exclude the residual terms: $\eta(\bs,t) + \gamma(t) +\alpha_i + \alpha_i(\bs)$. 
To preserve inference on $B$, the low-rank functional coefficients, we project the random effects of our model into the orthogonal column space of $\bM_s\bX \otimes \bK_\beta$. We define $\bP$ to be the projection into the column space of $\bM_s\bX \otimes \bK_\beta$, 
\begin{equation}
\begin{aligned}
\bP &= \left(\bM_s\bX \otimes \bK_\beta \right)\left( \left(\bM_s\bX \otimes \bK_\beta \right)^T \left(\bM_s\bX \otimes \bK_\beta \right) \right)^{-1} \left( \bM_s\bX \otimes \bK_\beta \right)^T, \\
&= \left( (\bM_s\bX) \left((\bM_s\bX)^T(\bM_s\bX) \right)^{-1} (\bM_s\bX)^T \right) \otimes \left( \bK_\beta \left(\bK_\beta^T \bK_\beta\right)^{-1} \bK_\beta^T \right), \\
&=\bP_X \otimes \bP_K,
\end{aligned}
\end{equation}
where $\bP_X$ and $\bP_K$ are projections into the column space of $\bM_s\bX$ and $\bK_\beta$, respectively. We define the orthogonal transformation,
\begin{equation}
\bP^\perp = \bI - \bP.
\end{equation}

Using this, \eqref{eq:block_model} can be rewritten as 
\begin{equation}\label{eq:block_model2}
\begin{aligned}
\text{vec}(\bY) &=  \alpha\bone + \left(\bM_s\bX \otimes \bK_\beta \right) \text{vec}(B) + 
\bP\left(\bI \otimes \bK_\eta \bA \right)  \text{vec}(\bM_s\bw)  + \bP^\perp\left(\bI \otimes \bK_\eta \bA \right)  \text{vec}(\bM_s\bw)+  \\
&\bP (\bone \otimes \bgamma) + \bP^\perp (\bone \otimes \bgamma) + 
\bP(\bM_{sg}\balpha_s\otimes \bone)  + \bP^\perp(\bM_{sg}\balpha_s \otimes \bone) +  
 \text{vec}(\bepsilon) \\
 &=\alpha\bone +   \left(\bM_s\bX \otimes \bK_\beta \right) \text{vec}(B^*) +  \bP^\perp\left(\bI \otimes \bK_\eta \bA \right)  \text{vec}(\bM_s\bw)+  \bP^\perp(\bone \otimes \bgamma) +  \\ & \bP^\perp(\bM_{sg}\balpha_s\otimes \bone) +  
 \text{vec}(\bepsilon). 
\end{aligned}
\end{equation}
Importantly, as in the manuscript, \eqref{eq:block_model2} illustrates the correspondence between $B$ and $B^*$, the unconfounded and confounded regression coefficients.
In fact, if the model in \eqref{eq:block_model} is fit using MCMC initially, as we do, posterior samples of $B^*$, as well as orthogonalized random effects, can be recovered after model fitting. 

As a final comment, we treat $\gamma(t)$ as a heterogeneous random effect rather than a fixed effect in this analysis. However, we acknowledge that viewing $\gamma(t)$ as a fixed effect is a reasonable choice. Specifically, if the global wavelength function is treated as a fixed effect, we would include a column of ones in the design matrix $\bX$ defining the projection. We would also update $\gamma(t)$ as we do with the functional coefficients $\bbeta(t)$. Because this selection represents a different base model, this would change the inference on ``unconfounded'' coefficients.

\section{Extended Analysis of Results}\label{sec:results}

To illustrate the degree of confounding, we show that the \emph{confounded} coefficients are near zero for all wavelengths for Asteraceae. This is similar for all three families. For all families, we plot the functional regression coefficients (after orthogonalization) with the regression coefficient functions found through least-squares using a model without any random effects. 

%
%

\subsection{Asteraceae}

\begin{figure}[H]
\begin{center}
\includegraphics[width=.35\textwidth]{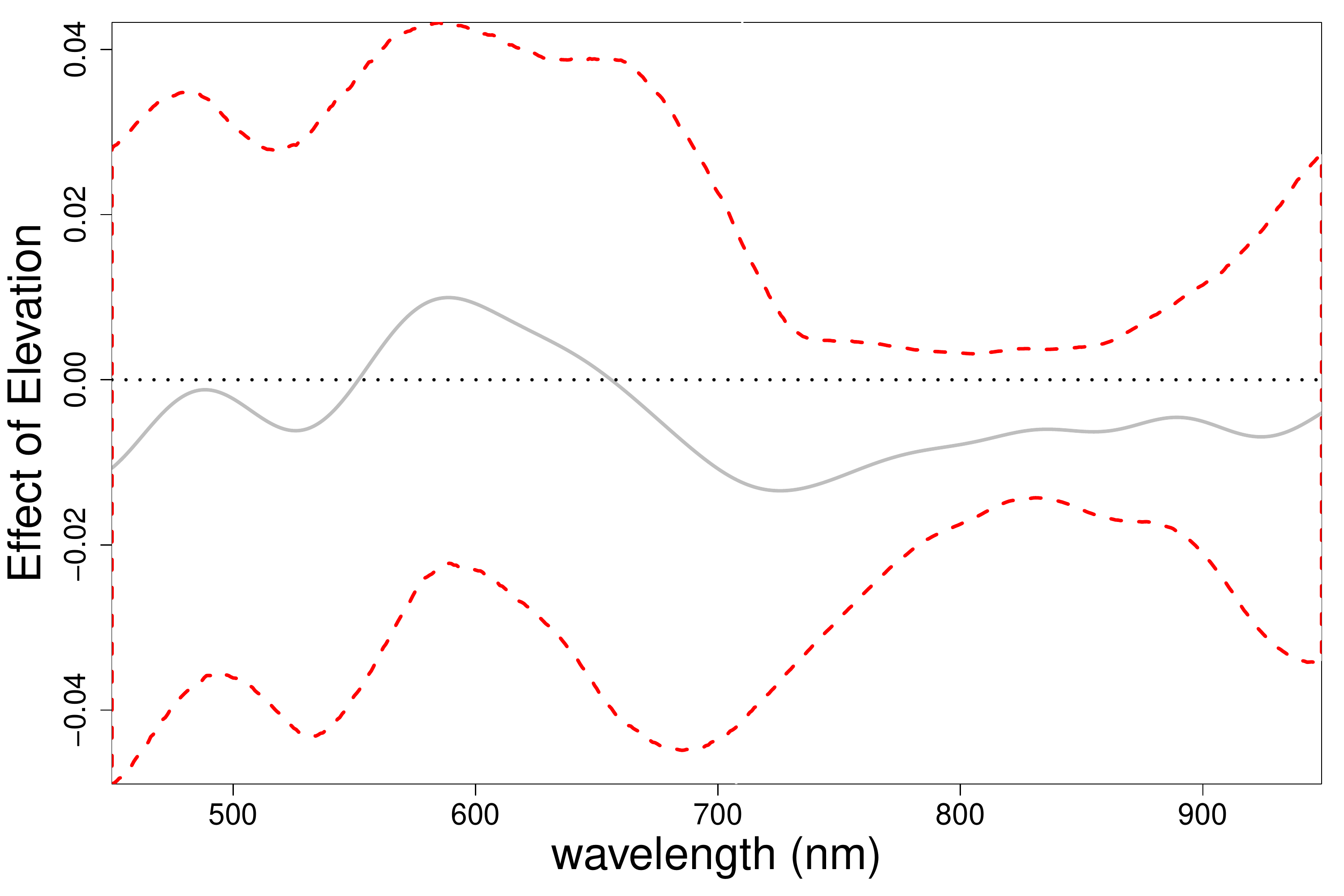}
\includegraphics[width=.35\textwidth]{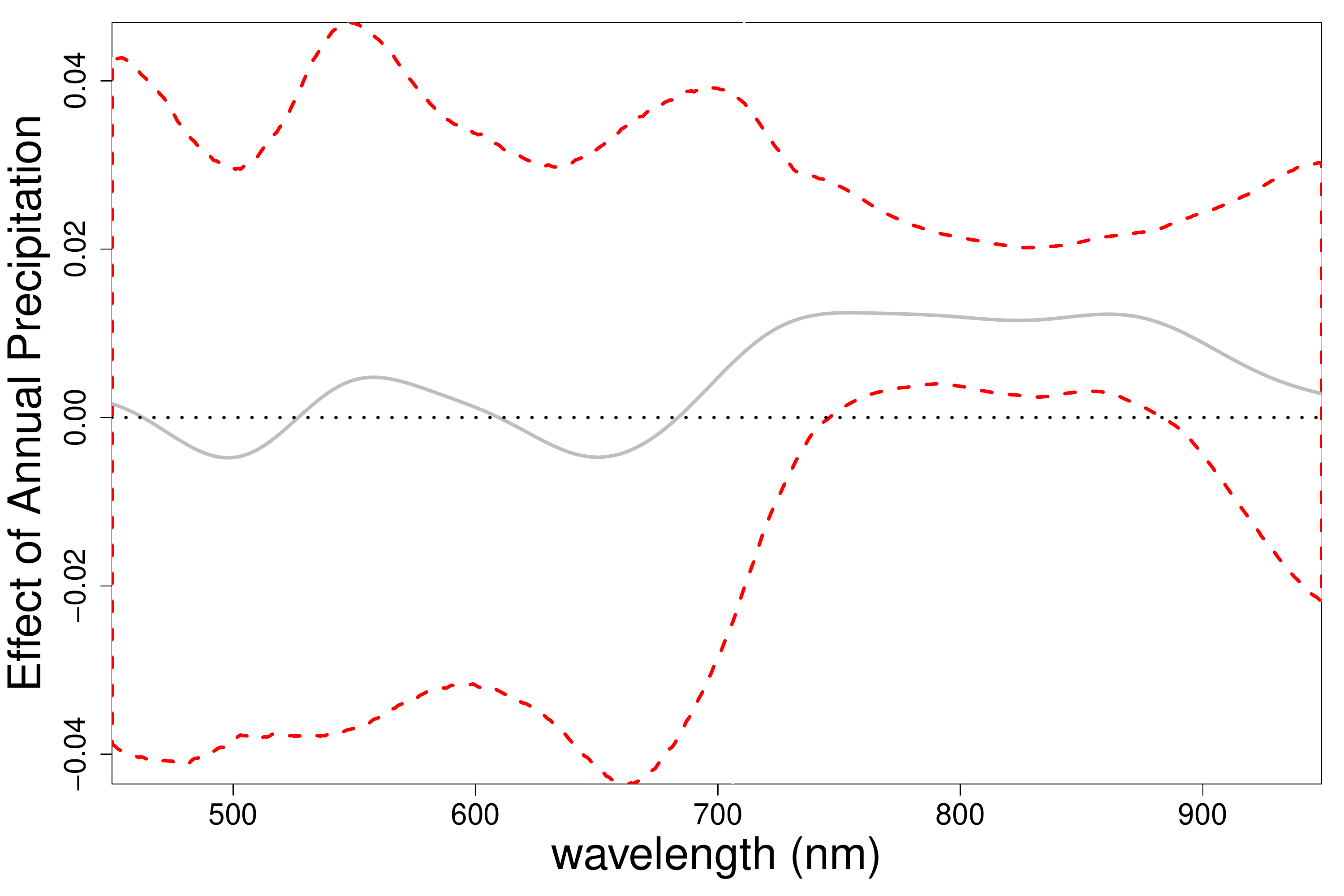}
\includegraphics[width=.35\textwidth]{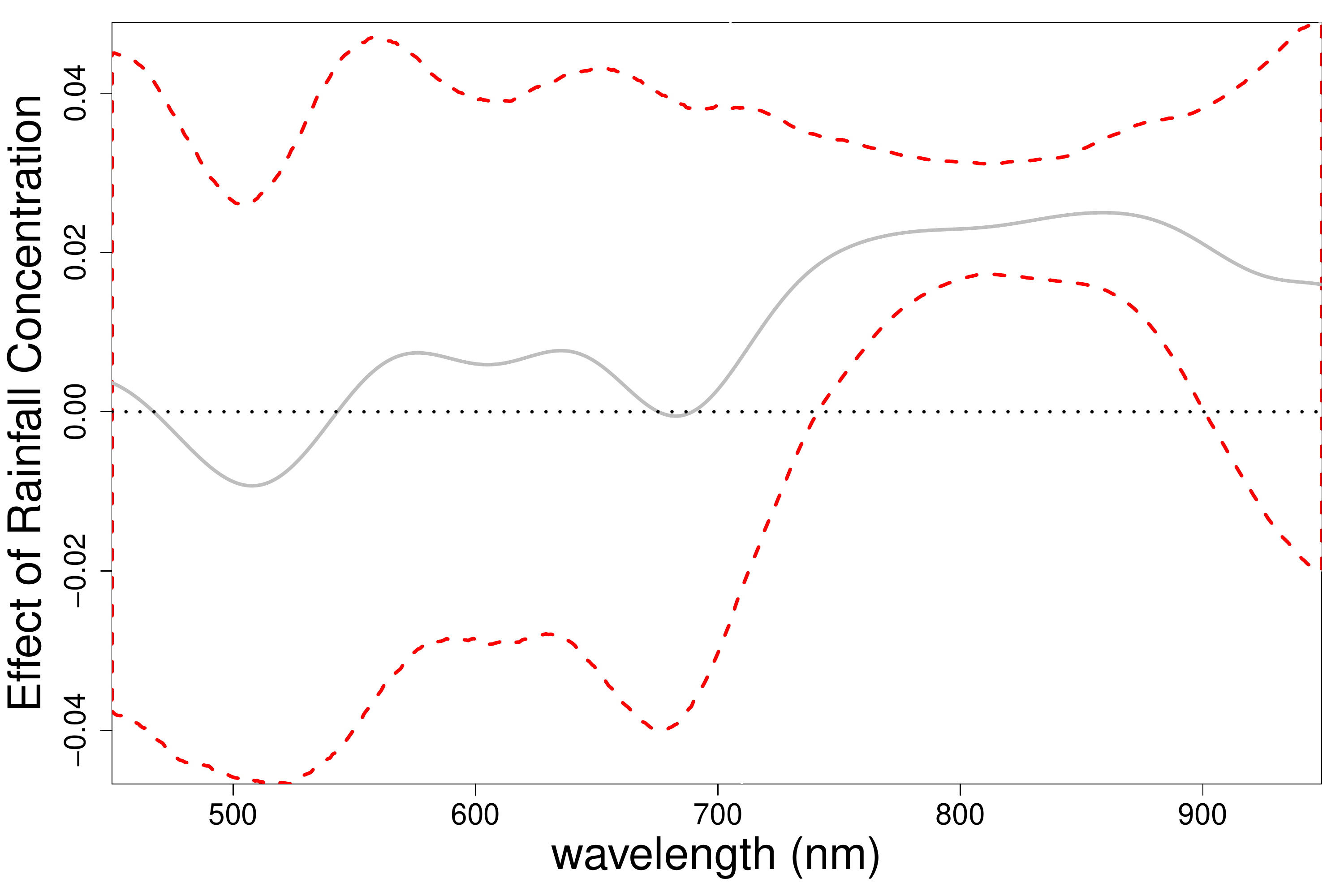}
\includegraphics[width=.35\textwidth]{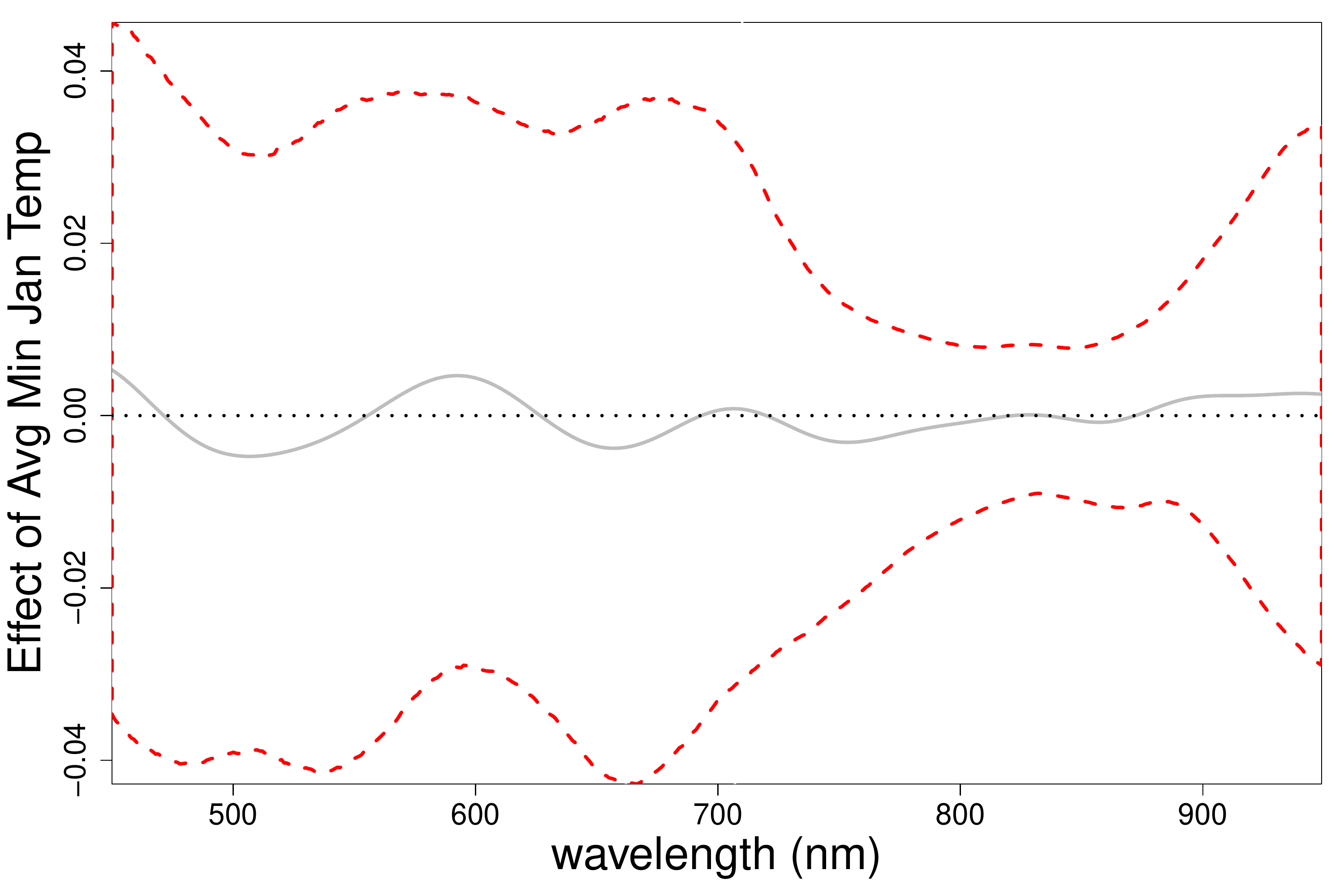}

\end{center}
\vspace{-4mm}

\caption{Confounded coefficient functions for Asteraceae.}\label{fig:beta_t_confounded}
\end{figure}

To demonstrate the degree of confounding between the random effects (genus, wavelength, and spatial) and covariates, we plot the posterior mean and credible interval for regression coefficients without orthogonalization. The confounding pushes $\bbeta(t)$ to zero (See Figure \ref{fig:beta_t_confounded}), obliterating any significant inference with regard to the effect of environmental variables on log-reflectance. Although we only present these results for the Asteraceae family, similar patterns are present for the Aizoaceae and Restionaceae families.

\begin{figure}[H]
\begin{center}
\includegraphics[width=.4\textwidth]{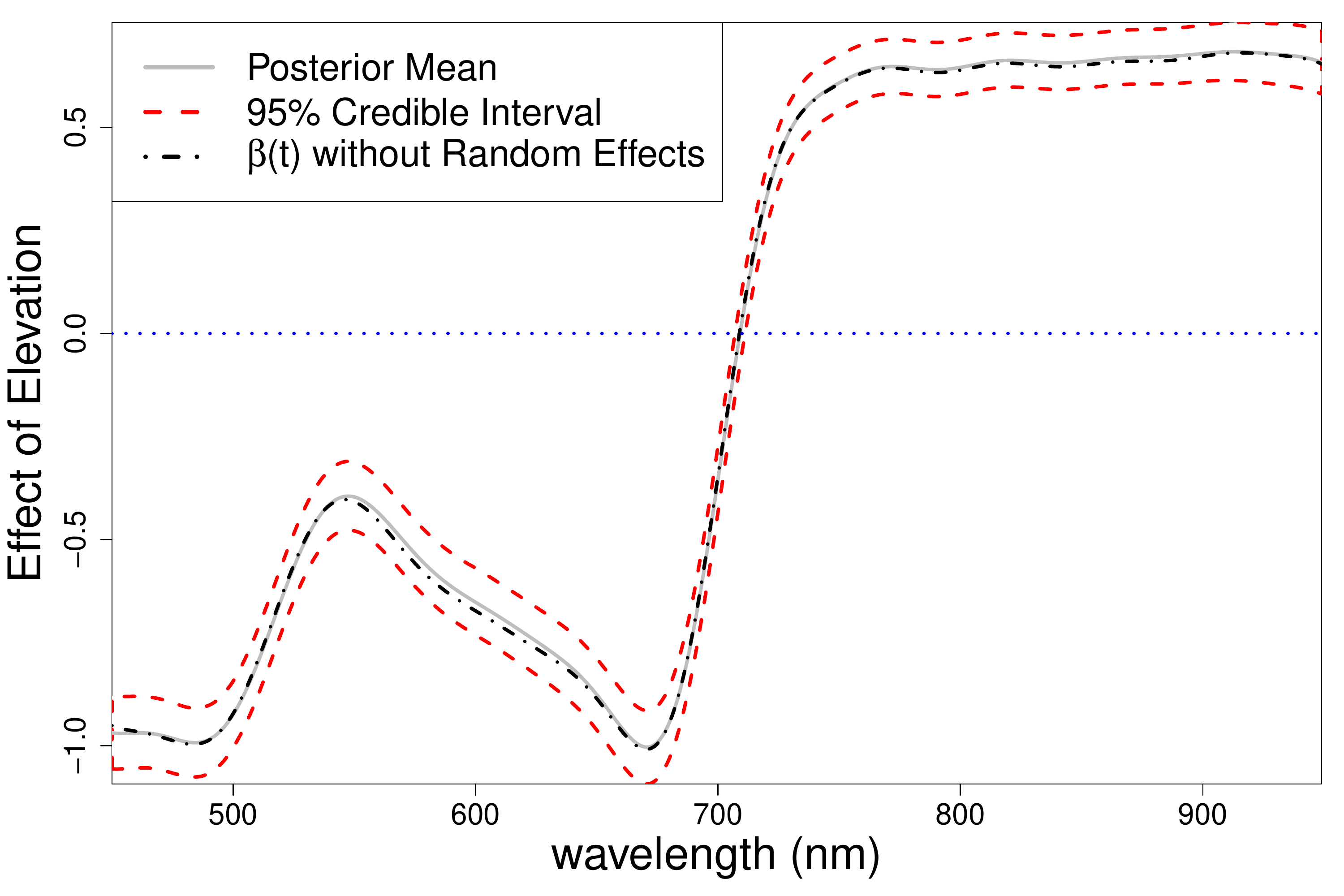}
\includegraphics[width=.4\textwidth]{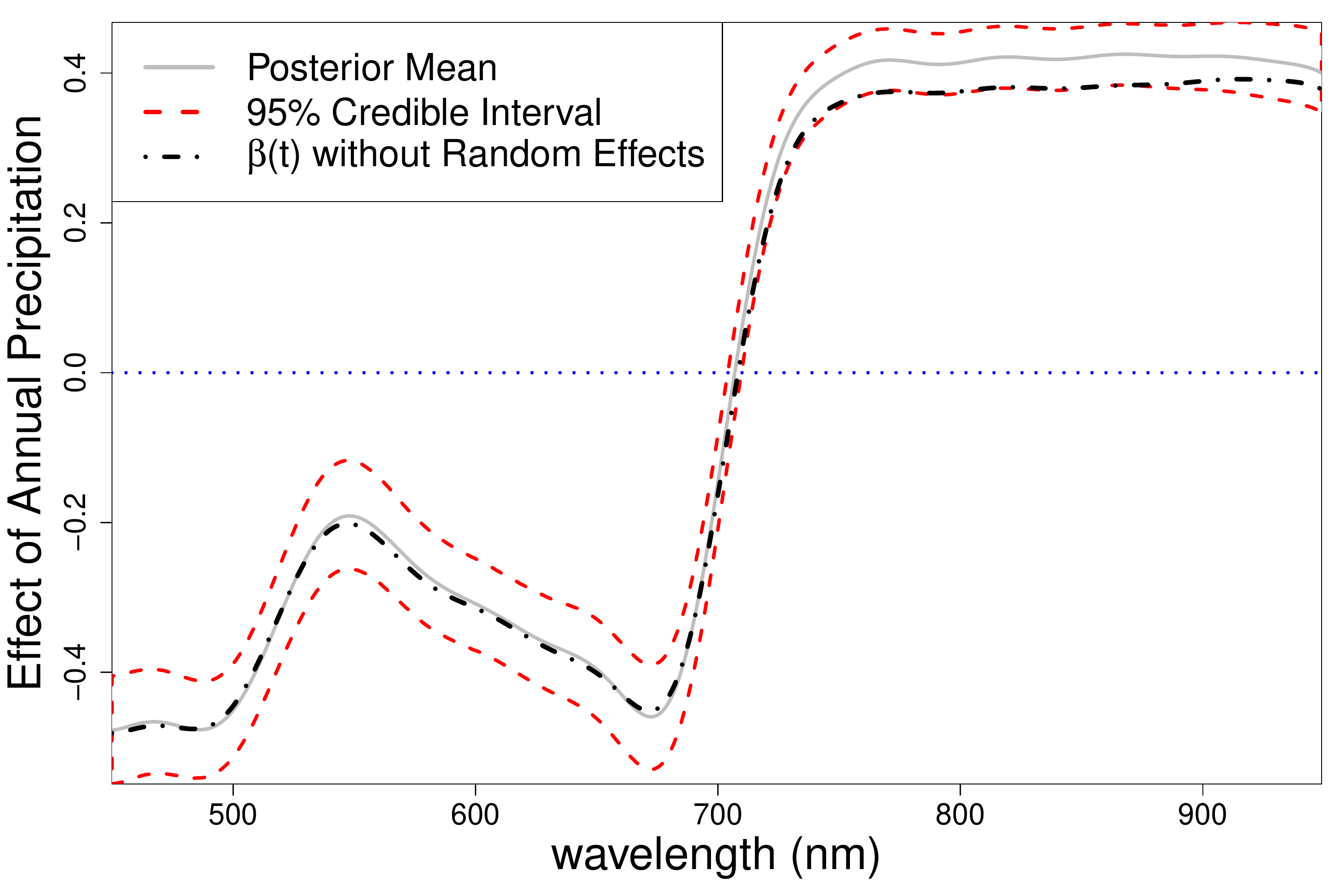}
\includegraphics[width=.4\textwidth]{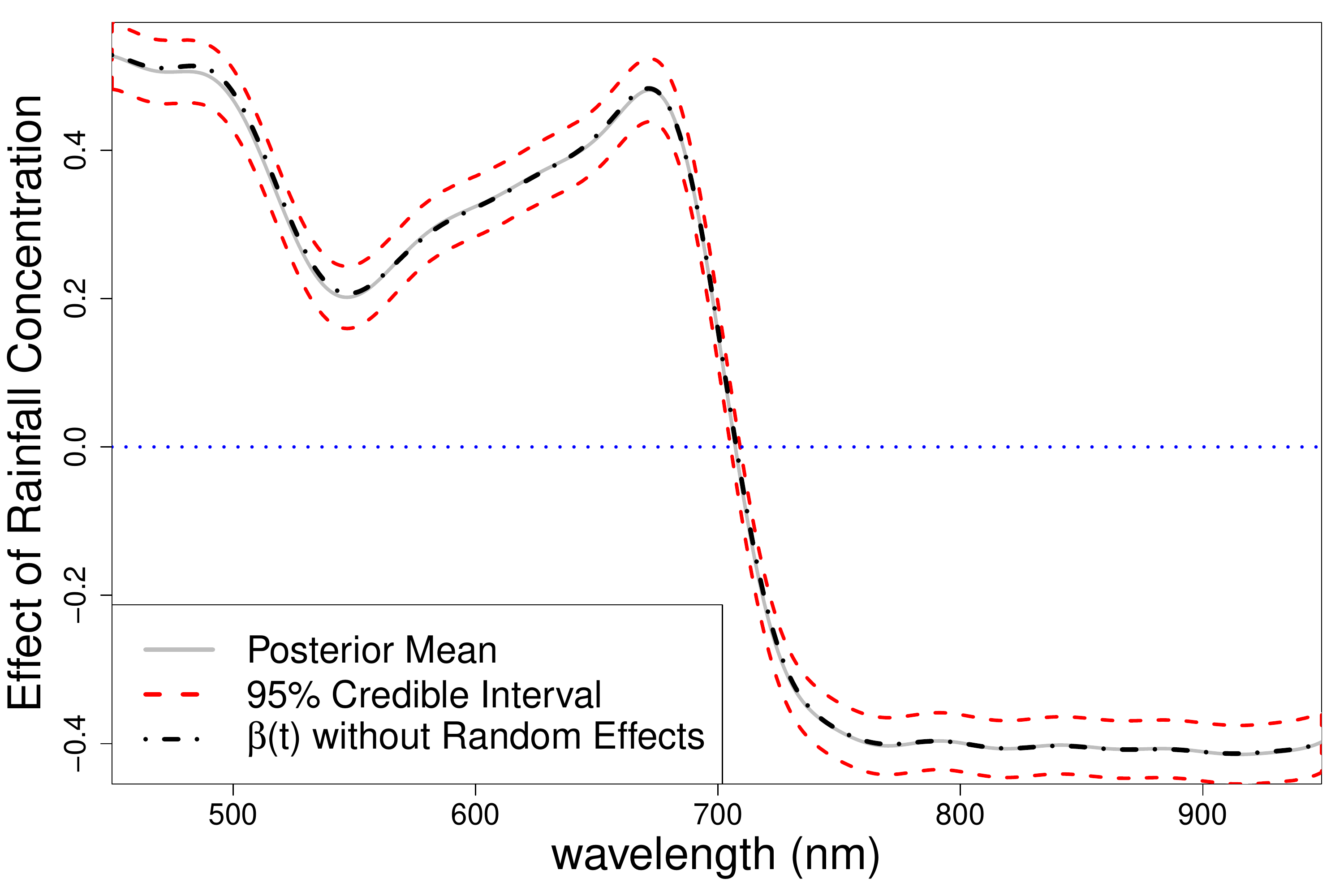}
\includegraphics[width=.4\textwidth]{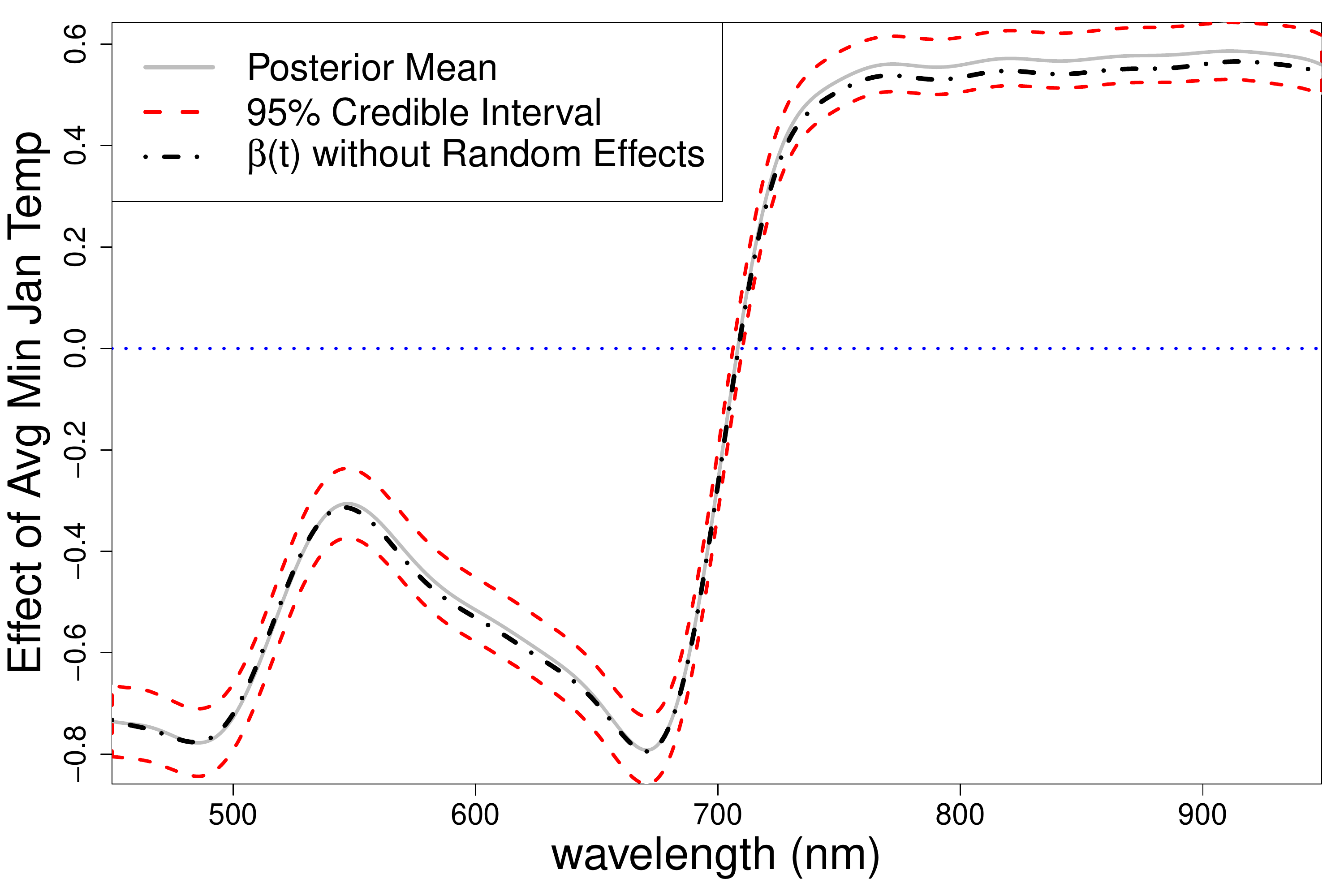}

\end{center}
\vspace{-4mm}

\caption{Unconfounded coefficient functions for Asteraceae.}\label{fig:beta_t}
\end{figure}

After updating $\bbeta(t)$ using the projection of the random effects onto the column spaces of $\bX$ and $\bK_\beta$, we present our inference on covariates. In Figure \ref{fig:beta_t}, we present the posterior mean, 95\% credible interval for each $\beta(t)$, as well as the least-squares estimate of $\beta(t)$ with no random effects. Thus, we demonstrate that we effectively recover the unconfounded coefficient functions.

%
%
%
%

%
%
%

\subsection{Aizoaceae}

\begin{figure}[H]
\begin{center}
\includegraphics[width=.4\textwidth]{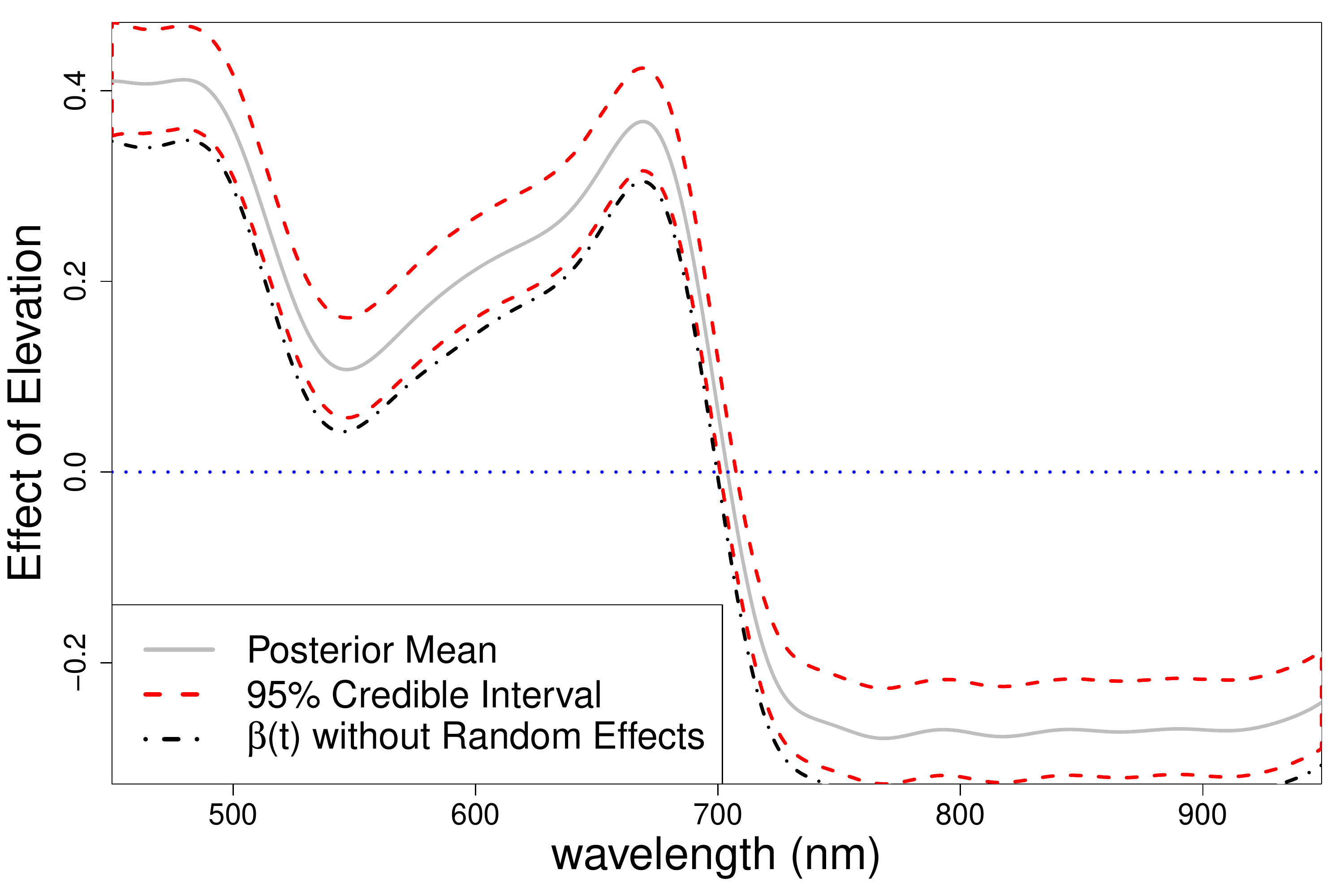}
\includegraphics[width=.4\textwidth]{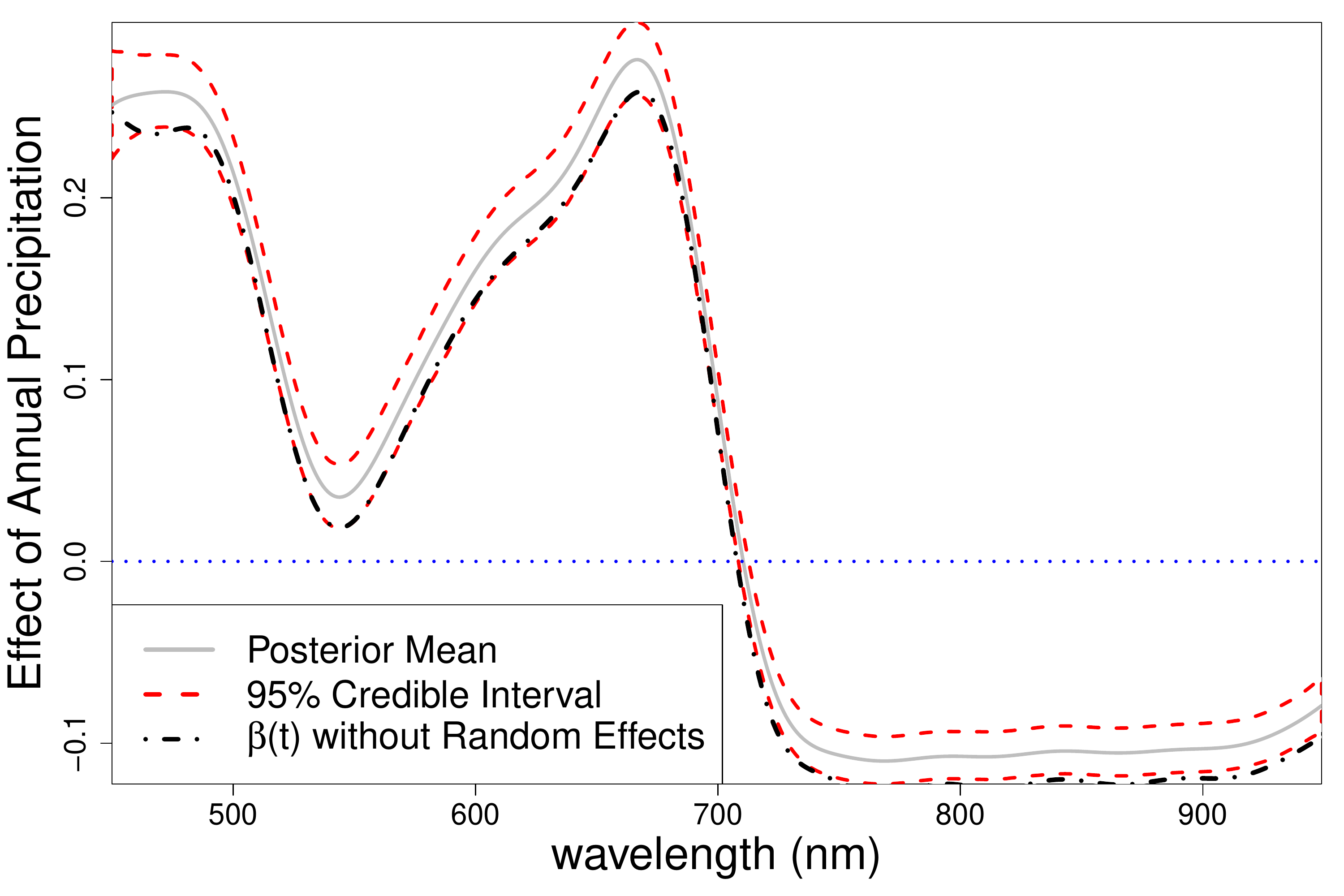}
\includegraphics[width=.4\textwidth]{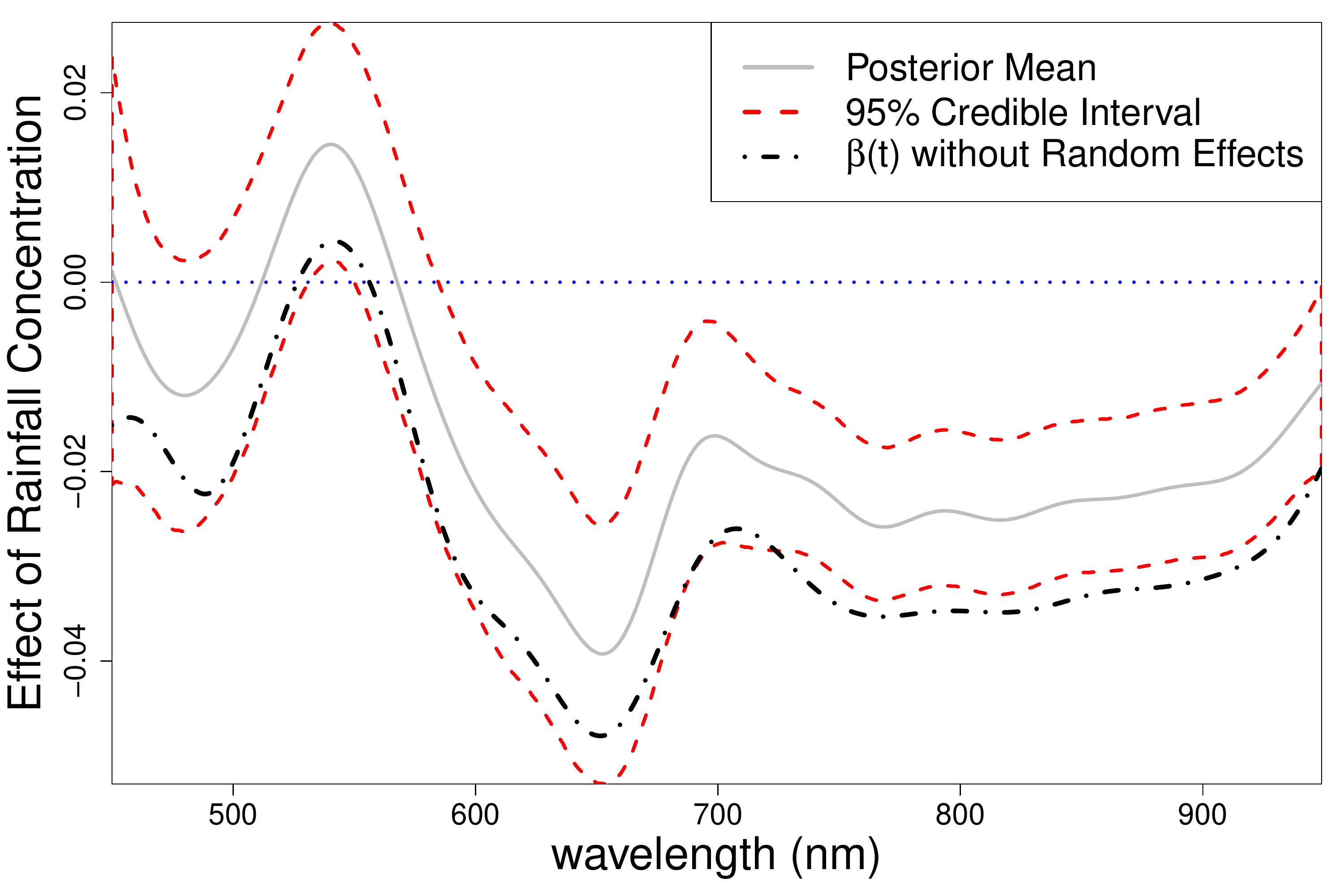}
\includegraphics[width=.4\textwidth]{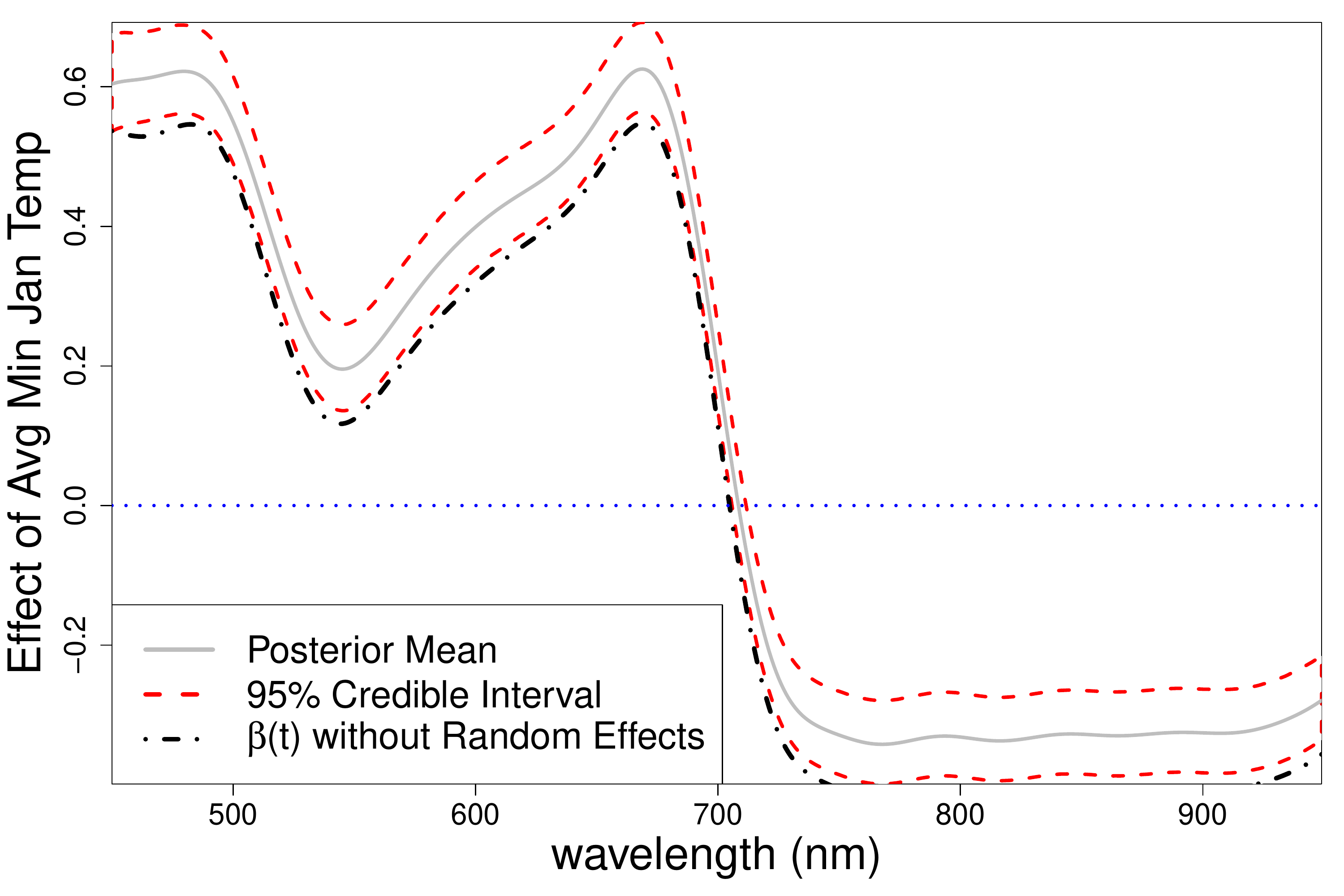}
\end{center}
\vspace{-4mm}

\caption{Posterior mean and 95\% credible intervals for functional regression coefficients for Aizoaceae.}\label{fig:aizo_beta_t}
\end{figure}

In the manuscript, we compared the unconfounded functional regression coefficients for Asteraceae (the most prevalent family in our dataset) to the coefficients with not random effects. Here, we present the posterior mean and pointwise 95\% credible intervals for the function regression coefficients for the effects of environmental variables on log-reflectance for Aizoaceae (See Figure \ref{fig:aizo_beta_t}). The coefficient functions for elevation, precipitation, and temperature all show a similar relationship to log-reflectance curves and appear to have a pattern similar to an inverted log-reflectance specturm. In general, increases in elevation, precipitation, and temperature correspond to higher log-reflectance at low-wavelengths and lower log-reflectance at higher wavelengths. The effect of rainfall concentration is quite small over the wavelength domain.



%

\subsection{Restionaceae}

\begin{figure}[H]
\begin{center}
\includegraphics[width=.4\textwidth]{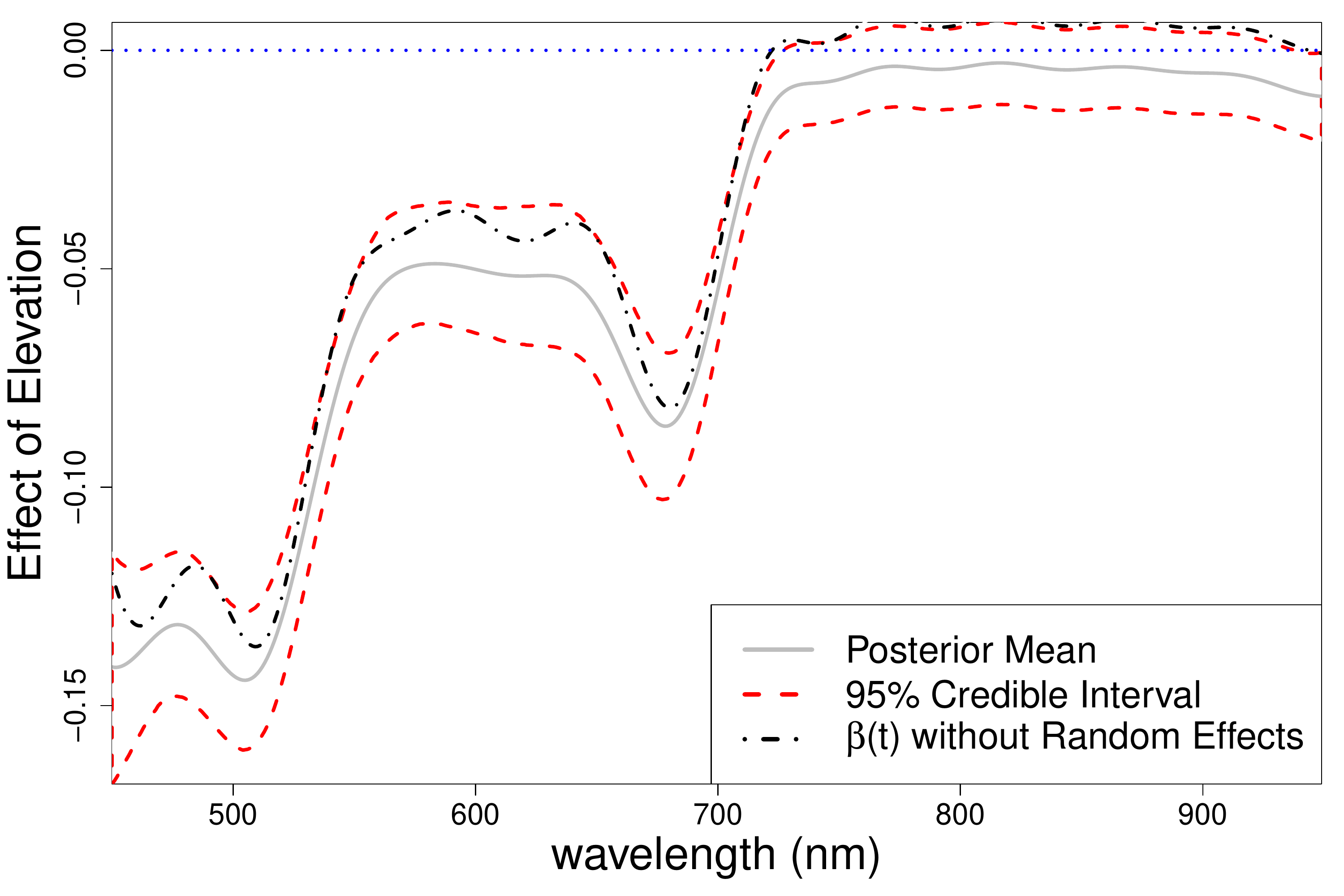}
\includegraphics[width=.4\textwidth]{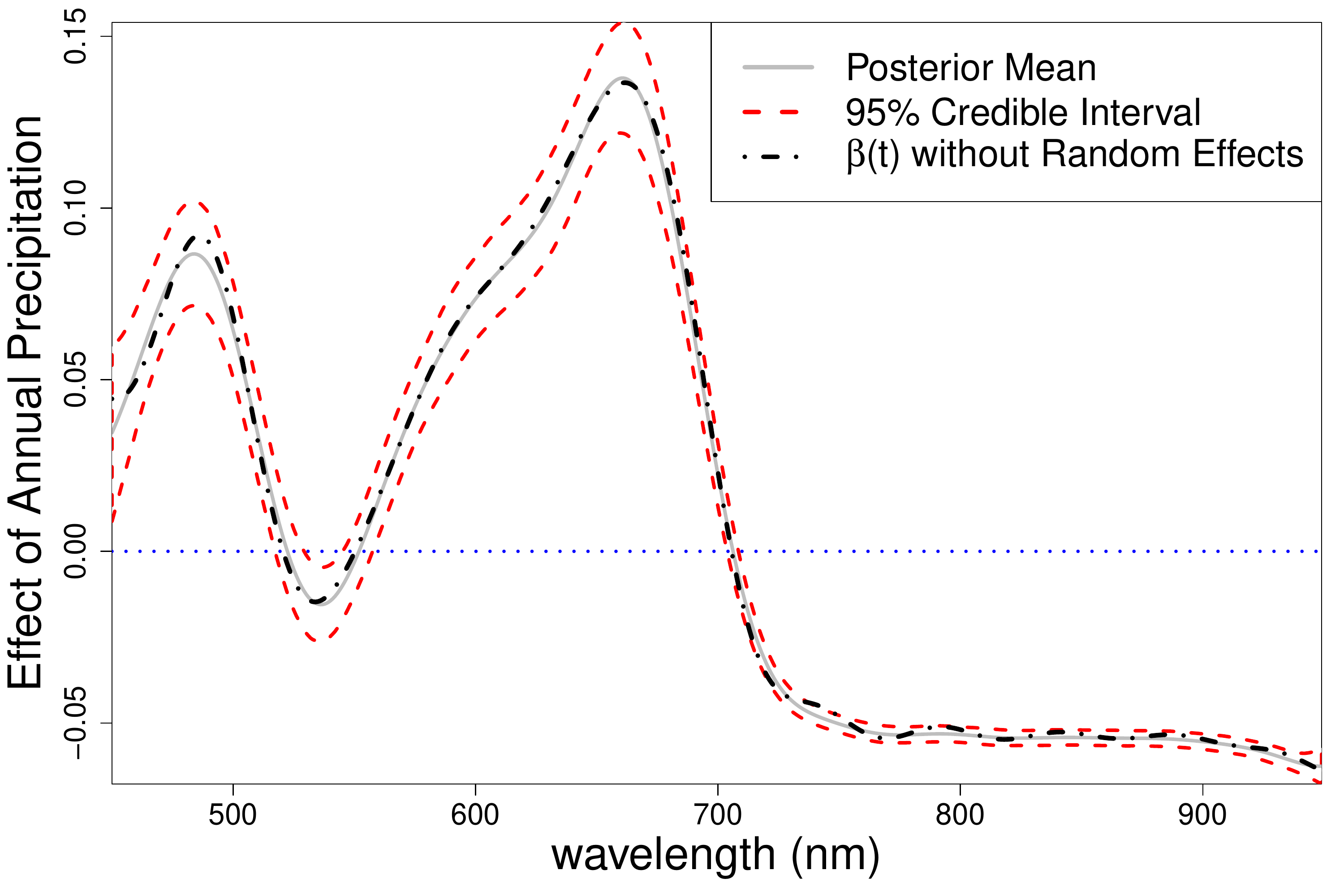}
\includegraphics[width=.4\textwidth]{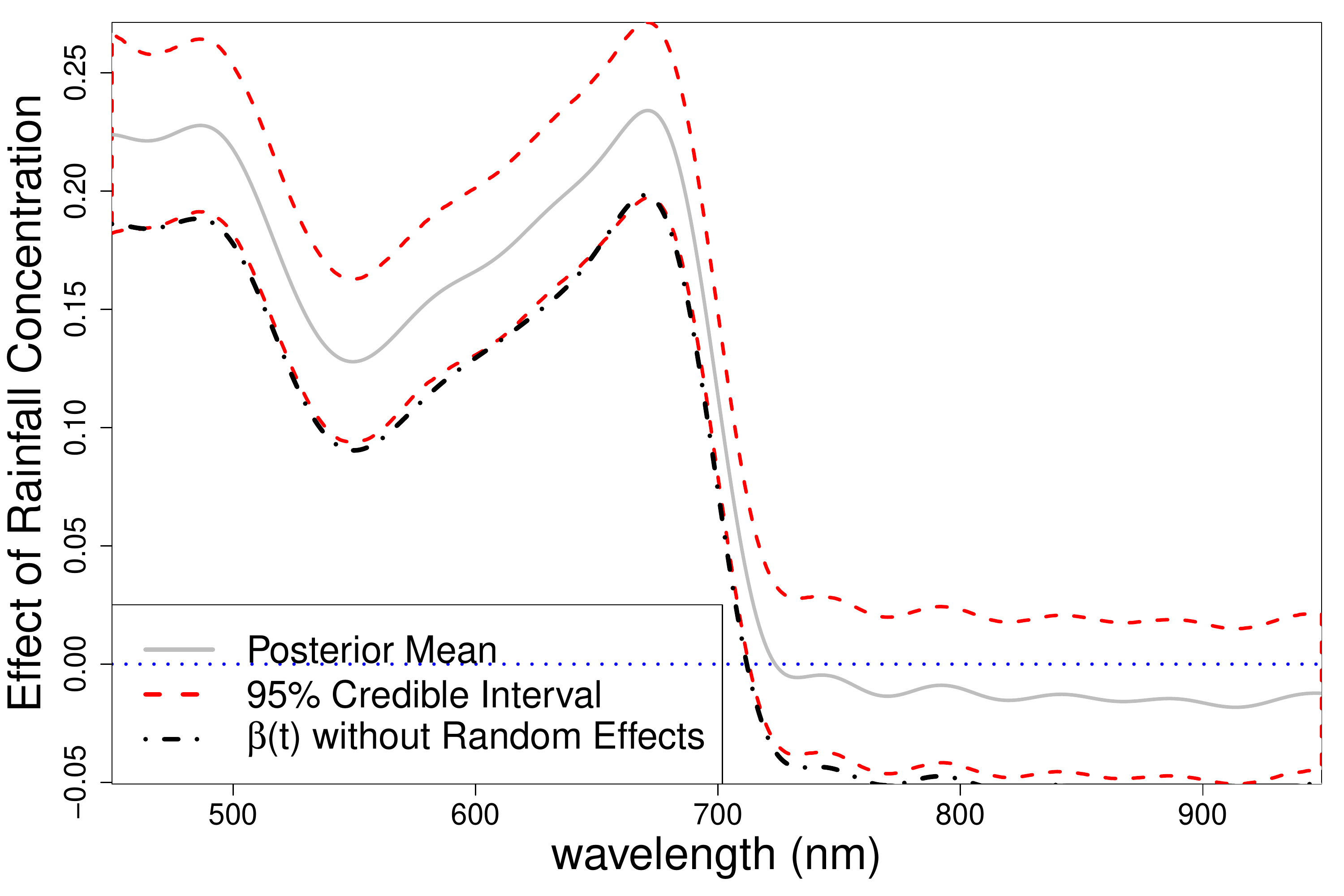}
\includegraphics[width=.4\textwidth]{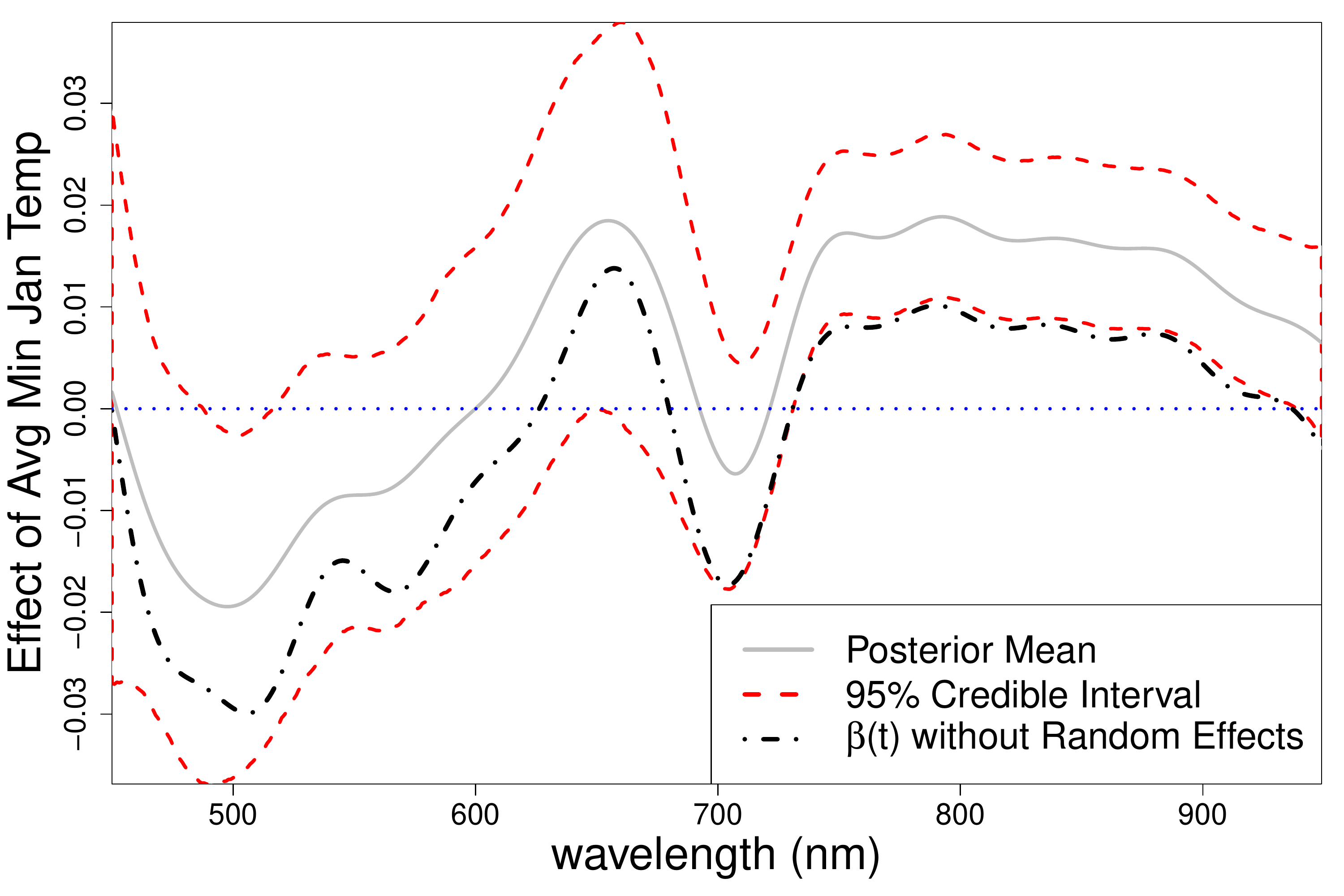}

\end{center}

\vspace{-6mm}

\caption{Posterior mean and 95\% credible intervals for functional regression coefficients for Restionaceae.}\label{fig:restio_beta_t}
\end{figure}

As we did with Aizoaceae, we present the posterior mean and pointwise 95\% credible intervals for the function regression coefficients for the effects of environmental variables on log-reflectance for Restionaceae (See Figure \ref{fig:restio_beta_t}). Unlike both Asteraceae and Aizoaceae, the coefficient functions follow less closely the form of the reflectance curve. The coefficient functions for elevation are negative for low wavelengths but near 0 for higher wavelengths. The coefficent functions for precipitation and rainfall concentrations are positive for most (or all) wavelengths less than 700 nm but negative or near 0 for wavelengths greater than 750 nm. The effect of temperature is week for all wavelengths.  


\bibliographystyle{apalike}
\bibliography{ref}